\pdfoutput=1
\documentclass[journal=mamobx,manuscript=article]{achemso}
\setkeys{acs}{articletitle = true}

\usepackage{amsmath}
\usepackage{graphicx}
\usepackage{mciteplus}
\usepackage{xcolor}

\newcommand{\Angle}{0} %0 for pdf and 270 for ps

\newcommand{\Comment}[1]{ {\bf \textcolor{red}{ #1 }}}

\SectionNumbersOn

\makeatletter

\title{Characteristic time and length scales in melts of Kremer-Grest bead-spring polymers with wormlike bending stiffness}
\author{Carsten Svaneborg}
\email{science@zqex.dk}
\affiliation{University of Southern Denmark, Campusvej 55, DK-5230 Odense M, Denmark}

\author{Ralf Everaers}
\affiliation{ Univ Lyon, ENS de Lyon, Univ Claude Bernard, CNRS, Laboratoire de Physique and Centre Blaise Pascal, F-69342 Lyon, France}

\begin{document}

\maketitle

{\huge \center Table of content graphic\\}

\begin{center}
{%
\setlength{\fboxrule}{5pt}
\setlength{\fboxsep}{0pt}
\fbox{\includegraphics[width=8.3cm]{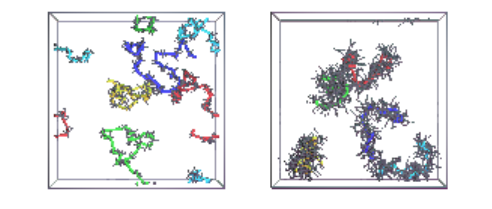}}
}%
\end{center}

\newpage 

\begin{abstract}
% 140 words max 150.
The Kremer-Grest (KG) model is a standard for studying generic polymer
properties. Here we have equilibrated KG melts up to and beyond $200$
entanglements per chain for varying chain stiffness.
We present methods for estimating the Kuhn length corrected for
incompressibility effects, for estimating the entanglement length corrected
for chain stiffness, for estimating bead frictions and Kuhn times
taking into account entanglement effects. These are the key parameters
for enabling quantitative, accurate, and parameter free comparisons
between theory, experiment and simulations of KG polymer models with
varying stiffness.  
We demonstrate this for the mean-square monomer displacements in
moderately to highly entangled melts as well as for the shear relaxation
modulus for unentangled melts, which are found to be in excellent
agreement with the predictions from standard theories of polymer dynamics.
%\Comment{write something stronger?}
\end{abstract}

\section{Introduction}

%%% 	Characteristic scales 
Polymeric materials share universal properties~\cite{Flory_53,flory1969statistical,degennes79,DoiEdwards86,RubinsteinColby,HassagerBird}, which depend on atomistic details only through a small number of interrelated characteristic time and length scales and which vary in a characteristic manner with the contour length, $L$, the molecular weight, $M \sim L$, or the number of monomers, $N \sim M \sim L$, of linear chains~\cite{Flory_53,flory1969statistical,degennes79,DoiEdwards86,RubinsteinColby,HassagerBird}.
For example, the mean-square end-to-end distance in a melt varies like, $\langle R^2 \rangle =l_K L$, as chains with a contour length exceeding the Kuhn~\cite{Kuhn} length, $l_K$, adopt random walk conformations~\cite{flory49}. 
%while the longest intra-chain relaxation time, $\tau_{max}$, corresponds to the time required by the chains to diffuse over a distance of their size~\cite{DoiEdwards86}.
%Topological constraints dominate the chain dynamics beyond the entanglement scale, which is defined by a spatial distance, $d_T$, or ``tube diameter'', a corresponding entanglement time, $\tau_e$, required by monomers to diffuse over this scale, and a characteristic entanglement (contour) length or weight, $L_e \sim M_e \sim N_e$.
%In particular, two polymer melts are expected to show identical rheological behaviour, if the (linear) chains have the same {\em effective} length, $Z=L/L_e=M/M_e=N/N_e$.
%% with a longest intra-chain relaxation time, $\tau_{max} \sim Z^3 \tau_e$ \cite{DoiEdwards86}.
%The smallest characteristic scale is the Kuhn scale \cite{Kuhn}. The Kuhn length, $l_K$, is implicitly defined by a mapping of the chain statistics to a freely-jointed chain model composed of $N_K$ Kuhn steps, which reproduces the contour or end-to-end distance at full extension, $L=l_K N_K$, and the mean-square end-to-end distance, 
%%
%\begin{equation}\label{eq:Kuhnlengththeory}
%\langle R^{2}\rangle=l_{K}^2 N_K,
%\end{equation}
%of the target polymers. 
%Intrinsically flexible polymers exhibit {\em universal} behavior~\cite{degennes79,DoiEdwards86,RubinsteinColby} beyond the Kuhn scale, while behavior on smaller scales is material specific and dependent on atomic details. 

%%% 	Analytic theory
Analytical theory~\cite{Flory_53,flory1969statistical,degennes79,DoiEdwards86,RubinsteinColby,HassagerBird} distills understanding of the complex emergent behavior in terms of simplifying phenomenological models of the large scale conformational statistics and dynamics of densely packed, interpenetrating, random walk-like chains. 
Computer simulations\cite{binder1995monte,padding2002time,muller2002coarse,tzoumanekas2006atomistic,MultiscalePeterKremerSOftMatter2009,MultiscalePeterKremerFaradayDiss2010,maurel2015prediction}  play an increasingly important, complementary role 
%to theory and experiment 
by providing insight~\cite{PPA} and by validating phenomenological theories~\cite{hou2010stress} for well-defined model systems. 
%with similar {\em emergent} universal properties as real commodity polymer melts.
%With theory being essential in guiding the setup and analysis of simulations, an important qualitative difference becomes clear in the study of melts of non-concatenated ring polymers, whose {\em static} properties differ from their linear counterparts in a non-trivial way~\cite{NechaecKhokhlov85,Rubinstein86}.
%This difference naturally emerges in simulations of ring polymers without the need to adjust the chain model as such~\cite{references}. In contrast, a corresponding phenomenological description needs to be developed~\cite{NechaecKhokhlov85,Rubinstein86,Grosberg_SoftMatter14,SmrekGrosberg_dynamics}  and validated \cite{rosa_prl_2014,Schram_SoftMatter_inPress} from scratch.
%
Here we are interested in the standard Molecular Dynamics (MD) model for simulating the structure and dynamics of polymer melts, the bead-spring model of Kremer and Grest (KG)~\cite{grest1986molecular,kremer1990dynamics}. 
In the KG model approximately hard sphere beads are connected by strong non-linear springs, generating the connectivity and the liquid-like monomer packing characteristic of polymer melts. 
The model is formulated in terms of the microscopic energy scale, $\epsilon$, bead diameter, $\sigma$,  mass, $m_b$, and time $\tau=\sigma\sqrt{m_{b}/\epsilon}$ of the Lennard-Jones interactions between the beads.
The parameters are tuned to energetically prevent polymer chains from passing through each other. 
Since the model reproduces the local topological constraints, which dominate the dynamics of long-chain polymers~\cite{kremer1990dynamics}, non-trivial large scale entanglement properties {\em emerge} through the exact same mechanisms as in real polymer melts.
The KG model was devised to study the generic properties of polymers with an emphasis on simplicity and computational efficiency.
It has been used to investigate the effects of polymer entanglements, branching, chain polydispersity, and/or chemical cross-linking, see Refs. \cite{grest1989relaxation,grest90a,duering91,duering94,svaneborg2004strain,svaneborg2008connectivity,Halverson2011molecular,rosa2014ring} for examples. %%They recognized that real polymers have chemistry specific local interactions
%%between monomers that gives rise to a varying degree of stiffness of the polymer
%%chains. Hence they they augmented the KG model with an additional potential
%%depending on the angle between neighboring bonds, which allows the stiffness
%%of KG model polymers to be varied.
The utility of generic models is not limited to bulk materials. KG-like models have also been used to study universal aspects of the behaviour of  tethered and spatially confined polymers, of welding of polymer interfaces or of composite materials formed by adding filler particles to a polymer melt or solid, see Refs. \cite{MuratGrest,murat89a,grest95,grest96a,aoyagi2001molecular,gersappe2002molecular,pierce2009interdiffusion,yagyu2009coarse,sussman2014entanglement,ge2014healing} for examples.
%Given the increase in available computing power, KG simulations are now for many purposes a viable {\em alternative} to experiments.
%All of these problems are described by suitably generalized phenomenological models, which help to guide the setup and analysis of simulations. These can in turn serve to develop and validate theories, since {\em emergent} phenomena can be studied without adjusting the chain model as such.

%
% and validated from scratch
%
%Clearly theory is essential in guiding the setup and analysis of numerical studies. 
%
%The advantage of simulations is that the investigated phenomena naturally {\em emerge} in the above studies with no need to adjust the chain model as such, while suitably generalized phenomenological description need to be developed and validated from scratch.

%
%
%With theory being essential in guiding their setup and analysis, the point is that the investigated phenomena naturally {\em emerge} in simulations with no need to adjust the chain model as such, while suitably generalized phenomenological description need to be developed and validated from scratch.

Here we present an accurate characterization and parameterization of the key characteristic time and length scales of the KG model: the Kuhn length, $l_K$, the number of Kuhn segments, $N_{eK}$, between entanglements, the Kuhn segmental friction, $\zeta_K$ and the associated Kuhn time, $\tau_K$, and entanglement time $\tau_e$. 
In particular, we study their variation with the strength of an additional bending potential introduced by Faller and M\"uller-Plathe\cite{faller1999local,faller2000local,faller2001chain}.
Our results are based on the analysis of well equilibrated melts of KG chains of varying stiffness, which cover the entire range from unentangled to highly entangled systems with up to and beyond $Z=200$ entanglements per chain. 
They allow us to address three different types of questions:
\begin{enumerate}
\item With respect to the KG model itself, we seek to establish and better understand how the additional bending term affects the emergent length and time scales. 
%In particular, we provide reliable interpolations for the stiffness dependence of the Kuhn length, $l_K$, the Kuhn friction, $\zeta_K$, the Kuhn time, $\tau_K$, the entanglement length, $N_{eK}$, and the entanglement time, $\tau_e$.
%We discuss the rapid closure of the gap between the Kuhn time, $\tau_K$, and the entanglement time,  $\tau_e$, with increasing chain stiffness; we ask, if the effective bending stiffness can be understood by only taking into account local effects; and we find, that the viscosity $\eta_K$ at the Kuhn scale hardly varies across the considered systems. 
%This is similar to the experimental interest in carefully characterizing suitable reference systems in understanding the consequences of, say, the replacement of a hydrogen atom by a methyl group {\bf Better idea for defining a family of chemically similar systems?}.  
This is similar to the experimental interest in carefully characterizing suitable reference systems or families of chemically similar
systems with tunable microscopic structure\cite{lutz2013sequence,gody2013rapid} to allow for systematic experimental studies of structure-property relations.

%
%such as random copolymers["Architecturally Complex Polymers with Controlled Heterogeneity", Krzysztof Matyjaszewski, Science  2011:
%Vol. 333, Issue 6046, pp. 1104-1105] or sequence-controlled polymers[Sequence-Controlled Polymers Jean-François Lutz1,*, Makoto Ouchi2, David R. Liu3, Mitsuo %Sawamoto, Science, Vol. 341, Issue 6146, 1238149] where the microscopic structure can be tuned to some extent. 
%

\item With respect to experiment, a follow-up article~\cite{svaneborgKGmapping} introduces ``Kuhn scale-mapped KG models'' as minimal material-specific polymer models by matching (i) the Kuhn number, $n_K=\rho_K l_K^3$, as a dimensionless measure of density and (ii) the number of Kuhn segments per chain, $N_K$, as a dimensionless measure of chain length. We show that the family of models investigated here covers the full range relevant for melts of commodity polymers. The standard KG model maps onto the intrinsically most flexible experimental systems (PDMS and poly-isoprene), which are characterized by the smallest Kuhn number. These systems come closest to the theoretical ideal of an infinite number of Kuhn segments, $N_{eK}$, between entanglements. Melts of stiffer polymers like poly-ethylene or poly-carbonate have higher Kuhn numbers and only a finite number of Kuhn segments per entanglement length. Corresponding Kuhn scale-mapped KG models account for this effect and can be expected to reproduce deviations from idealised theoretical predictions. As a simple example, the reader may think of the finite maximal elongation of entanglement segments. 

\item With respect to polymer theory, we note that expressions for experimental observables are usually formulated in units defined by the relevant time and length scales in combination with $k_BT$ as the typical energy scale for entropy dominated phenomena. Knowledge of these fundamental scales thus puts us in a position to carry out parameter-{\em free} tests of theoretical predictions for a wide and representative range of generic polymer models. Theoretical predictions are often derived for limiting cases. Deviations (such as static and dynamic effects due to chain stiffness, finite number of Kuhn segments between entanglements, incompressibility, as well as end and entanglement effects on chain friction) observed for the present model systems do {\em not} constitute artifacts of an arbitrarily defined computational polymer model. Given the mapping to experimental systems they should rather be thought of as representative for real polymers.
%Firstly, we can explore the relation between independently measured scales within a homogeneous family of models. In particular, we consider how the number of Kuhn segments per entanglement length, $N_{eK}$, varies with the Kuhn number, $n_K=\rho_K l_K^3$, of Kuhn segments per Kuhn length cube. 
%Secondly, we can test predictions of polymer theory for independent observables like the shear-relaxation modulus, $G(t)$, of unentangled chains, or the monomer mean-square displacement as a function of time, $MSD(t)$, of entangled chains. 
\end{enumerate}

%To improve the readability, we have divided the material into two separate papers.
%The accompanying article~\cite{svaneborgKGmapping} focuses on the second aspect. 
%%and introduces ``Kuhn scale-mapped KG models'' as minimal material-specific polymer models. 
%%To make the article self-contained, we have included a concise theory section, where we summarize well-known key relations between the characteristic time and length scales in polymer melts, which are needed to understand the idea behind our approach and to judge the applicability range of the resulting models. 
%%As a first test, we consider how the number of Kuhn segments per entanglement length, $N_{eK}$, varies with the Kuhn number, $n_K=\rho_K l_K^3$, of Kuhn segments per Kuhn length cube. 
%%
%Here we concentrate on the first and third, intimately entwined aspects of the list above.

The paper is structured as follows. 
In Sec.~\ref{sec:theory} we introduce the Kremer-Grest model and provide the necessary theoretical background: we (i) define the targeted characteristic time and length scales, (ii) relate them to computational observables and (iii) discuss corrections to the ideal behavior assumed in their definition. 
Section~\ref{sec:Methods} presents the methods we use to set up and simulate our systems and to analyze the raw data.
For example, our estimator for the Kuhn length accounts for long-range correlations induced by the melt incompressibility~\cite{Wittmer_Meyer_PRL04,semenov2010bond} and
our PPA~\cite{PPA} estimator for the entanglement length~\cite{uchida2008viscoelasticity} takes into account the finite number of Kuhn segments per entanglement. 
A key point is our strategy to estimate the effective bead friction, which uses theoretical guidance to disentangle the influence of end effects~\cite{pearson1994viscosity,fuchs1997polymer,fuchs1997polymer2,lodge1999reconciliation}, inertia, local chain stiffness, the correlation hole~\cite{Semenov2012PRE}, viscoelastic hydrodynamic interactions~\cite{Semenov2011PRL,Semenov2012PRE2,Semenov2012JPhys}, and entanglements~\cite{DoiEdwards86}.
In Sec.~\ref{sec:results} we present our results. In particular, we extract the Kuhn and entanglement lengths as well as the effective bead friction in KG melts. 
In the following Discussion in Sec.~\ref{sec:Discussion}, we derive and discuss the corresponding Kuhn and entanglement times. 
The phenomenological interpolations for the stiffness dependence of the Kuhn length, $l_K$, the Kuhn friction, $\zeta_K$, the Kuhn time, $\tau_K$, the entanglement length, $N_{eK}$, and the entanglement time, $\tau_e$ form the basis of our parameterization of ``Kuhn scale-mapped KG models'' in the accompanying paper \cite{svaneborgKGmapping}.
Here we focus on the third point of the above list and present {\em parameter-free} comparisons between our data and the predictions of the Rouse and tube models of polymer melts.
The two models work remarkably well on scales, where KG chains can be described as Gaussian chains.
% A bit too strong:
% In particular, they {\em quantitatively} 
% predict the amplitude of the characteristic $t^{1/4}$-regime in the monomer diffusion.
On smaller scales, the Rouse model describes the dynamics of unentangled chains only qualitatively. 
Deviations become more pronounced for stiffer chains, as the gap between the Kuhn and the entanglement scales closes rapidly with increasing bending rigidity.  
%
%We ask, if the effective bending stiffness can be understood by only taking into account local effects; we find, that the viscosity $\eta_K$ at the Kuhn scale hardly varies across the considered systems; and we 
%discuss the rapid closure of the gap between the Kuhn time, $\tau_K$, and the entanglement time,  $\tau_e$, with increasing chain stiffness.
%Furthermore, we present parameter-free tests of predictions for 
%the entanglement length, $N_{eK}$ as a function of the Kuhn number, $n_K=\rho_K l_K^3$, for the shear-relaxation modulus, $G(t)$, of unentangled chains, and for the monomer mean-square displacement as a function of time, $MSD(t)$, of entangled chains. 
In the final Sec. \ref{sec:Conclusion}, we summarise and present our conclusions.

\section{Model and background\label{sec:theory}}

Polymeric materials share universal properties, which depend on atomistic details only through a small number of interrelated characteristic time and length scales. In the following we present and define these scales together with the Kremer-Grest (KG) model.  The material is essentially standard with the exception of the phantom KG simulations (Sec.~\ref{sec:phantom KG model}) and the {\em definition} of the Kuhn time, $\tau_K$, based on the Rouse model (Sec.~\ref{sec:Rouse Dynamics}).
A detailed discussion of our estimators for the various time and length scales can be found in the following Methods section.

%Polymeric materials share universal properties, which depend on atomistic details only through a small number of interrelated characteristic time and length scales. In the following, we (i) present these scales,  (ii) relate them to computational observables and (iii) propose corrections to these relations, which account for deviations from the ideal behavior assumed in the scales' definitions. To set the stage, we recall the definition of the Kremer-Grest model, for which we want to determine these scales.

\subsection{The Kremer Grest Model\label{sec:KGModel}}

The KG model~\cite{grest1986molecular,kremer1990dynamics} is a bead-spring model,
where the mutual interactions between all beads are given by the truncated Lennard-Jones or
Weeks-Chandler-Anderson (WCA) potential, 
\begin{equation}\label{eq:WCA}
U_{WCA}(r)=4\epsilon\left[\left(\frac{\sigma}{r}\right)^{-12}-\left(\frac{\sigma}{r}\right)^{-6}+\frac{1}{4}\right]\quad\mbox{for}\quad r<2^{1/6}\sigma,
\end{equation}
where $\epsilon=k_{B}T$ and $\sigma$ are chosen as the simulation units of energy and distance, respectively. 
Bonded beads additionally interact through the
finite-extensible-non-linear elastic spring (FENE) potential,
\begin{equation}\label{eq:FENE}
U_{FENE}(r)=-\frac{kR^{2}}{2}\ln\left[1-\left(\frac{r}{R}\right)^{2}\right].
\end{equation}
We adopt the standard choices for the Kremer-Grest model of $R=1.5\sigma$ for the maximal bond length, $k=30\epsilon\sigma^{-2}$ for the spring constant, and $\rho_{b}=0.85\sigma^{-3}$ for the bead density. 
Here and below we use subscript ``b'' to denote bead specific properties to distinguish these from Kuhn units. 
Following Faller and M\"u{}ller-Plathe~\cite{faller1999local,faller2000local,faller2001chain} we add an entropic ``wormlike''~\cite{KratkyPorod} bending potential defined by
\begin{equation}\label{eq:Ubend}
U_{bend}(\Theta)=\kappa\, k_B T \left(1-\cos\Theta\right),
\end{equation}
where $\Theta$ denotes the angle between subsequent bonds and $\kappa$ is a dimensionless 
bending stiffness. Temperature-independent bending potentials of this form are routinely used to model semi-flexible polymers. For other possible
choices of bending potentials, see e.g. Ref.~\cite{rosa2010looping}. 
The average bond length is $l_{b}=0.965\sigma$, so that the contour length of a linear chain composed of $N_b$ beads is given by $L=(N_b-1) l_b$.
The variation of $0.2\%$ of the bond length over the considered $\kappa$-range is so small, that we may neglect the effect for the present purposes.
%The WCA interactions between next-nearest beads define a minimum bending angle between subsequent bonds. 
%%% You energy is minimal, when $\Theta=0$. 
The WCA interactions between next-nearest beads define a maximum bending angle between subsequent bonds. 
In simulations with negative $\kappa$ this effect is partially compensated, resulting in KG melts that are somewhat more flexible than the standard KG model.

As in the original KG papers~\cite{grest1986molecular,kremer1990dynamics}, we
integrate Langevin equations of motion
\begin{equation}\label{eq:Langevin}
m_{b}\frac{\partial^{2}{\bf R}_{i}}{\partial t^{2}}=-\nabla_{{\bf R_{i}}}U-\Gamma\frac{\partial{\bf R}_{i}}{\partial t}+{\boldsymbol \xi}_{i}(t)
\end{equation}
where ${\bf R}_{i}$ denotes the position of bead $i$ and $U$ the total potential
energy. ${\boldsymbol \xi}_{i}(t)$ is a Gaussian distributed random vector with
$\langle{\boldsymbol \xi}_{i}(t)\rangle=0$ and $\langle{\boldsymbol \xi}_{i}(t)\cdot{\boldsymbol \xi}{}_{j}(t')\rangle=6\Gamma k_{B}T\delta(t-t')\delta_{ij}$.
The mass of a bead is denoted $m_{b}$, and we choose this as our mass scale
for the simulations. From the units defined so far, follows that time in KG simulations is measured in units of $\tau=\sigma\sqrt{m_{b}/\epsilon}$.

\subsection{Large scale structure and Kuhn length}\label{sec:Kuhn length}

The most important measure of the overall chain size is the mean-square end-to-end distance, $\langle R^2 \rangle$. For many polymeric systems,  it varies in a characteristic manner,  $\langle R^2 \rangle \sim L^{2\nu} \sim M^{2\nu} \sim N^{2\nu}$,
with the contour length, $L$, the molecular weight, $M$, the number of monomers, $N$, or any other measure of the length of linear chains~\cite{Flory_53,flory1969statistical,degennes79,DoiEdwards86,RubinsteinColby,HassagerBird}. 
In the melt state, polymers adopt random walk conformations~\cite{flory49} with $\nu=1/2$. In this case, the gyration radius is given by $\langle R_g^2 \rangle = \langle R^2 \rangle/6$.

%\subsection{Kuhn length\label{sec:Kuhn length}}

%The smallest characteristic scale is the Kuhn scale \cite{Kuhn}. 
For chains characterized by $\nu=1/2$, the Kuhn length~\cite{Kuhn}, $l_K$, is defined by a mapping to a freely-jointed chain model composed of $N_K$ Kuhn steps, which reproduces the mean-square end-to-end distance and the end-to-end distance at full extension, 
\begin{eqnarray}
\label{eq:R2 Kuhn}
\langle R^{2}\rangle&=&l_{K}^2 N_K\\
\label{eq:L Kuhn}
L &=&l_{K} N_K\ ,
\end{eqnarray}
of the target polymers. 
Beyond the Kuhn scale the behavior of polymer chains is dominated by thermal fluctuations and is {\em universal}~\cite{degennes79,DoiEdwards86,RubinsteinColby}. 
Below the Kuhn scale chains are rigid, $\langle R^{2}\rangle\propto L^2$. Their behavior is material specific and dependent on atomic details. 
Given $l_K$, it is straightforward to obtain the number of Kuhn segments per chain, $N_{K}=\langle R^{2}\rangle/l_{K}^{2}=L^{2}/\langle R^{2}\rangle$. The number density of Kuhn segments is $\rho_K=N_K\rho_c$ where $\rho_c$
denotes the number density of chains.
%The Kuhn length is the fundamental length scale characterizing chain
%configurations beyond the monomer scale. Above the Kuhn length the chains
%adopt random walk conformations, on shorter length scales they behave as
%rigid rods.

For freely rotating (bead-spring) chains with the bending potential, Eq.~(\ref{eq:Ubend}), the bare Kuhn length in the absence of excluded volume interactions is given by~\cite{flory1969statistical,auhl2003equilibration}:
\begin{eqnarray}
\label{eq:lK_definition}
l_{K}&=&l_b \frac{1+\langle \cos(\theta)\rangle}{1-\langle \cos(\theta) \rangle}\\
\label{eq:lK_bare}
l_{K}^{(0)}(\kappa)& =& l_b \times
 \begin{cases}
  \frac{2 \kappa +e^{-2\kappa }-1}{1-e^{-2 \kappa } (2 \kappa +1)} & \text{if $\kappa \neq 0$} \\
  1 & \text{if $\kappa=0$}
\end{cases}
\end{eqnarray}
The Kratky-Porod wormlike chain expression~\cite{KratkyPorod}:
\begin{equation}\label{eq:R2 WLC}
\frac{\langle R^{2}(N_K)\rangle}{l_K^2 N_K } = 1 -\frac{1}{2N_K}\left( 1-e^{-2 N_K} \right)\ ,
\end{equation}
provides a convenient interpolation between the rigid rod and random walk limits. For chains following random walk statistics on larges scales, equations~(\ref{eq:R2 Kuhn}) and (\ref{eq:L Kuhn}) suggests to define the Kuhn length as
\begin{equation}\label{eq:Kuhnlengththeory}
l_{K}=\lim_{L\rightarrow\infty}\frac{\langle R^{2}\rangle}{L}.
\end{equation}
Note, however, that $l_K$ can in general not be estimated by fitting Eq.~(\ref{eq:R2 WLC}) to (simulation) data for chains in a melt, since this expression neglects long-range bond orientation correlations due to the incomplete screening of excluded volume interactions. ~\cite{Wittmer_Meyer_PRL04,wittmer2007polymer,wittmer2007intramolecular,beckrich2007intramolecular,semenov2010bond} (see Sec.~\ref{sec:Kuhn length estimator} for a more detailed discussion).

\subsection{Langevin dynamics}\label{sec:single chain Langevin dynamics}

%The standard Rouse and tube models for flexible polymers are based on two {\em independent} approximations: the Gaussian chain model and the assumption, that the (local) dynamics can be described by a single-chain Langevin equation, Eq.~(\ref{eq:overdampedLangevin}). 
%

The standard theory of polymer dynamics in the melt state~\cite{DoiEdwards86} describes the coarse-grain chain motion through an overdamped {\em single-}chain Langevin equation, 
\begin{equation}\label{eq:overdampedLangevin}
0=-\nabla_{{\bf R_{i}}}\tilde{\cal H}-\frac{\zeta_{cm}}N\frac{\partial{\bf R}_{i}}{\partial t}+{\boldsymbol \xi}_{i}(t)\ .
\end{equation}
for the retained $N$ spatial degrees of freedom, ${\bf R}_{i}$.  $\tilde{\cal H}$ denotes a corresponding effective single-chain Hamiltonian to be inferred from the chain statistics in the fully interacting system, i.e. the potential of mean force felt by a chain\cite{kirkwood1935statistical}. Eq.~(\ref{eq:overdampedLangevin}) implies a constant effective friction $\zeta_{cm}/N$ per bead. 
${\boldsymbol \xi}_{i}(t)$ are normally distributed random vectors with statistics characterized by
$\langle{\boldsymbol \xi}_{i}(t)\rangle=0$ and $\langle{\boldsymbol \xi}_{i}(t)\cdot{\boldsymbol \xi}{}_{j}(t')\rangle=6\frac{\zeta_{cm}}N k_{B}T\delta(t-t')\delta_{ij}$. Within this description, the chain centers-of-mass ${\bf R}_{cm}(t)$ exhibit simple diffusion, 
%On scales beyond the gyration radius chains exhibit simple diffusion, i.e. the CM mean-square displacements are given by
%
\begin{equation}\label{eq:cmdiffusion}
\lim_{t\rightarrow\infty} \langle \left( {\bf R}_{cm}(t) -  {\bf R}_{cm}(0) \right)^2\rangle
\equiv \lim_{t\rightarrow\infty} g_3(t)
= 6 D_{cm} t = 6 \frac{k_BT}{\zeta_{cm}} t\ ,
\end{equation}
at {\em all} times, i.e. $\zeta_{cm}/N$ is the effective friction per retained degree of freedom in the fully interacting system.
% with a CM friction coefficient, $\zeta_{cm} = N_b \zeta_b$. 
The largest conformational relaxation time, $\tau_{max} \propto R_g^2/D_{cm}$, is of the order of the time required by the chains to diffuse over a distance comparable to their size.

 \subsection{Phantom KG chains as single-chain reference model for the dynamics}\label{sec:phantom KG model}

In the present context, it is natural to retain the bead degrees of freedom of the target KG chains, and to study the Langevin dynamics (including the inertia term from Eq.~(\ref{eq:Langevin})) of ``phantom KG chains'':
%
%
%  WHY NOT phantom KG model?
%
\begin{equation}\label{eq:PhantomKGLangevin}
m_{b}\frac{\partial^{2}{\bf R}_{i}}{\partial t^{2}}=-\nabla_{{\bf R_{i}}}\tilde{\cal H}-\zeta_{b}\frac{\partial{\bf R}_{i}}{\partial t}+{\boldsymbol \xi}_{i}(t)\ .
\end{equation}
We define the effective Hamiltonian, $\tilde{\cal H}$,  through the same functional form, Eqs.~(\ref{eq:WCA}) to (\ref{eq:Ubend}), for the bonded interactions as for the full KG model.
In contrast, non-bonded intra- and interchain excluded volume interactions, which are largely screened in melts~\cite{flory49}, are neglected.  
The effective bending stiffness, $\tilde\kappa(\kappa)$, of the phantom KG chains is defined through the condition, that their bare Kuhn length, Eq.~(\ref{eq:lK_bare}), has to reproduce the measured Kuhn lengths of the KG chains in the target melts, $l_{K}^{(0)}(\tilde\kappa)\equiv l_{K}(\kappa)$. The value of the effective bead friction, $\zeta_b=\zeta_{cm}/N_b$, essentially sets the time scale of the polymer motion. The crossover from the initial ballistic to the diffusive regime occurs at $t = m_b/\zeta_b$ for both the beads and the chain CM, whose diffusion constant is given by $D_{cm}=k_BT/\zeta_{cm}$ with $\zeta_{cm}=\zeta_b N_b$.
In KG melts, $\zeta_b\gg \Gamma$, i.e. the effect of the thermostat is small compared to the friction, which arises from the interactions between the beads~\cite{kremer1990dynamics}. One of our tasks is to determine appropriate values of $\zeta_b$ as a function of the stiffness parameter, $\kappa$.

\subsection{Rouse dynamics\label{sec:Rouse Dynamics}}

The Rouse-model~\cite{rouse1953theory} describes the Brownian dynamics, Eq.~(\ref{eq:overdampedLangevin}), beyond the Kuhn scale. 
Flexible polymers are described as ``Gaussian'' chains,
\begin{equation}\label{eq:HRouse}
\tilde{\cal H} = \frac12 k_R  \sum_{i=1}^{N_R} \left({\bf R}_{i+1}-{\bf R}_{i} \right)^2
\end{equation}
composed of $N_R+1$ beads with $N_R \sim L \sim M \sim N_K$. The beads are connected by harmonic springs with spring constant $k_R=3 N_R k_BT/\langle R^2\rangle$. 
%With $\zeta=\zeta_{cm}/(N_R+1)$ (or, more intuitively, $\zeta_i=\zeta_{cm}/N_R$ for $2\le i \le N_R$ and $\zeta_1 = \zeta_{N_R+1}=\frac12 \zeta_{cm}/N_R$), the total friction for the CM motion is uniformly distributed over the chain.
%The bead friction is assumed to be independent of chain length, $\zeta_{cm}  \sim L \sim M \sim N_K$.

Rouse theory represents the internal chain dynamics in terms of a set of $p=1,\dots,N_R$ Rouse modes. 
While the Rouse model is a coarse-grain description, $N_R\ll N_K$, it is typically solved in the $N_R\rightarrow\infty$ continuum limit, where each mode has a characteristic relaxation time $\tau_p=\tau_R/p^{2}$. 
%The mode with $p=N_K$ has the shortest relaxation time equal to the Kuhn time, $\tau_K$. 
The relaxation time of the mode with $p=1$ defines the Rouse time, 
\begin{equation}\label{eq:Rouse-time}
\tau_R=\frac1{3\pi^2} \frac{\zeta_{cm} \langle R^2 \rangle}{k_BT}\sim N_K^2\ .
\end{equation}
%According to the Rouse model, the chain end-to-end vector, $\bf R$, essentially decorrelates on scale of the Rouse time,
%
%\begin{eqnarray}\label{eq:P(t) Rouse}
%P(t)\equiv \langle {\bf R}(t)\cdot{\bf R}(0) \rangle &=& \langle R^2 \rangle \sum_{p=1,3,\ldots}^\infty \frac{8}{\pi^2 p^2} \exp\left(-\frac{t}{\tau_p}\right)\ ,
%\end{eqnarray}
%as
%
%\begin{eqnarray}
%\tau_{e2e} &\equiv& \int_0^\infty dt \frac{\langle {\bf R}(t)\cdot{\bf R}(0) \rangle}{ \langle R^2 \rangle} = \frac{\pi^2}{12} \tau_R\ .
%\end{eqnarray}

By construction, the model reproduces the overall size, $\langle R^2 \rangle$, of target chains.
Furthermore, within the Rouse model Eq.~(\ref{eq:cmdiffusion}) for the CM diffusion holds at all times and can be expressed as
\begin{equation}\label{eq:g3 Rouse}
g_3(t) = \frac2{\pi^2}  \langle R^2 \rangle \frac{t}{\tau_R}\ .
\end{equation}
%so that $g_3(\tau_{e2e}) = \frac16 \langle R^2 \rangle = \langle R_g^2 \rangle$.

In particular, the Rouse model predicts sub-diffusive monomer motion, ${\bf \delta R}_i(t) \equiv {\bf R}_i(t)- {\bf R}_{cm}(t)$,
\begin{eqnarray}\label{eq:rouse g2}
g_2(t) &\equiv & \left< \left[{\bf \delta R}_i(t)-{\bf \delta R}_i(0) \right]^2 \right>\\
   &=&  \frac{2}{\pi^2} \langle R^2 \rangle \sum_{p=1}^{\infty} p^{-2} \left[ 1-\exp\left(-\frac{t}{\tau_p}\right)  \right],\\
   &\approx& \langle R^2 \rangle \left\{ \begin{array}{cl}
   		\frac{2}{\pi^{3/2}} \sqrt{ t/\tau_R } 	& \mbox{ for $t\ll \tau_R$} \label{eq:rouse g2 asymptotic}\\
   		\frac13							& \mbox{ for $\tau_R\gg t$}
           \end{array}\right.\ ,
\end{eqnarray}
$g_2(t)$ measures the monomer motion relative to the CM, $g_2(t)$, which levels off on approaching the Rouse time, $\tau_R$. Beyond this time, the total monomer motion,
\begin{equation}\label{eq:g1}
g_1(t)\equiv \left< \left[{\bf R}_i(t)-{\bf R}_i(0) \right]^2 \right>=g_2(t)+g_3(t)\ ,
\end{equation}
is dominated by the CM diffusion, Eq.~(\ref{eq:g3 Rouse}).

%%
%\begin{equation}\label{eq:rousemsdapprox}
%\approx \frac{2l_K^2}{\pi^{3/2}}\sqrt{ \frac{t}{\tau_K} },
%\end{equation}
The sub-diffusive monomer motion corresponds to an extended power-law decay of the shear relaxation modulus,
\begin{eqnarray}\label{eq:G(t) Rouse}
G_{R}(t) &=&  \rho_c k_BT \sum_{p=1}^{\infty} \exp\left(- \frac{2t}{\tau_p} \right)\\
   &\approx& k_BT\rho_c \left\{ \begin{array}{cl}
   		\sqrt{\frac{\pi}{8}} \sqrt{\tau_R/t } 	& \mbox{ for $t\ll \tau_R$}\\
   		\exp\left(- \frac{2t}{\tau_R} \right)							& \mbox{ for $\tau_R\gg t$}
           \end{array}\right.\ 
\end{eqnarray}
%%
%\begin{equation}\label{eq:rouseshearrelaxationapprox}
%  \approx  k_BT\rho_K \left(\frac{\pi}{8}  \frac{\tau_K}{ t} \right)^{1/2} \exp\left(- \frac{2t}{\tau_R} \right),
%\end{equation}
which implies a macroscopic viscosity of
\begin{equation}
\eta_R = \int_0^\infty \mbox{d}t G_{R}(t) = \frac{k_BT\rho_c}{2} \sum_{p=1}^{\infty} \tau_p
= \frac{\pi^2}{12} \rho_c k_BT \tau_R
= \frac1{36} \zeta_{cm} \langle R^2 \rangle\ . 
\end{equation}

\subsection{Defining Kuhn time, friction and viscosity through the Rouse model\label{sec: Definition Kuhn time and friction}}

Discretizing the chains at the Kuhn scale, the friction coefficient of Kuhn segments is given by
\begin{equation}\label{eq:zetaK}
\zeta_K = \frac{\zeta_{cm}}{N_K}\ .
\end{equation}
We {\em define} the Kuhn time (including prefactors) as 
\begin{equation}\label{eq:Kuhn-time}
\tau_{K}\equiv \frac{1}{3\pi^{2}}\frac{\zeta_{K}l_{K}^{2}}{k_{B}T} \ ,
\end{equation}
by identifying it with the relaxation time of the $p=N_K$ Rouse mode. Furthermore, it is convenient to define an effective viscosity at the Kuhn scale as
\begin{equation}\label{eq:etaK}
\eta_K = \frac{1}{36} \frac{\zeta_K}{l_K}\ 
\end{equation}
by interpreting $\zeta_K$ as a viscous Stokes drag, $\zeta_K \propto \eta_K l_K$.
Using these Kuhn units, the predictions of the Rouse model take the simple dimensionless form:
\begin{eqnarray}
\label{eq:taurouse}
\frac{\tau_{R}}{\tau_K} &=&N_{K}^{2} \\
\label{eq:etarouse}
\frac{\eta}{\eta_K} &=& n_K N_K\ ,
\end{eqnarray}
where the Kuhn number,
\begin{equation}\label{eq:nkdef}
n_{K}=\rho_{K}l_{K}^{3},
\end{equation}
is a dimensionless measure of density. 
A second advantage is that Kuhn units naturally indicate the validity limit of the Gaussian chain and the Rouse model for flexible polymers.
Furthermore, they reveal that except for the CM motion,
\begin{eqnarray}
\label{eq:g3 Rouse Kuhn units}
\frac{g_3(t)}{l_K^2} &=& \frac2{\pi^2}  \frac{1}{N_K} \frac{t}{\tau_K}\ , 
\end{eqnarray}
the characteristic dynamics below the Rouse time is independent of chain length:
\begin{eqnarray}
\label{eq:g2 Rouse Kuhn units}
\frac{g_2(t)}{l_K^2} &=& \frac{2}{\pi^{3/2}}\sqrt{\frac t{\tau_K} }   , \\
\label{eq:G(t) Rouse Kuhn units}
\frac{G_{R}(t) l_K^3}{k_BT}&=&  \sqrt{\frac{\pi}{8}} n_K  \sqrt{\frac{\tau_K}t } 
\end{eqnarray}

\subsection{Tube model for loosely entangled polymers}

Diffusing polymers can slide past each other, but since the chain backbones cannot cross, their Brownian motion is subject to transient topological constraints, which dominate the long-time chain dynamics.
Within the tube model~\cite{Edwards_procphyssoc_67, degennes71,DoiEdwards86}, the effective Hamiltonian in Eq.~(\ref{eq:overdampedLangevin}) and the corresponding static properties are thought to remain unchanged. The same holds for the isotropic short-time dynamics described by Eq.~(\ref{eq:overdampedLangevin}), while the effect of the topological constraints can be illustrated through the presence of a jungle gym-like array of obstacles~\cite{khokhlov1985polymer}. 
These constraints affect the polymer dynamics beyond a characteristic entanglement (contour) length or weight, $L_e \sim M_e \sim N_e$  \cite{DoiEdwards86}.
Continuing with a description at the Kuhn scale, we denote by $N_{eK}\equiv L_e/l_k$ the number of Kuhn segments per entanglement length.
For our present purposes $N_{eK} > 1$, i.e. the chains exhibit flexible chain behavior on the entanglement scale or ``loosely entangled''~\cite{morse1998viscoelasticity}.

Within the tube model~\cite{DoiEdwards86}, Eq.~(\ref{eq:overdampedLangevin}) is assumed to remain valid up to the entanglement time, $\tau_e$. 
The standard tube model~\cite{DoiEdwards86} for loosely entangled polymers builds upon the Rouse model, so that 
% and the assumption, that the chains exhibit Gaussian behavior on the entangement scale. 
%
\begin{equation}\label{eq:tau_e}
\tau_e= N_{eK}^2 \tau_K\ .
\end{equation}
Beyond the entanglement time, 
chains are expected to behave as if confined to a tube~\cite{degennes71}
of diameter $d_T\sim l_K N_{eK}^{1/2}$, Kuhn length $a_{pp} = d_T$, and  overall length $(L/L_e) d_T  \equiv Z d_T$, which follows their coarse-grain contour. 
%{\bf Need to explain the relation to $d_e$ in Sec. IV.B!!!!}
With the chain dynamics reduced to a one-dimensional diffusion within the tube (```reptation'' \cite{degennes71}), full equilibration requires the chain centers of mass to diffuse over the entire length of the tube. Assuming the same effective bead friction as for the local dynamics, the terminal relaxation time is thus given by 
\begin{equation}\label{eq:tau_d}
\tau_{d}=3Z \tau_R = 3Z^3 \tau_e\ . 
\end{equation}

According to the tube model the monomer mean-square displacements exhibit cross-overs at the Kuhn time $\tau_K$, the entanglement
time $\tau_e$, the Rouse time $\tau_R$, and the terminal relaxation time
$\tau_{max}$. All regimes are characterized by a particular power laws~\cite{DoiEdwards86}.
%: Before the Kuhn time, monomers display free diffusion independently of
%each other ($MSD(t)\sim t$). After the Kuhn time, monomers feel the bonds to
%neighboring monomers and display the characteristic sub-diffusive behavior of the Rouse model ($MSD(t)\sim t^{1/2}$).
%At times above the entanglement time, this motion continues but is restricted
%to the direction parallel to the primitive path, which itself adapts random
%walk statistics ($MSD(t)\sim t^{1/4}$). Above the Rouse time, the internal
%dynamics ceases to be relevant and the polymer displays collective reptation
%motion forwards and backwards along the (random walk like) primitive path
%($MSD(t)\sim t^{1/2}$). Finally above the terminal relaxation time, the
%topological constraints have decayed, and the polymer displays free unhindered
%diffusion ($MSD(t)\sim t$).
%
The mean-square displacements in  the various tube regimes can be calculated by projecting {\em one}-dimensional Rouse motion along the tube into three dimensions (see Eq. 6.37 in Ref.~\cite{DoiEdwards86})
\begin{equation}\label{eq:projection}
\langle \delta {\bf R}^2(t) \rangle = \langle a_{pp} | \delta s(t)| \rangle = a_{pp} \sqrt{\frac2\pi \langle \delta s(t)^2 \rangle}\ ,
\end{equation}
where the first equal sign follows since the primitive-path adopts a random walk conformation with step-length $a_{pp}$ and $\delta s(t)$
denotes the curvilinear displacement of a bead along the primitive-path curve. The second equal sign follows from the relation between
the mean magnitude and the root-mean-square moment for a Gaussian distributed random variable.
Using $\langle \delta s^2 \rangle=g_1(t)/3$ with Eqs.~(\ref{eq:g2 Rouse Kuhn units}) and (\ref{eq:g3 Rouse Kuhn units}) and applying
continuity between the different power-law regimes, Hou~\cite{hou2017note} determined the prefactors Doi and Edwards~\cite{DoiEdwards86} had neglected in assembling the various scaling laws and crossovers: 
\begin{equation}\label{eq:entangledmsd}
g_1(t) = \frac{2}{\pi^{3/2}} l_K^2\times
\begin{cases}
    \frac{t}{\tau_K}                                            & t\ll \tau_K \\
    \left( \frac{t}{\tau_K} \right)^{1/2}                       & \tau_K \ll t\ll \frac\pi9 \tau_e \\
    N_{eK} \left( \frac\pi9 \right)^{1/4} \left( \frac{t}{\tau_e} \right)^{1/4}    & \frac\pi9\tau_e \ll t\ll \pi\tau_R \\
     N_K \left( \frac{t}{\tau_{max}} \right)^{1/2}      &\pi \tau_R \ll t\ll \pi\tau_{max} \\
    \frac1{\sqrt{\pi}} N_K \frac{t}{\tau_{max}}                           & \pi\tau_{max} \ll t\\
\end{cases}\ .
\end{equation}
%For quantitative comparison between theory and simulation or experimental data we also need to identify the prefactors of
%and cross-over times between these powerlaw regimes.
%As already noted by Likhtmann (see Appendix B in Ref. \cite{likhtman2002quantitative}), the powerlaw cross-over from
%$t^{1/2}$ to $t^{1/4}$ power laws does not occur exactly at the entanglement time. rather he predicted that it should
%occur at $\tau_e^*=\pi^3/36 \tau_e$. This is due to the inclusion of the correct Rouse theory prefactors for $g_1(t)$.
%Furthermore, accounting for the prefactor in eq. (\ref{eq:projection}), Hou\cite{hou2017note} modified that estimate
%to $\tau_e^*=\pi/9 \tau_e$, and also found a similar shift at the  cross-over from the $t^{1/4}$ to $t^{1/2}$ power-laws:
%$\tau_R^* = \pi \tau_R$. 
%These prefactors and shifts are of order unity and appear insignificant. 
In particular, he demonstrated that accounting for these numerical constants is essential for quantitative estimates of the entanglement time~\cite{kremer90,puetz00a,likhtman2002quantitative,hsu2016static} and quantitative comparisons between PPA based predictions of the tube model and dynamical measurements. Finally, we obtain for $g_3(t)$ 

\begin{equation}\label{eq:entangledg3}
g_3(t) = \frac{2}{\pi^{3/2}} l_K^2\times
\begin{cases}
    \frac{N_{eK}^2}{ \sqrt{\pi}N_K}       \left( \frac{t}{\tau_e} \right)             & t \ll \frac\pi9 \tau_e \\
    \frac{N_{eK}}{3}           \left( \frac{t}{\tau_{R}} \right)^{1/2}                & \frac\pi9 \tau_e \ll t \ll \pi\tau_{R} \\
    \frac{N_{eK}}{3\sqrt{\pi}} \frac{t}{\tau_{R}}                                     & \pi\tau_R \ll t\\
\end{cases}\ .
\end{equation}

%
% Not including such prefactors\cite{hsu2016static} can lead to significant errors.
%

\section{Methods\label{sec:Methods}}

The present, more technical section describes the methods we have employed. Besides providing details on our Molecular Dynamics simulations (Sec.~\ref{sec:KGModel Simulations}) and the way we set up the systems (Sec.~\ref{sec:MD setup}), we define the estimators we use to extract the Kuhn length (Sec.~\ref{sec:Kuhn length estimator}), describe the primitive path analysis used to estimate the entanglement length (Sec.~\ref{sec:PPA methods}), and our strategy for extracting the effective bead friction in KG melts  (Sec.~\ref{sec:zeta_b methods}).

\subsection{Molecular Dynamics Simulations of the Kremer Grest Model\label{sec:KGModel Simulations}}

We have carried out two types of KG simulations. For our fully interacting reference melts, we use the standard KG~\cite{kremer1990dynamics} choice of $\Gamma=0.5m_{b}\tau^{-1}$ for the friction of the Langevin thermostat. As a complement, we have also simulated non-interacting ``phantom'' bead-spring polymers (Sec.~\ref{sec:phantom KG model}) with a suitably renormalized bending stiffness, $\tilde\kappa(\kappa)$, using the same approach. In this case, we employ the effective bead friction in the fully interacting systems, $\Gamma=\zeta_b$. Note that Eq.~(\ref{eq:Langevin}) reduces to the overdamped limit, Eq.~(\ref{eq:overdampedLangevin}), for $t>m_b/\zeta_b$. For a time step of $\Delta t$, the noise amplitude is given by $\langle{\boldsymbol \xi}_{i}(t)\cdot{\boldsymbol \xi}{}_{j}(t')\rangle=\frac{6\Gamma k_{B}T}{\Delta t}\delta_{t,t'}\delta_{ij}$.

For integrating the Langevin dynamics of our systems, we use the Gr\o{}nbech-Jensen/Farago (GJF)
integration algorithm\cite{gronbech2013simple,gronbech2014application} as
implemented in the Large Atomic Molecular Massively Parallel Simulator (LAMMPS).\cite{plimpton1995fast}
This integrator has the feature that the conformational temperature is exact,
while the kinetic temperate is correct to $O(\Delta t^2)$ accuracy.
In practice, this means that positions are exactly Boltzmann distributed as
expected from the force field and the preset temperature of the thermostat, while
the velocity distribution is only approximately the Maxwell-Boltzmann distribution
expected at the preset temperature. The systematic error in velocities are of the
order of $O(\Delta t^2)$. Most thermostats add or subtract heat in response to
the instantaneous kinetic temperature, since this observable is trivial to
obtain and use in a feed-back control loop. As a result, such a thermostat will
instead make an unknown time step dependent integration error in the
conformational temperature.\cite{gronbech2013simple} The effects of such an
error is quite difficult to detect and correct for, whereas an error in the
kinetic temperature is easy to detect, and can be fixed trivially by decreasing
the time step. In practice, this error also reflected in dynamic properties
for instance the diffusion coefficient estimated from mean-square displacements
(of positions) or from the integrated autocorrelation function (of velocities)
would incur a time step size dependent systematic Langevin integrator error
depending on the design of the integrator. With our choice of $\Delta t=0.01\tau$
time step, the average kinetic temperature is $T=0.979\epsilon$ for KG melt
states indicating a Langevin integrator error of the order of $2\%$.

To accurately estimate the time dependent shear relaxation modulus we follow
the approach of Ramirez et al.~\cite{ramirez2010efficient}. We use the correlator
algorithm that they implemented in LAMMPS and include both the autocorrelation of
the off-diagonal elements of the stress tensor as well as the autocorrelation
of the normal stresses. In simulations where we aim to study static properties
we aim for a single system system of sizes of $\sim 5 \times 10^6$ beads. However,
for sampling dynamic properties, we instead run five statistical independent
samples of systems of typical size $5-10\times 10^4$ beads.

%During a simulation we continuously sample the virial stress tensor
%\begin{equation}\label{eq:virialstresstensor}
%\sigma_{\alpha\beta}= \frac{m_b}{V} \sum_i v^\alpha_i v^\beta_i+\frac{1}{V}\sum_{<i,j>} r^\alpha_{ij} F^\beta_{ij},
%\end{equation}
%
%where $r^\alpha_i, v^\alpha_i, F^\alpha_i,$ denotes the $\alpha$'th Cartesian
%component of the $i$'th particles position, velocity and total force, respectively.
%The force contains contributions from both pair, bond and angular interactions.
%The shear relaxation modulus is obtained from the virial stress tensor using
%the Green-Kubo relation
%
%\begin{equation}\label{eq:gtgreenkubo}
%G(|t-t'|)=\frac{V}{k_B T} \left< \sigma_{xy}(t') \sigma_{xy}(t) \right>
%\end{equation}
%
%where we also average over the $yz$ and $xz$ off-diagonal elements of the stress
%tensor. We sample the virial stresses at every time step, and use the correlator
%algorithm\cite{ramirez2010efficient} to effectively sample the stress
%autocorrelation function for lag times $|t-t'|$ ranging from the time step
%up to the full duration of the simulation.

% For the highly entangled systems $Z>80$, we
%have integrated the dynamics for $4-6\times 10^8$ steps. For the moderately
%entangled systems $Z=10,20$, we have integrated the dynamics for $2-6\times 10^9$ steps.

\subsection{System setup and equilibration\label{sec:MD setup}}

We have generated equilibrated entangled KG model melt states both for
unentangled or weakly entangled systems as well as for moderately to highly
entangled melts. We used brute force equilibration methods for the weakly
entangled systems, while we set up the moderately and highly entangled
melts using a sophisticated multiscale process, which we developed recently.\cite{SvaneborgEquilibration2016}

\subsubsection{Brute-force equilibration of weakly entangled systems}
In the case of weakly entangled melts, the oligomers were randomly inserted in
the simulation domain, and the energy was minimized using the force-capped
interaction potential and subsequently transferred to the KG force-field as in
Ref. \cite{SvaneborgEquilibration2016}. The resulting conformations were then simulated
using Molecular Dynamics as above, while also performing double bridging
Monte Carlo moves\cite{karayiannis2002novel,karayiannis2002atomistic}
as implemented in LAMMPS\cite{sides2004effect}. During these moves,
bonds are swapped between different chains in such a way that the melt remains
monodisperse. Such connectivity altering moves are known to accelerate the
equilibration by side stepping potential barriers to the physical dynamics.

To characterize the Kuhn length dependence of stiffness, we performed brute
force simulations of weakly entangled melts with nominal chain length $N_K=10,20,40,80$
for varying chain stiffness using the approximate Kuhn length estimate from
Ref. \cite{SvaneborgEquilibration2016}. During the production run, we
continued to perform double bridging moves.
To estimate the Kuhn friction dependence of stiffness, we equilibrated oligomer
melts with nominal length $N_K=1,3,\dots,30$ and varying stiffness as above.
Our production runs were up to $2-10\times10^{5}\tau$ steps long and performed
without the double bridging moves. The computational effort was about $60$
core years of simulation time. 

A good measure for the length of the simulation where double bridging takes
place is the number of displacement times. A displacement time is the time it
takes the mean-square bead displacement to match the chain mean square
end-to-end distance. This measure is well defined in our case 
where we apply connectivity altering Monte Carlo\cite{karayiannis2002novel,karayiannis2002atomistic,sides2004effect} moves during
the simulations, whereas the chain center of mass can not be defined
in this case. During the analysis, the data obtained within the first
displacement time was discarded, and trajectory data for subsequent displacement times
were binned and averaged assuming statistical independent results for
each displacement time bin. We report mean-square internal distances
sampled over at least five displacement times for the longest and stiffest
chains and well in excess of $1000$ displacement times for shorter and less stiff chains. 

\subsubsection{Multi-scale equilibration of highly entangled systems}
For estimating the number of Kuhn segments between entanglements and the Kuhn time,
we required highly entangled melts. 
To prepare such states, used an effective multiscale equilibration process\cite{SvaneborgEquilibration2016}
which transfers melt states between three computationally different, but physically
equivalent polymer models. We initially equilibrate a
lattice melt where density fluctuations are removed using Monte Carlo simulated
annealing. We choose the tube diameter as the lattice
parameter, such that approximately $19$ entanglement segments occupy the
same lattice site. The lattice melt state is then transferred to an auxiliary
KG model, where WCA pair-forces are capped. The pair-forces are chosen such
that local density fluctuations are further reduced compared to the lattice
melts, while also partially allowing chains to move through each other.
This auxiliary model produce Rouse dynamics, and we simulate the dynamics
for sufficiently long to equilibrate the random walk chain structure
beyond the lattice length scale. Finally we transfer the auxiliary model melt
state to the KG force field, and equilibrate local bead packing. The three
models were designed to reproduce identical large scale chain statistics,
and the two MD models were designed to produce the same local chain statistics,
which required using a renormalized bending stiffness in the auxiliary model. 
The first set of data we prepared using this approach comprises systems
with fixed numbers of beads $N_b=10000$ and varying chain stiffness. The effective chain
length, measured in numbers of entanglements per chain, varies between
$Z(\kappa=-1)=85$ to $Z(\kappa=2.5)=570$.
To check for finite length effects for the flexible melts with $-1<\kappa<0$, we 
generated additional melts with fixed number of entanglements $Z=200$ and
varying number of beads. The total computational effort of preparing these
entangled melts was about $12$ core years.
As a quality check, we monitor 
(i) density fluctuations and
(ii) the single chain statistics~\cite{auhl2003equilibration}.
The melt structure factor, $S(q)$, is constant on all scales above the bead size, documenting the absense of all long-wave length density fluctuations as expected from a nearly incompressible system (data not shown).  Fig. \ref{fig:msid} shows excellent agreement between the reduced mean-square internal distance, $\langle \left( {\bf R}_i - {\bf R}_{i+n} \right)^2 \rangle/n$ for brute force equilibrated melts of chains of medium length and for the highly entangled melts with $Z\geq 200$, which we have prepared by our multiscale equilibration process.

\subsection{Estimating the Kuhn length\label{sec:Kuhn length estimator}}

The incompressibility of polymer melts induces weak long range repulsive interaction along the chain, which manifest themselves in long-range bond orientation correlations~\cite{Wittmer_Meyer_PRL04,wittmer2007polymer,wittmer2007intramolecular,beckrich2007intramolecular,semenov2010bond}. 
If $\langle R^{2}(l)\rangle_{L}$ denotes the mean-square spatial distance of beads separated by a contour distance $l$ along a chains in a monodisperse melt of chains of length $L \ge l$, then $\langle R^{2}(l)\rangle_{L} < \lim_{L\rightarrow\infty} \langle R^{2}(l)\rangle_{L}$.
With local estimators risking to underestimate the true Kuhn length, we propose to use a global estimator based on eq. (\ref{eq:Kuhnlengththeory}),
\begin{eqnarray}
\label{eq:lK_asymptotic}
l_K &=& \lim_{l \rightarrow\infty} \lim_{L\rightarrow\infty} l_{K}(L^{-1/2},l^{-1/2})\\
\label{eq:lK_observable}
l_{K}(L^{-1/2},l^{-1/2}) &=&\frac{\langle R^{2}(l)\rangle_{L}}{l},
\end{eqnarray}
and to fit data for finite internal distances, $l$, on chains of finite length, $L$, to a polynomium of the form
\begin{equation}
\label{eq:lK_l_L}
l_K( L^{-\frac{1}{2}}, l^{-\frac{1}{2}} )
= l_K + c_{10} L^{-\frac{1}{2}} + c_{01} l^{-\frac{1}{2}}
+ c_{20} L^{-1}+ c_{11} L^{-\frac{1}{2}}l^{-\frac{1}{2}}
 +c_{02}l^{-1} 
\end{equation}
motivated by the theory of Wittmer et al.\cite{wittmer2007polymer}.

\subsection{PPA for estimating entanglement lengths\label{sec:PPA methods}}

The entanglement length can be estimated via a primitive path analysis~\cite{everaers2004rheology} of the microscopic topological state of entangled melts. The idea~\cite{Edwards_tube_procphyssoc_67,RubinsteinHelfand85} is to fix the chain ends and to convert the chains into rubber bands, which contract without being able to cross. 
This is achieved by minimizing the potential energy for fixed chain ends and disabled intrachain excluded volume interactions. 
We performed the minimization by using the steepest descent algorithm implemented in LAMMPS. 
The minimization is followed by dampened Langevin dynamics as in the standard PPA algorithm.
We performed a PPA that preserves self-entanglements by only disabling  interactions between pairs of beads within a chemical distance of $2N_{eK}$ Kuhn segments along the chain. The computational effort for the PPA analysis is insignificant in comparison to the other observables.

Primitive paths are locally smooth, kinked at entanglement points between different chains, and have the same large scale random walk statistics as the original chains~\cite{DoiEdwards86,everaers2012topological} (see Fig. \ref{fig:visualization}).
In particular, primitive paths can be characterized by a contour length, $L_{pp}<L$, and a Kuhn length, $a_{pp}>l_K$. 
From the relation $ l_K L = \langle R^{2}\rangle = a_{pp} L_{pp}$, it follows that $a_{pp}/l_K = L/L_{pp}$, i.e. during PPA the Kuhn length increases in inverse proportion to the shortening of the contour length. 
Assuming random walk statistics between entanglement points, $a_{pp}^2=l_K^2 N_{eK}$, one obtains the classical estimator \cite{DoiEdwards86,everaers2004rheology,everaers2012topological} 
\begin{equation}\label{eq:neclassical}
N_{eK}^{classical}=\frac{a^2_{pp}}{l_{K}^{2}}
=\frac{L^2}{L_{pp}^{2}},
\end{equation}
of the number of Kuhn segments per entanglement length as function of a convenient PPA observable, the ratio of the contour length of the primitive paths and of the original chains.

$N_{eK}^{classical}$ suffers from finite $Z$ effects, when applied to the
relatively short chain melts, that can be equilibrated by brute-force molecular
dynamics. Hoy et al.\cite{HoyPRE09} showed that eq. (\ref{eq:neclassical})
provides a lower bound on the true entanglement
length and suggested another estimator that provides an upper bound
\begin{equation}\label{eq:hoy}
N_{eK}^{Hoy}=N_K \left( \frac{ L_{pp}^2 }{\langle R^2 \rangle } -1 \right)^{-1}.
\end{equation}

For flexible chains the Kuhn length of the PPA path, $a_{pp}$, and the contour distance, $l_e$, between entanglements along the primitive path are proportional to each other.\cite{everaers2012topological}.
However, as chain stiffness increases, the assumption of random walk statistics between entanglements becomes questionable (see e.g. Fig. \ref{fig:visualization}). 
To correct for this effect~\cite{uchida2008viscoelasticity}, one can assume WLC statistics in equating $l_e$ with the root mean-square {\em spatial} distance between entanglement points 
\begin{equation}\label{eq:de}
l_e^2 = \langle R^2 (N_{eK};l_K)\rangle \approx  l_K^2N_{eK}\left(1 -[1-\exp(-2N_{eK})]/[2 N_{eK}] \right)\ .
\end{equation}
Since the PPA contour length is given by $L_{pp} =Z l_e$ and the contour length of the original chain by $L=l_K N_K=Z N_{eK} l_K$, this suggests a different relation between $N_{eK}$ and the square of the contour length contraction ratio,

\begin{equation}\label{eq:nenew}
\left( \frac{L_{pp}}{L} \right)^2 =
\frac{2 N_{eK}+\exp\left(-2 N_{eK}\right)-1}{2 N_{eK}^2}\ .
\end{equation}
Eq.~(\ref{eq:nenew}) is straightforward to invert numerically for any measured contraction ratio.
In the limit of $N_{eK}\gg1$, it reduces to $N_{eK}=L^2/L_{pp}^2$ consistent with Eq.~(\ref
{eq:neclassical}). In the limit, $N_{eK}\ll 1$, eq.~(\ref{eq:nenew}), remains
valid and converges to the Semenov expression for entanglement length in
tightly entangled semi-flexible chains~\cite{Semenov86,Morse01,uchida2008viscoelasticity}.

\subsection{Estimating the effective bead friction\label{sec:zeta_b methods}}

In the logic of the tube model~\cite{DoiEdwards86} we are interested in $\zeta_b=\zeta_{cm}/N_b$  in Eq.~(\ref{eq:overdampedLangevin}) for {\em un}entangled chains.
The fact, that the asymptotic diffusion of long chains is slowed down by entanglement effects~\cite{degennes71,DoiEdwards86}, is in itself not a major obstacle, because according to the tube model we should still be able to use Eq.~(\ref{eq:cmdiffusion}) for chains of arbitrary length as long as we limit the analysis to short enough times compared to the entanglement time. 
However, the Rouse and tube models discussed in the previous section are so well established~\cite{DoiEdwards86}, that it is easy to forget how drastic the approximations are, which are underlying the description. 
%Some of the neglected effects come back to haunt us, when we try to extract $\zeta_b$. 
In particular, experiments and simulations\cite{kolinski1987does,kremer1990dynamics,binder1997monte,smith2000comparison,padding2001zero,kreer2001monte,doxastakis2003chain,paul2004structure,brodeck2009study,abou2010rouse}
%
% pick a few references from https://iopscience.iop.org/article/10.1088/0953-8984/24/28/284105/meta
% KG, richter, W. Paul, ....
%
show an accelerated early-time CM diffusion, which can be explained \cite{Semenov2011PRL,Semenov2012PRE,Semenov2012PRE2,Semenov2012JPhys}
%Farago J, Meyer H and Semenov A N 2011 Anomalous diffusion of a polymer chain in an unentangled melt Phys. Rev. Lett. 107 178301
%
%Crossref
%[20]
%Farago J, Semenov A N, Meyer H, Wittmer J P, Johner A and Baschnagel J 2012 Mode-coupling theory of polymer diffusion in an unentangled melt: I. Density fluctuation approach Phys. Rev. E submitted
%[21]
%Farago J, Meyer H, Baschnagel J and Semenov A N 2012 Mode-coupling approach to polymer diffusion in an unentangled melt: II. The effect of viscoelastic hydrodynamic interactions Phys. Rev. E submitted
%
%https://iopscience.iop.org/article/10.1088/0953-8984/24/28/284105/meta
%
%That's the theory papers we'd have to compare to, if we want to make (more) sense of the force-cap dynamics
%
%
through correlation-hole effects and viscoelastic hydrodynamic interactions. 
As a consequence, 
(i) $\zeta_{cm}$ has to be extracted from long-term diffusion data and
(ii) $\zeta_{b}$ has to be inferred via a double extrapolation  beyond the monomer scale (to reduce end effects~\cite{ColbyFettersGraessleyMM1987}) and below the entanglement scale (to extract the effective friction constant for the local, topologically unrestricted motion). 

In contrast to the CM motion, the monomer dynamics is hardly affected by correlation-hole effects and viscoelastic hydrodynamic interactions~\cite{Semenov2011PRL,Semenov2012JPhys}. In particular, the local dynamics is independent of chain length and, on sufficiently small scales, entanglement effects. 
On the downside, (i) KG chains show ballistic motion at very short times, (ii) they are obviously not flexible below their respective Kuhn lengths, and, most importantly, (iii) the analysis of the chain statistics shows that commodity polymers (and corresponding KG models) are not even fully flexible on the entanglement scale~\cite{svaneborgKGmapping}.  
As a consequence, it is desirable to separately deal with the two {\em independent} approximations underlying the standard Rouse and tube models for flexible polymer dynamics: the Gaussian chain model and the assumption, that the (local) dynamics can be described by a single-chain Langevin equation, Eq.~(\ref{eq:overdampedLangevin}). 

%The standard Rouse and tube models for flexible polymers are based on two {\em independent} approximations: the Gaussian chain model and the assumption, that the (local) dynamics can be described by a single-chain Langevin equation, Eq.~(\ref{eq:overdampedLangevin}). 
%The analysis of the chain statistics shows that commodity polymers (and corresponding KG models) are not perfectly flexible on the entanglement scale~\cite{svaneborgKGmapping}.  As a consequence, we need to test the two approximations independently and to account for stiffness corrections, when we analyse the chain dynamics in terms of the standard theoretical models.
%
%To cut through the Gordian knot of 

Langevin simulations of phantom KG chains (Sec.~\ref{sec:phantom KG model}) account for bead inertia inertia and effective local chain stiffness. The only input required are estimates of the Kuhn length, $l_K$, and of the bead friction, $\zeta_b$, in the fully interacting systems. 
To validate the single-chain description {\em independently} of chain stiffness, it suffices to compare the dynamics of the fully interacting and the phantom systems at times up to the entanglement time.

\section{Results}\label{sec:results}
%\section{Kuhn characterization}\label{sec:results}

Below we present our simulation results for the chain structure and dynamics in KG melts. We start out from the conformational statistics (Sec.~\ref{sec:R2}) to extract the Kuhn length as a function of the bending rigidity  (Sec.~\ref{sec:lK results}). Next we turn to the primitive path analysis of the microscopic topological state and the entanglement length  (Sec.~\ref{sec:PPA results}). To estimate the bead friction in KG melts, we analyse the center-of-mass motion (Sec.~\ref{sec:CM motion results}). In a second step, we analyse the monomer diffusion and validate our choice of the bead friction through a comparison to the dynamics of Phantom KG chains  (Sec.~\ref{sec:Monomer motion results}).

\subsection{Conformational statistics\label{sec:R2}}

%Do we also need $R_g^2(N)$ and $R^2/R_g^2$? Because $g_2$ converges to $R_g^2$.

\begin{figure}
\includegraphics[angle=\Angle,width=0.5\columnwidth]{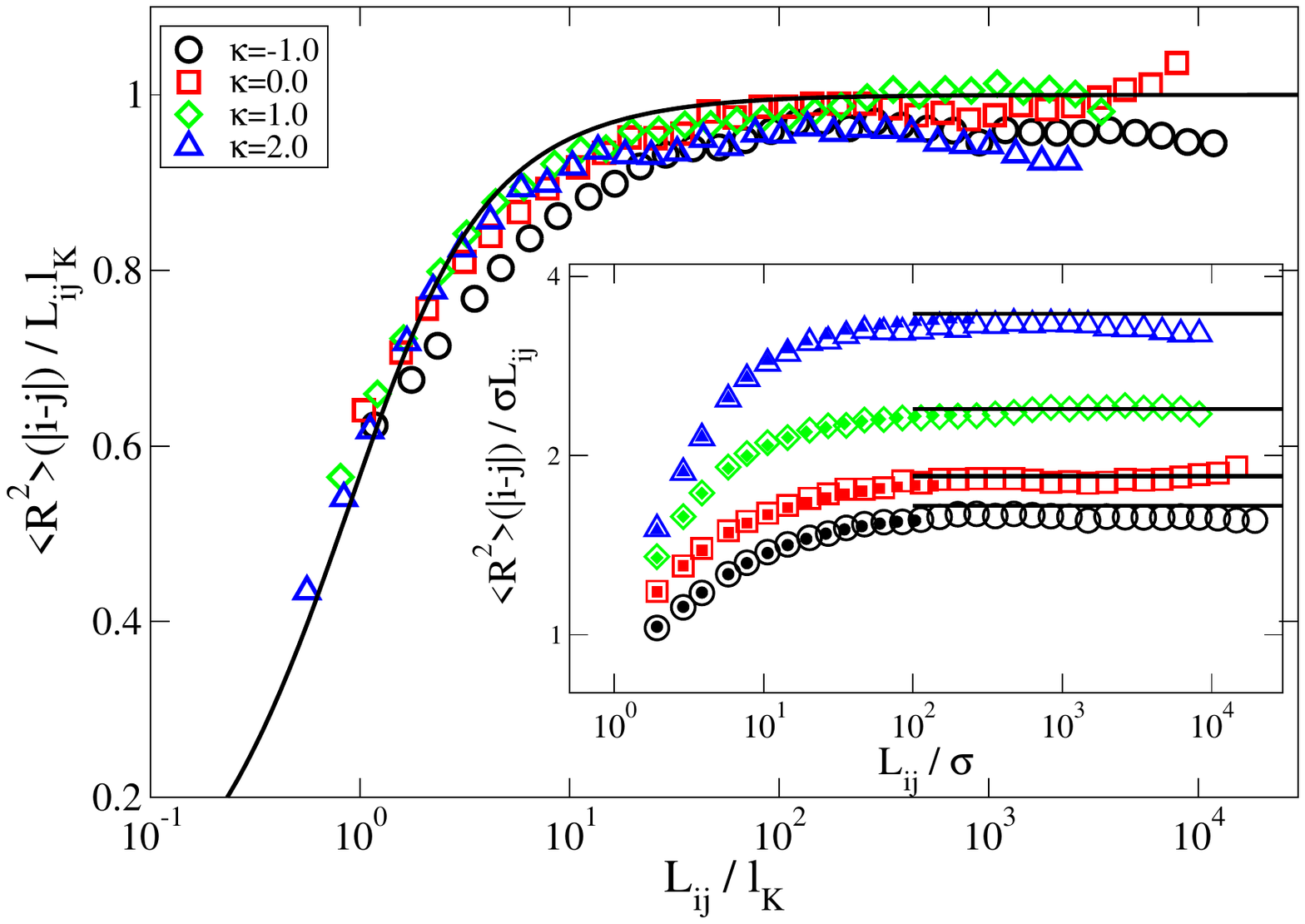}
\caption{\label{fig:msid}Reduced mean-square internal distances as a function of contour distance for different values of the effective bending stiffness ($(\kappa=-1)$/black, $(\kappa=0$)/red, $(\kappa=1$)/green, and $(\kappa=2$)/blue). Main figure: Kuhn units, inset: LJ units.
Small filled symbols represent data for brute force equilibrated, moderately entangled melts with $N_K=80$.
Large open symbols show data for highly entangled melts with $Z \geq 200$, which we generated by an efficient multi-scale equilibration procedure~\cite{SvaneborgEquilibration2016} . 
In the inset we show our estimates of the Kuhn lengths as black lines.
The black line in the main figure represents the wormlike chain expression, Eq.~(\ref{eq:R2 WLC}).
}
\end{figure}

The mean-square internal distances, $\langle ({\bf R}_i-{\bf R}_j)^2 \rangle$, between beads $i,j$ as a function of their contour distance, $L_{ij}=l_b|i-j|$, provide a convenient overview of the chain statistics on different length scales. 
The small filled symbols in the inset of Fig.~\ref{fig:msid} show the reduced internal distances, $\langle ({\bf R}_i-{\bf R}_j)^2 \rangle/L_{ij}$, for our brute force equilibrated KG melts of chains of medium length. The data exhibit a monotonous increase to a plateau, which is reached at chemical distances much larger than the Kuhn length.  A priori, the plateau height matches the Kuhn length of the chains, Eq.~(\ref{eq:R2 WLC}). However, when analyzing the data, care must be taken to account for slowly decaying power law corrections (see below).

A typical signature of insufficient equilibration is the appearance of swelling at intermediate length scales, which would be seen as a bump in the figure~\cite{auhl2003equilibration}.
Figure \ref{fig:msid} also contains results for highly entangled melts with $Z\geq 200$, which we have prepared by our multiscale equilibration process~\cite{SvaneborgEquilibration2016}. 
Their excellent agreement with the reference data suggest that these systems are properly equilibrated on all length scales.
In addition to validating the single-chain statistics, we have also sampled the the structure factor for these melts (see Ref.~~\cite{SvaneborgEquilibration2016}, data not shown for the present systems).
The structure factor is constant on all scales above the bead size. As expected for a nearly incompressible system, there are no  long-wave length density fluctuations.

\subsection{Kuhn length\label{sec:lK results}}

\begin{figure}
\includegraphics[angle=\Angle,width=0.5\columnwidth]{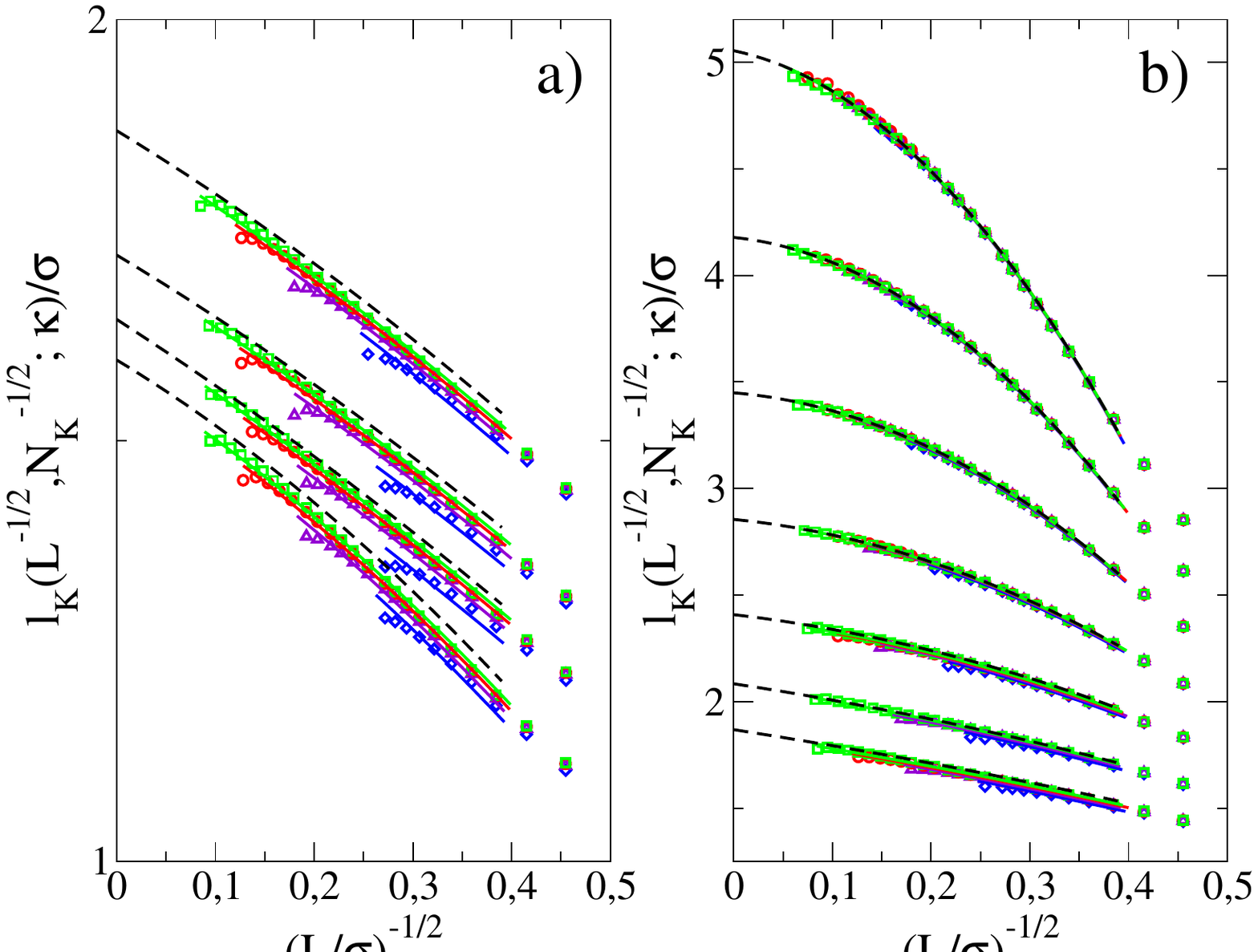}
\caption{\label{fig:Kuhn-length-extrapolation}
Kuhn length extrapolation for melts of nominal chain length $N_K=10,20,40,80$
(denoted by blue diamond, orange triangle up, red circle, and green box, respectively) 
for flexible melts with $\kappa=-2,-1,-0.5,0$ (bottom to top in panel a),
and semi-flexible melts with $\kappa=0,0.5,1.0,1.5,2.0,2.5,3.0$ (bottom to top in panel b).
Shown are also the results of the fits of eq. (\ref{eq:lK_l_L}) to individual melts (solid lines
matching the melt colors) and the extrapolation to infinite chain length
as function of contour length (black dashed line).
}
\end{figure}

%\begin{figure}
%\includegraphics[angle=270,width=0.5\columnwidth]{angles.ps}
%\caption{\label{fig:Kuhn-length}
%Extrapolated Kuhn length $l_{K}$ vs stiffness parameter $\kappa$ (blue filled circles)
%and local Kuhn lengths $l_K^{(1)}$  (open violet circles) as well as 
%literature data from Hoy et al.\cite{PhysRevE.72.061802} (red box)
%and Moreira et al.\cite{moreira2015direct} (green diamond).
%We also show our parameterizations of $l_K(\kappa)$ (black solid line),
%$l_K^{(1)}(\kappa)$ (black dotted line), and $l_K^{(0)}(\kappa)$ (black dashed line).
%}
%\end{figure}

We have estimated the asymptotic value of the Kuhn length, Eq. (\ref{eq:lK_asymptotic}), for all studied values of chain stiffness $\kappa$, on the basis of our ``gold standard'' reference data for brute-force equilibrated melts. 
To do so, we fitted mean-square internal distances, $\langle R^{2}(l)\rangle_{L}$, by the ansatz Eq.~(\ref{eq:lK_l_L}).
The $c_{nm}$ parameters account for the sub-dominant finite size effects in our data. During the fitting, $c_{01}$ and $c_{11}$ terms are restricted to be negative or zero. This ensures that the limit of infinite contour length is approached from below. 
Fits were performed on data with $l>6.25\sigma$. 
This was chosen to ensure some data points from the shortest melts still contribute to the analysis, while
discarding data necessitating the inclusion of further subdominant terms in our ansatz Eq.~(\ref{eq:lK_l_L}).

Fig. \ref{fig:Kuhn-length-extrapolation} shows data for the finite size
estimates of the Kuhn length along with the results of our extrapolation scheme
for flexible and semi-flexible KG systems (panel a and b, respectively). As the
contour length is increased the Kuhn length estimate increases monotonously
towards the limit given by the intercept with the y axis. For the semi-flexible
systems (panel b), we observe a good collapse of the data from melts with
different chain length, whereas for the flexible systems the data collapse is
less good indicating that finite size effects are more important in the limit
of flexible chains.  For the flexible melts, we also see a systematic downturn for the data
points corresponding to the longest contour lengths, which demonstrates that care should be taken
in choosing the range of data used for the extrapolation. Also shown in
the figure are predictions of finite size Kuhn lengths from the fits and
the limit of infinite chain length. We see good agreement of the fits to
our data. The extrapolation to infinite melt chain length is in very good
agreement with all data for the semi-flexible melts, but we observe that it is 
systematically above the data for the flexible melts. This illustrates the
importance estimating the asymptotic Kuhn length using data from melts with several
chain lengths simultaneously. For instance, extrapolating the $N_K=80$ data
towards infinite contour length i.e. extending the solid green lines to the
intercept with the axis would underestimate the true limiting Kuhn length
value for all the flexible melts.

\subsection{PPA and Entanglement length\label{sec:PPA results}}

\begin{figure}
\includegraphics[angle=\Angle,width=0.25\columnwidth]{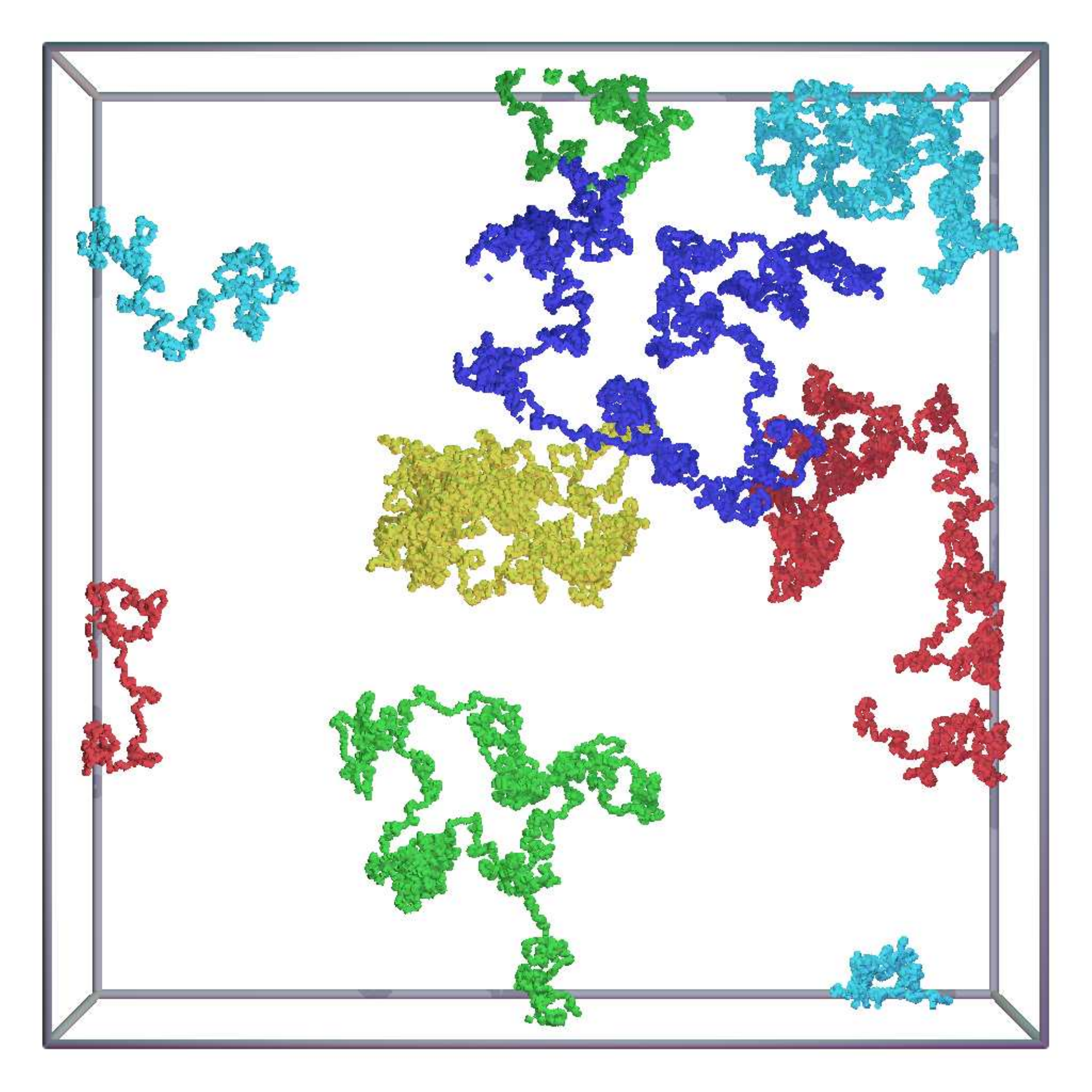}
\includegraphics[angle=\Angle,width=0.25\columnwidth]{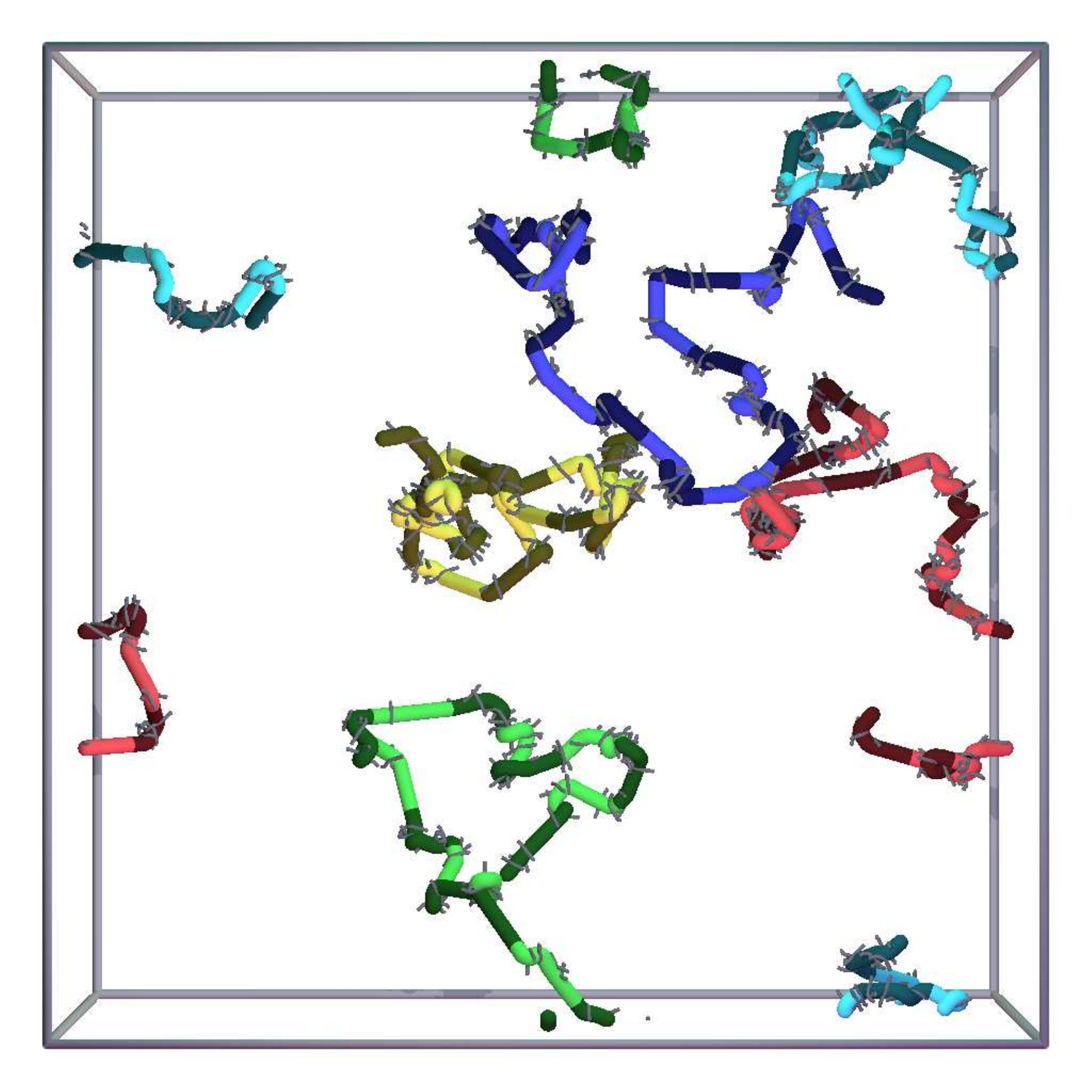}

\includegraphics[angle=\Angle,width=0.25\columnwidth]{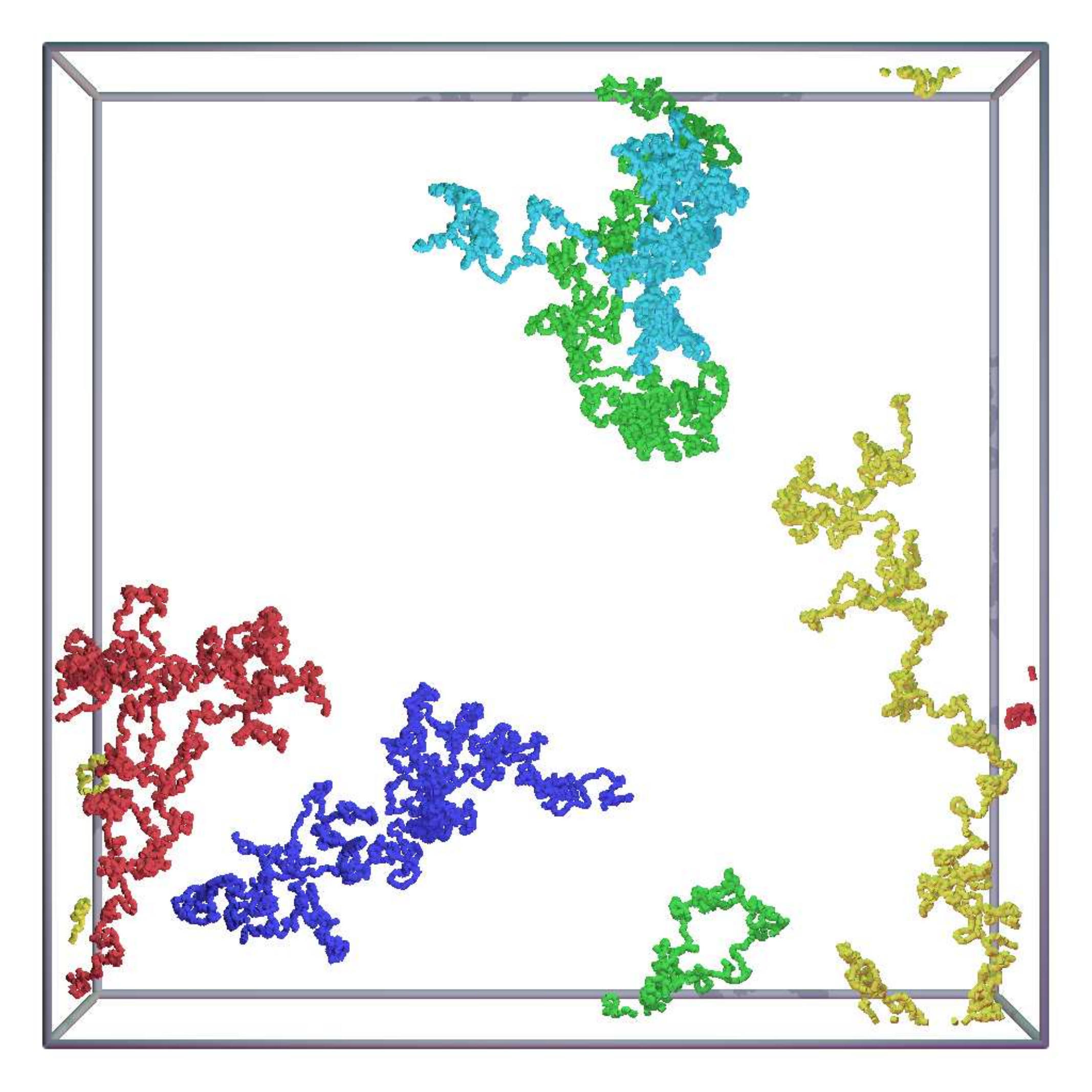}
\includegraphics[angle=\Angle,width=0.25\columnwidth]{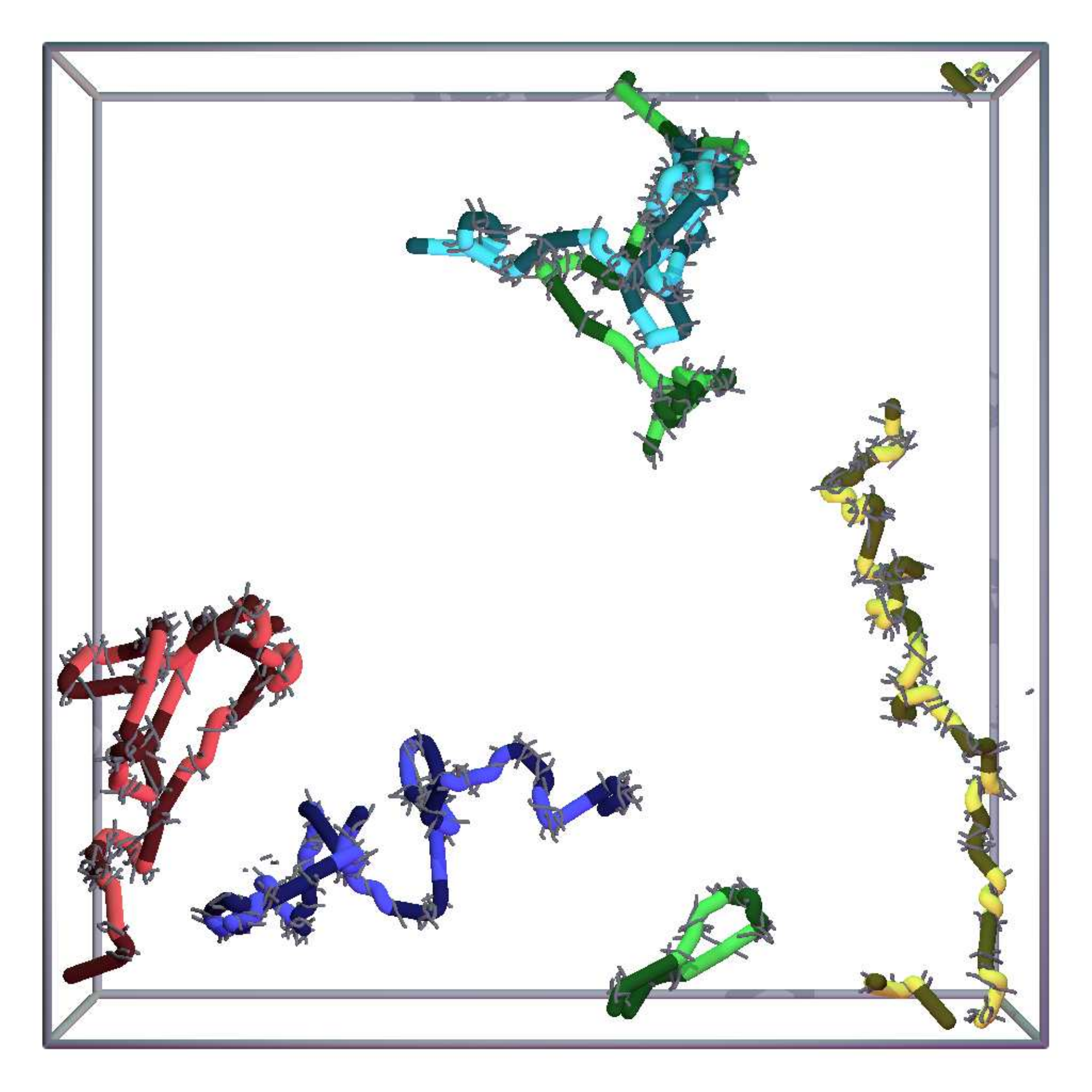}

\includegraphics[angle=\Angle,width=0.25\columnwidth]{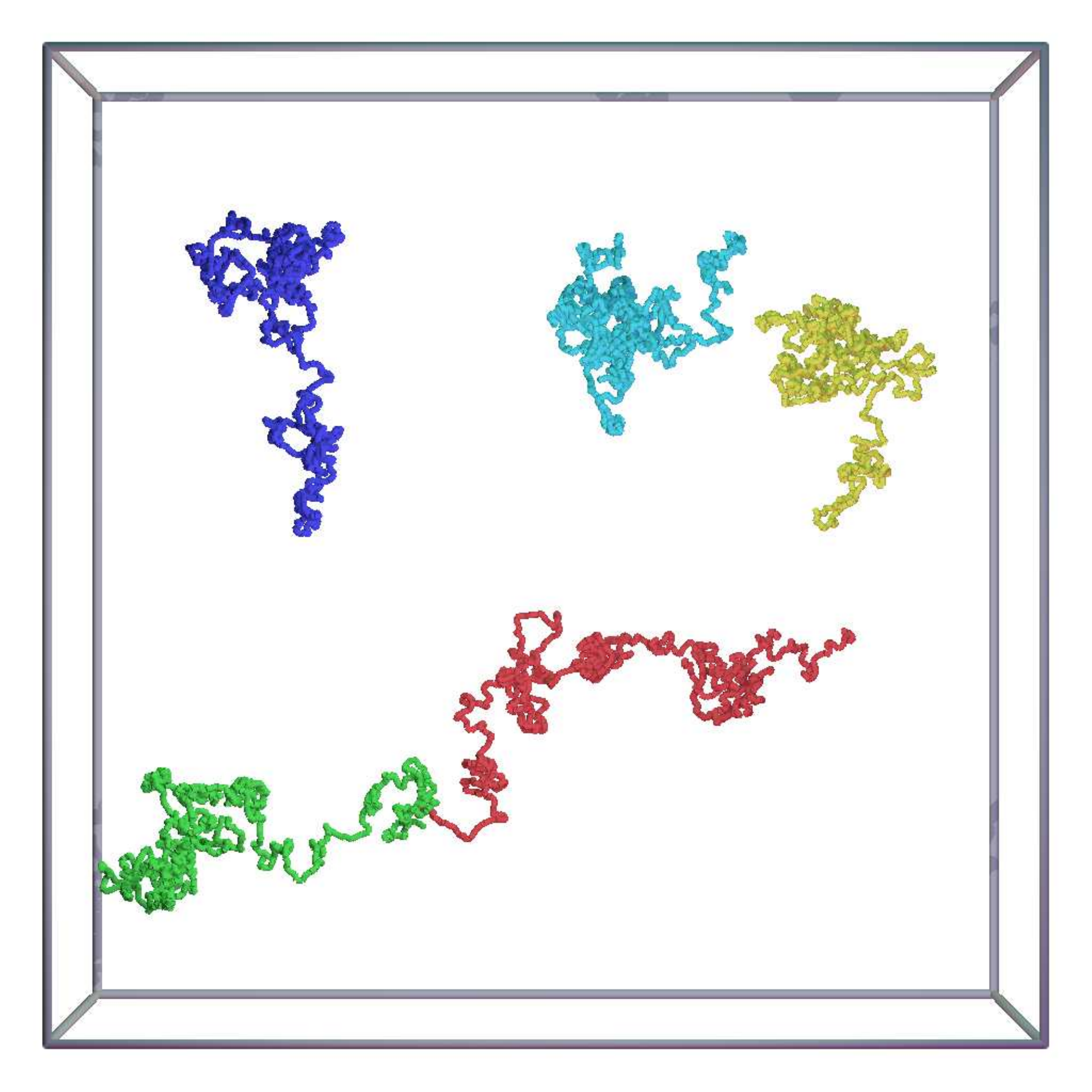}
\includegraphics[angle=\Angle,width=0.25\columnwidth]{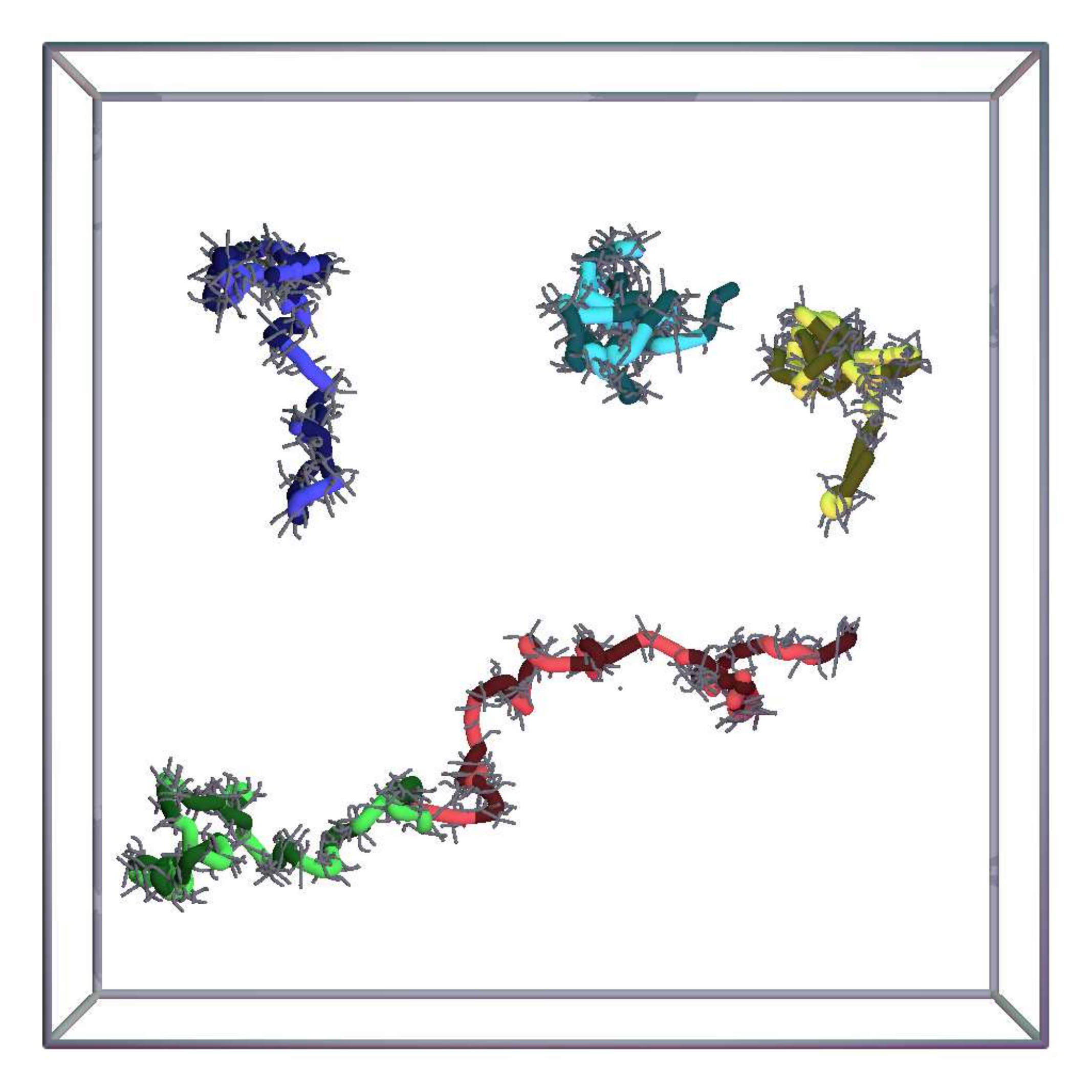}

\includegraphics[angle=\Angle,width=0.25\columnwidth]{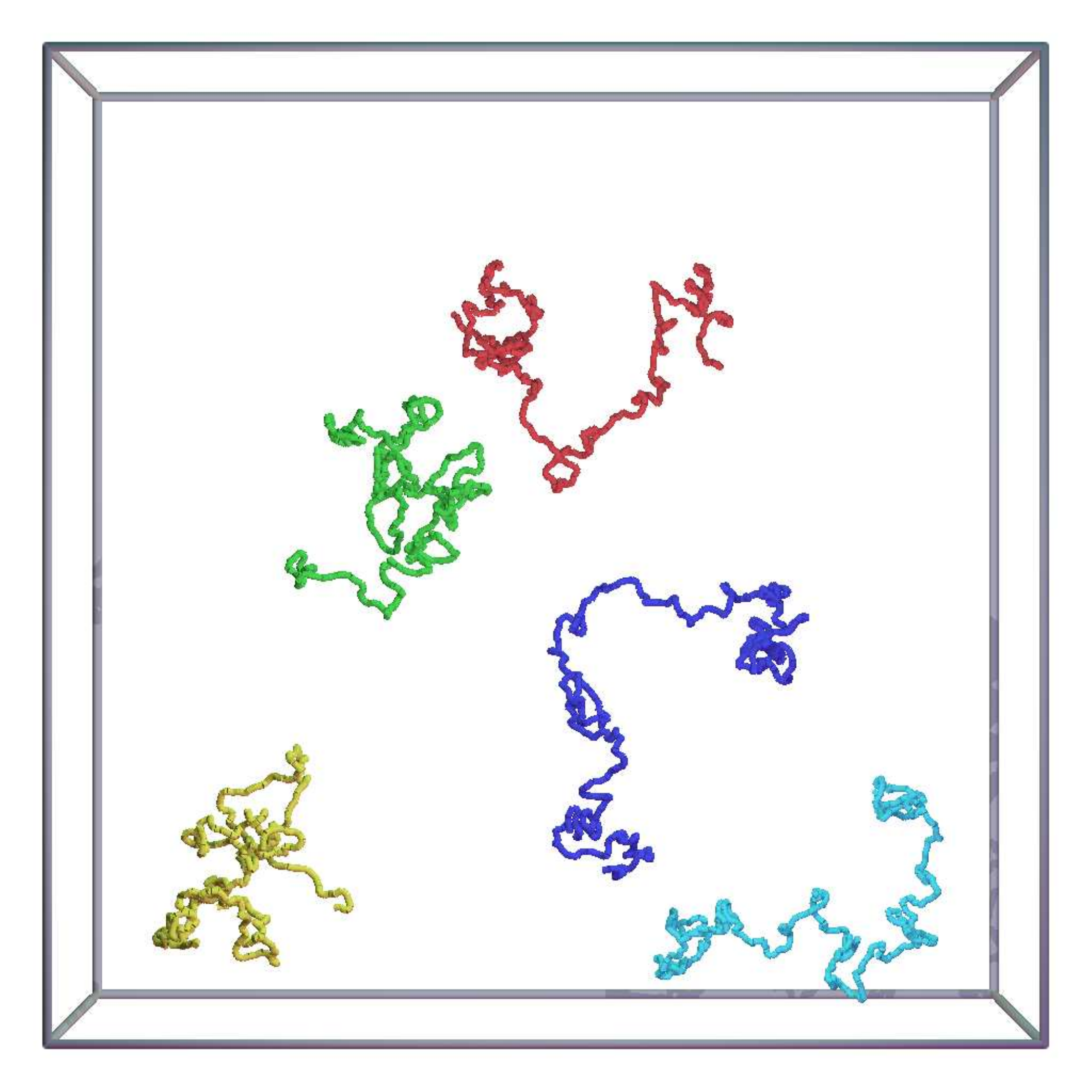}
\includegraphics[angle=\Angle,width=0.25\columnwidth]{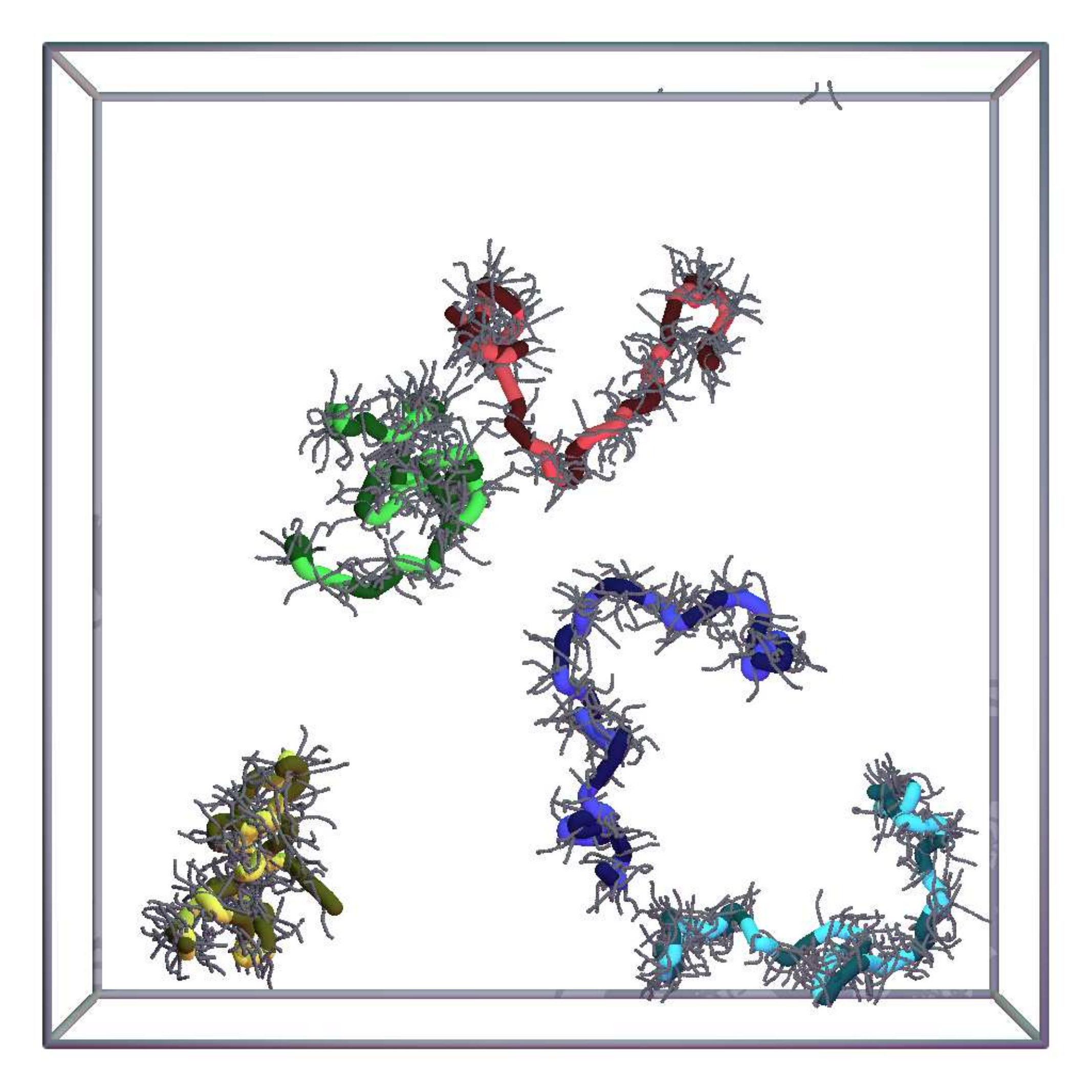}

\caption{\label{fig:visualization}Visualization of the same five chains in the melt state (left)
and their primitive-path (right) for stiffness $\kappa=-1,0,1,2$ (top to bottom), respectively,
for melts with constant number of entanglements $Z=20$. The entanglement length is illustrated
as an alternating color saturation along the primitive-paths. Short segments of
the entanglement partners are also shown along the primitive-paths (thin gray lines).
}
\end{figure}

\begin{figure}
\includegraphics[angle=\Angle,width=0.5\columnwidth]{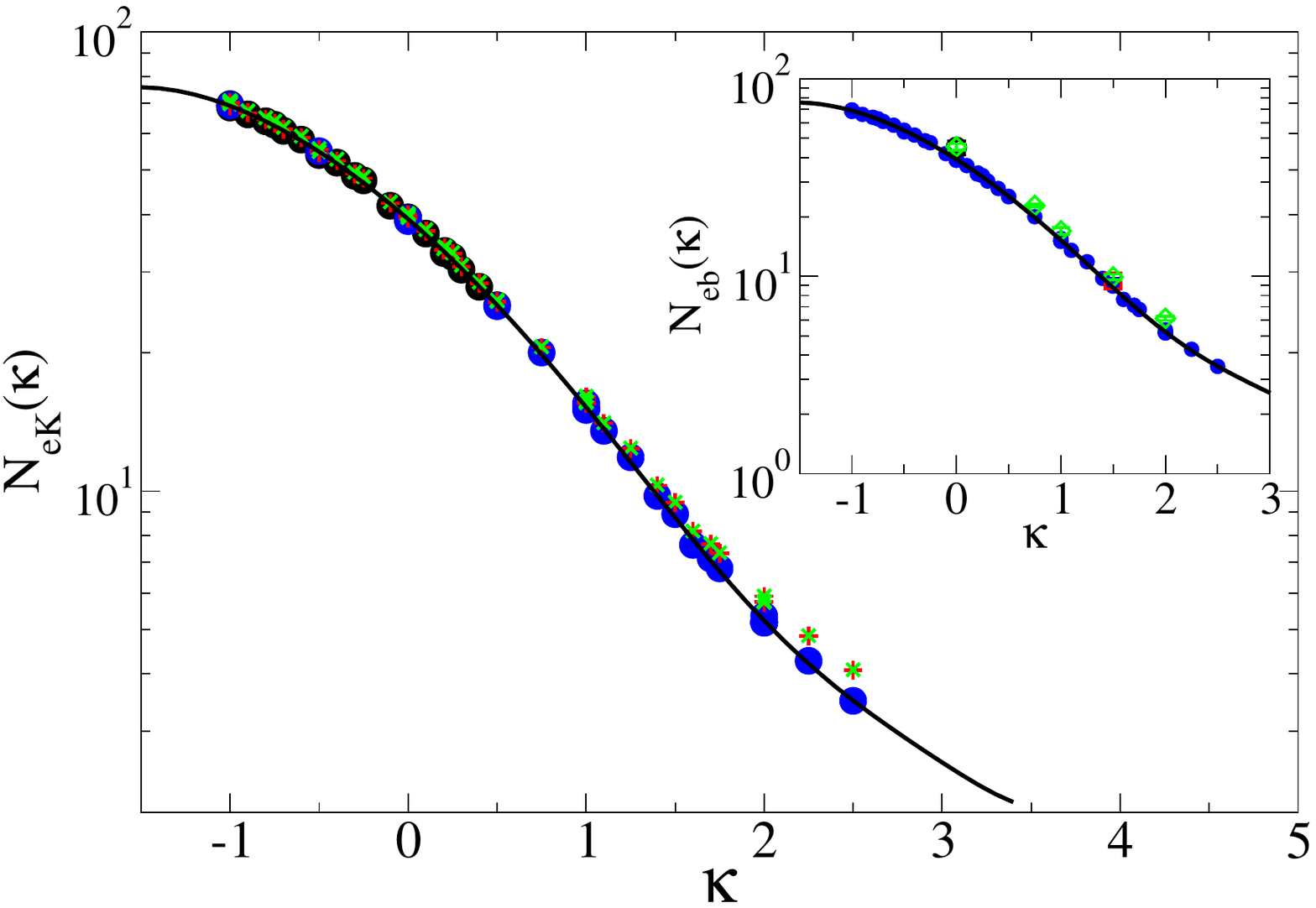}

\caption{\label{fig:Entanglementlength}Number of Kuhn segments between
entanglements $N_{eK}$ vs stiffness parameter for the KG polymer
melts. Our finite $N_{eK}$ corrected estimate eq. (\ref{eq:nenew}) $85<Z<200$ (black
$\circ$), and $Z>200$ (blue $\circ$), the classical estimator eq. (\ref{eq:neclassical})
(red $+$), the Hoy estimator eq. (\ref{eq:hoy}) (green $\times$).
The inset shows a comparison of our estimated $N_{eK}$ (small blue circles)
compared to literature results from Hoy et al.\cite{hoy2009topological} (black circle),
Hsu et al.\cite{hsu2016static} (red box), and Moreira et al.\cite{moreira2015direct}
(green diamonds). Our interpolation  eq. (\ref{eq:nekinterpolation}) is also
shown (solid black lines).
}
\end{figure}

Figure \ref{fig:Entanglementlength} shows our results for the dependence of the number of Kuhn segments between entanglements on chain stiffness. As chains
become stiffer, their spatial size increases, and hence the chains become more
strongly entangled as already expected from Fig. \ref{fig:visualization}.  This
leads to the observed progressive decrease in the number of Kuhn units between
entanglements. Note that if we continue to increase the stiffness far above
$l_K \gg 10\sigma$ i.e. $\kappa\gg 6$, then we expect the onset of
a isotropic to nematic transition.\cite{faller1999local}

We observe excellent agreement between the classical estimator eq. (\ref{eq:neclassical})
and the Hoy estimator eq. (\ref{eq:hoy}) indicating that our melts are sufficiently
long for finite-size effects to be irrelevant. We also observe good agreement
between our new estimator eq. (\ref{eq:nenew}) and the other estimators for 
flexible chains with $N_{eK}>10$, but for the stiffer and more entangled chains,
we can see that the previous estimators progressively overestimate the number
of entanglements by up to $20\%$ for the melts with the stiffest chains. The
solid line shown in Figure \ref{fig:Entanglementlength} is an empirical
interpolation given by eq. (\ref{eq:nekinterpolation}) to describe the
dependence of the entanglement length on stiffness.

% Due to the finite size of the melts we also
%expect a purely statistical sample-to-sample variation of the entanglement length.
%We have estimated it by generating at least six statistically independent and
%identically equilibrated melts for stiffness $\kappa=-1,0,1,2\epsilon$ and performed the PPA
%analysis of those melts. The relative sample variation of the entanglement
%length measured as the standard deviation divided by the mean is $0.4\%$ for
%$\kappa=-1\epsilon$ and drops down to $0.1\%$ for $\kappa=2\epsilon$. 

A potential source of error is the neglect of self- and image-entanglements
in the simplest PPA algorithm \cite{everaers2004rheology}, which disables
all intra-chain excluded volume interactions. 
The results reported here were obtained with a local version of
PPA~\cite{sukumaran2005identifying} which preserves self- and image-entanglements. 
While we had shown previously that self-entanglements may be safely
neglected~\cite{sukumaran2005identifying}, we briefly address the issue
of image-entanglements. The problem is easily understood for the extreme case
of a melt composed of a single chain in periodic boundary conditions,
which would appear to be unentangled in the simple version of the algorithm. 
In general, the importance of image-entanglements depends on the ratio
$r=V^{1/3}/(l_K(\kappa) \sqrt{N_K})$ where $V$ is the
volume of the cubic simulation box. 
Ideally, systems should be much larger than the individual chains i.e. $r > 1$
to limit self-interactions. However for very long chains, it is difficult to fulfil this condition. 
For the our melts with $N_b=10000$ the ratio varies from
$r(\kappa=-1)=1.8$ down to $r(\kappa=2)=1.3$ suggesting that
we should not expect significant finite box size effects.\cite{moreira2015direct}
Nonetheless, when we submitted our PPA meshes to a subsequent PPA analysis
disregarding all self- (and hence also image) entanglements, we found indeed
only a small but systematic increase of the entanglement length by $3.5\%$,
which we observed to be independent of chain stiffness.

% did \cite{moreira2015direct} introduce this ratio with respect to
% PPA or just any kind of interactions between images?}
%
% This was to study PPA, and based on the Sathish paper they conclude 
% that self-entanglements can be neglected when Sqrt<R^2> >> L
%

Figure \ref{fig:Entanglementlength} also contains previous PPA results
from the literature. Hoy et al.\cite{hoy2009topological} estimated the
asymptotic entanglement length for $\kappa=0$ using melts of
$N_b=100,\dots,3500$ for constant total numbers of beads. In this case,
while their longest chains reach $Z(\kappa=0)=46$, the sample
contained only as few as $M=80$ chains.  These melts were equilibrated
using double bridging.  Hoy et al. obtained $N_{eb}=86.1$. Hou et al.
estimated the systematic error due to various PPA algorithms and the
extrapolation schemes to be $\pm 7$.\cite{hou2010stress}  Recently,
Moreira et al.\cite{moreira2015direct} used more powerful equilibration
methods and were able to equilibrate melts up to $N_b=2000$ and $M=1000$
for stiffness $\kappa=0,0.75,1,1.5,2$ i.e. $Z(\kappa=0)=26$ up
to $Z(\kappa=2)=97$ entanglements.  Finally, Hsu et al.\cite{hsu2016static}
equilibrated melts up to $N_b=2000$ for $\kappa=1.5$ (i.e. $Z=74$). 
These results appears to be in good agreement with our data. We see that
Literature data are slightly but systematically above our data, which 
could be explained by the usage of a Kuhn length that has not been corrected
for incompressibility effects and/or the use of the old PPA estimators which
do not correct for stiffness.

\subsection{Center-of-mass motion}\label{sec:CM motion results}

\begin{figure}
\includegraphics[angle=\Angle,width=0.5\columnwidth]{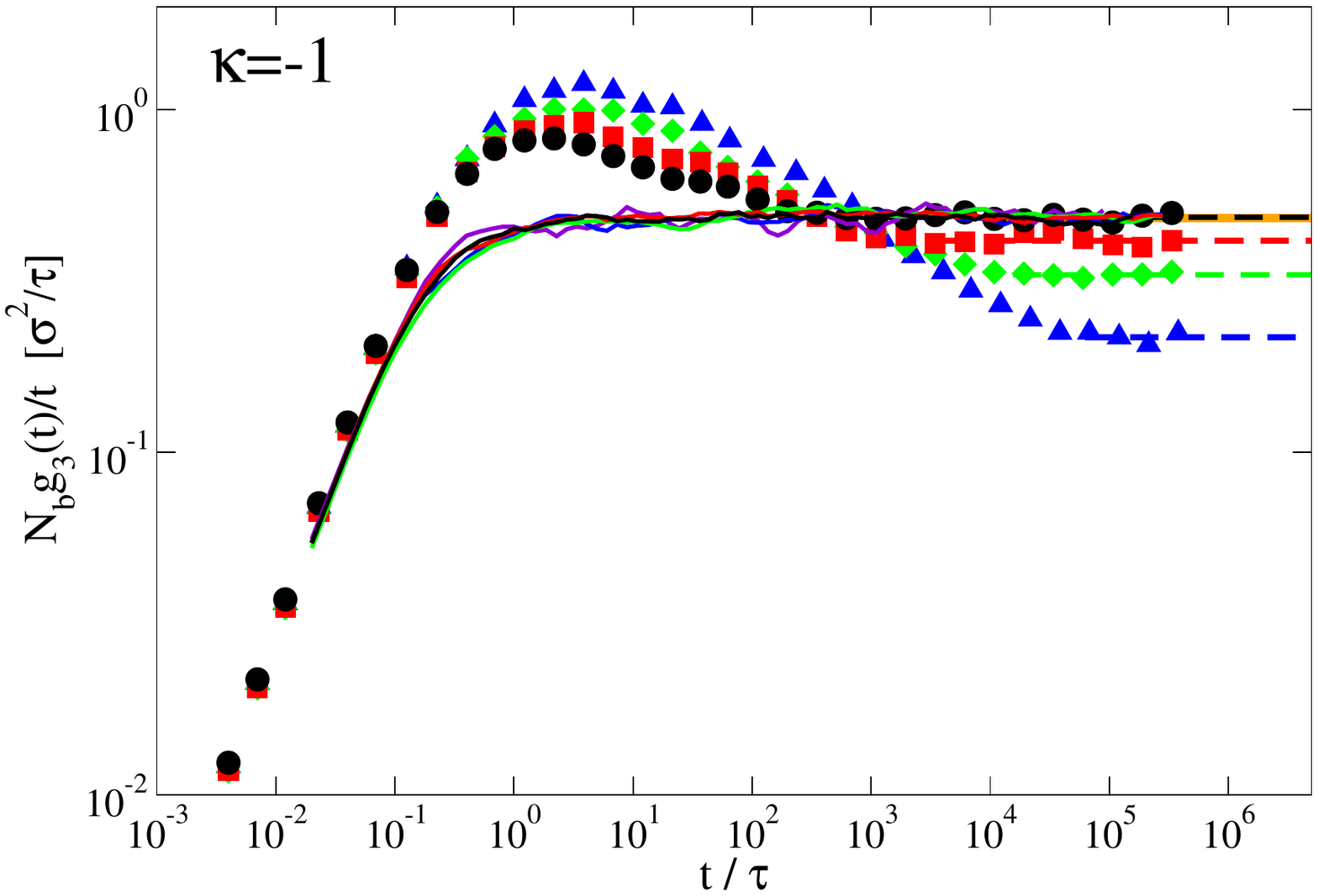}\includegraphics[angle=\Angle,width=0.5\columnwidth]{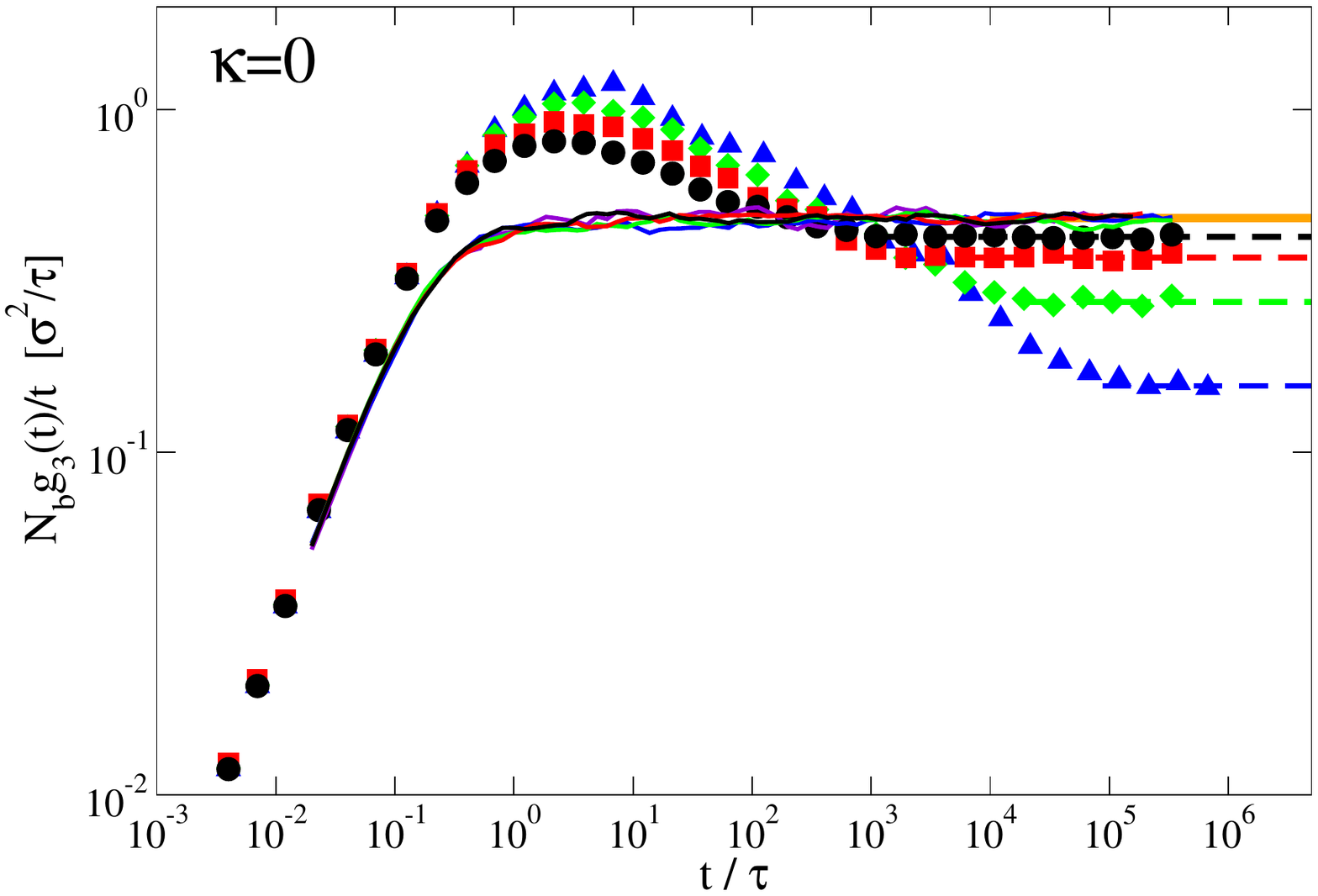}
\includegraphics[angle=\Angle,width=0.5\columnwidth]{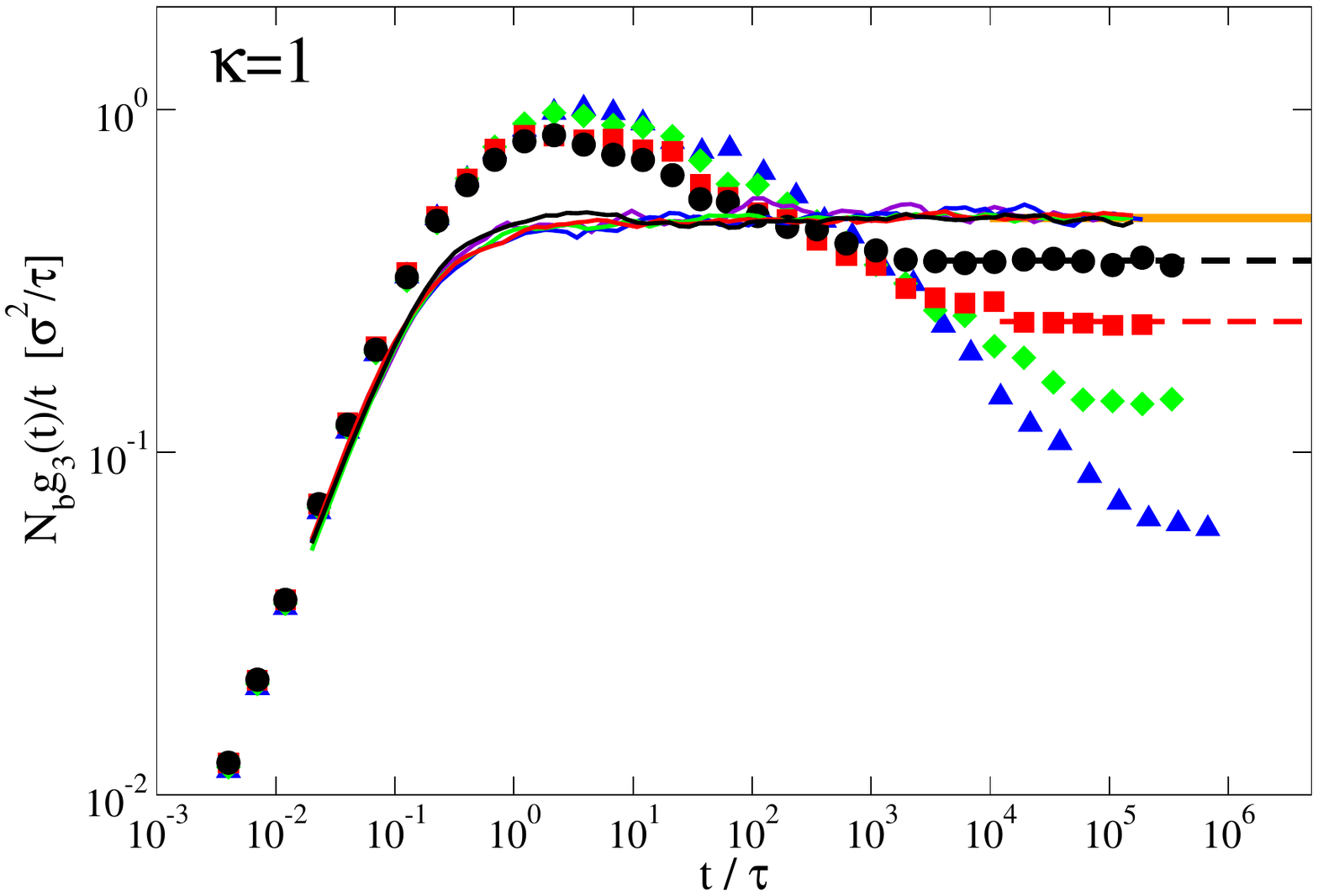}\includegraphics[angle=\Angle,width=0.5\columnwidth]{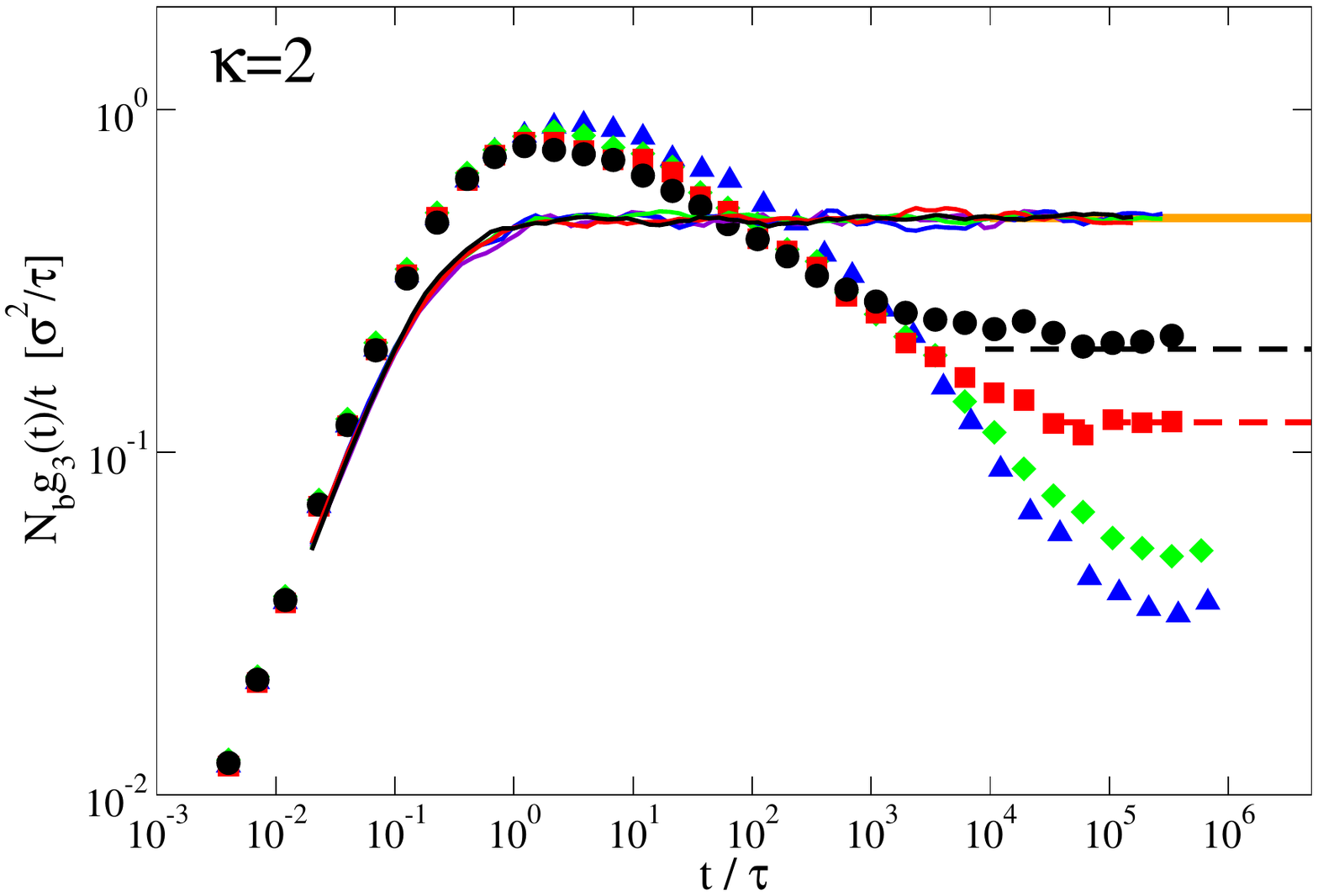}
\caption{\label{fig:g3red}
Normalized CM mean-square displacements, $g_3(t) N_b/t$ as a function of time. Colors indicate chain length ($N_K=10,20,40,80,160$ shown as black, red, green, blue, 
violet, respectively), panels show results for different values of chain stiffness. Symbols represent data for KG melts, solid lines results for corresponding phantom KG chains. Dashed lines indicate our fits of the long-time diffusion, Eq.~(\ref{eq:cmdiffusion}), in the fully interacting system. The theoretical phantom prediction $g_3(t) N_b/t = 6 k_B T/\zeta_b$ is shown as thick orange line.
}
\end{figure}

%\begin{figure}
%\includegraphics[angle=\Angle,width=0.5\columnwidth]{g3_kuhnred_kappa-1}\includegraphics[angle=\Angle,width=0.5\columnwidth]{g3_kuhnred_kappa0}
%\includegraphics[angle=\Angle,width=0.5\columnwidth]{g3_kuhnred_kappa1}\includegraphics[angle=\Angle,width=0.5\columnwidth]{g3_kuhnred_kappa2}
%\caption{\label{fig:g3redkuhn}  %notice KUHN here
%Normalized CM mean-square displacements, $g_3(t) N_b/t$ as a function of time. Colors indicate chain length ($N_K=10,20,40,80,160$ shown as black, red, green, blue, violet, respectively), panels show results for different values of chain stiffness. Symbols represent data for KG melts, solid lines results for corresponding phantom KG chains. Dashed lines indicate our fits of the long-time diffusion, Eq.~(\ref{eq:cmdiffusion}), in the fully interacting system. The theoretical phantom prediction $\zeta_{cm} g_3(t) / [6 kT t] = 1 $ is shown as thick orange line.
%}
%\end{figure}

\begin{figure}
\includegraphics[angle=\Angle,width=0.5\columnwidth]{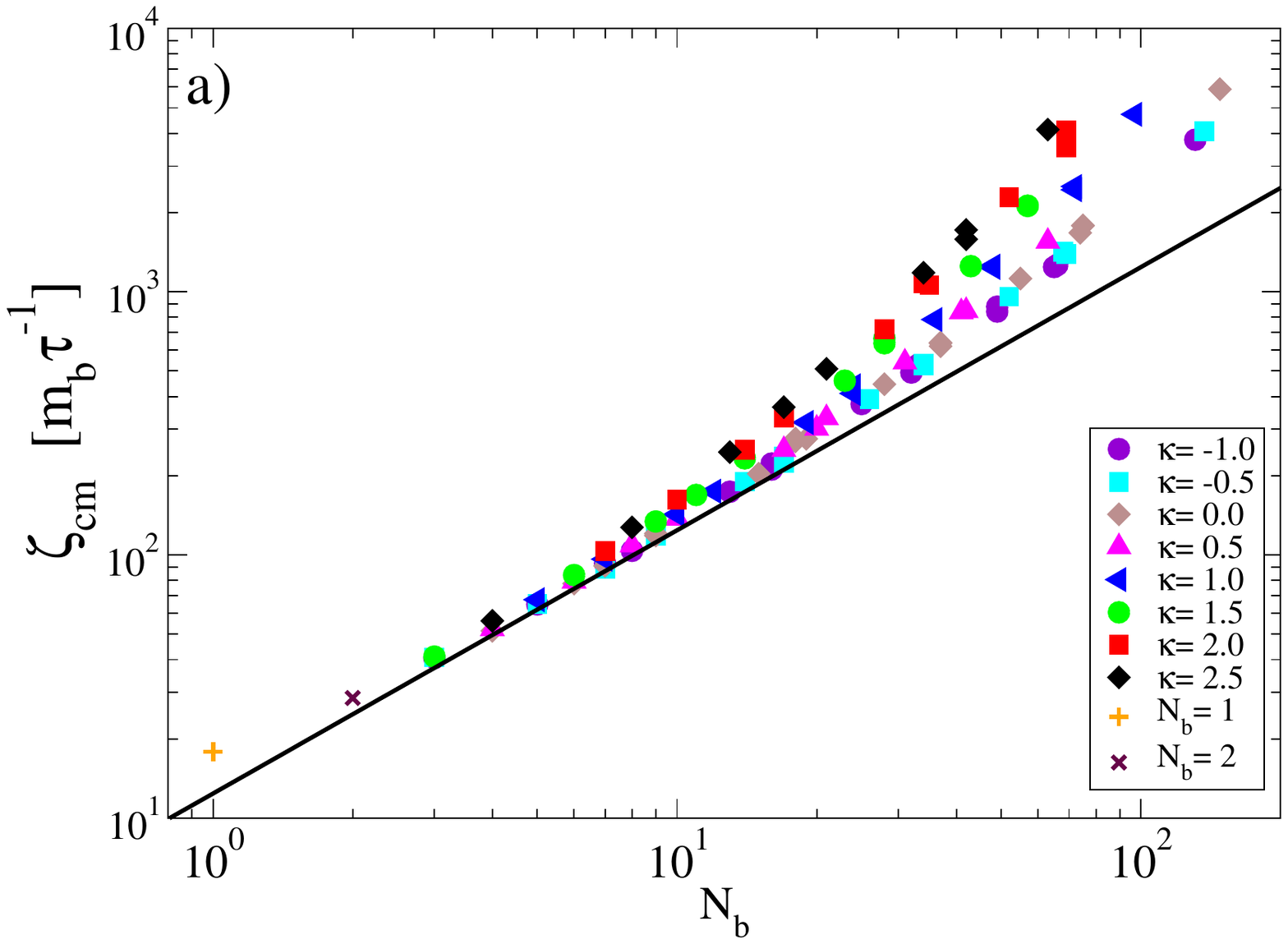}
\includegraphics[angle=\Angle,width=0.5\columnwidth]{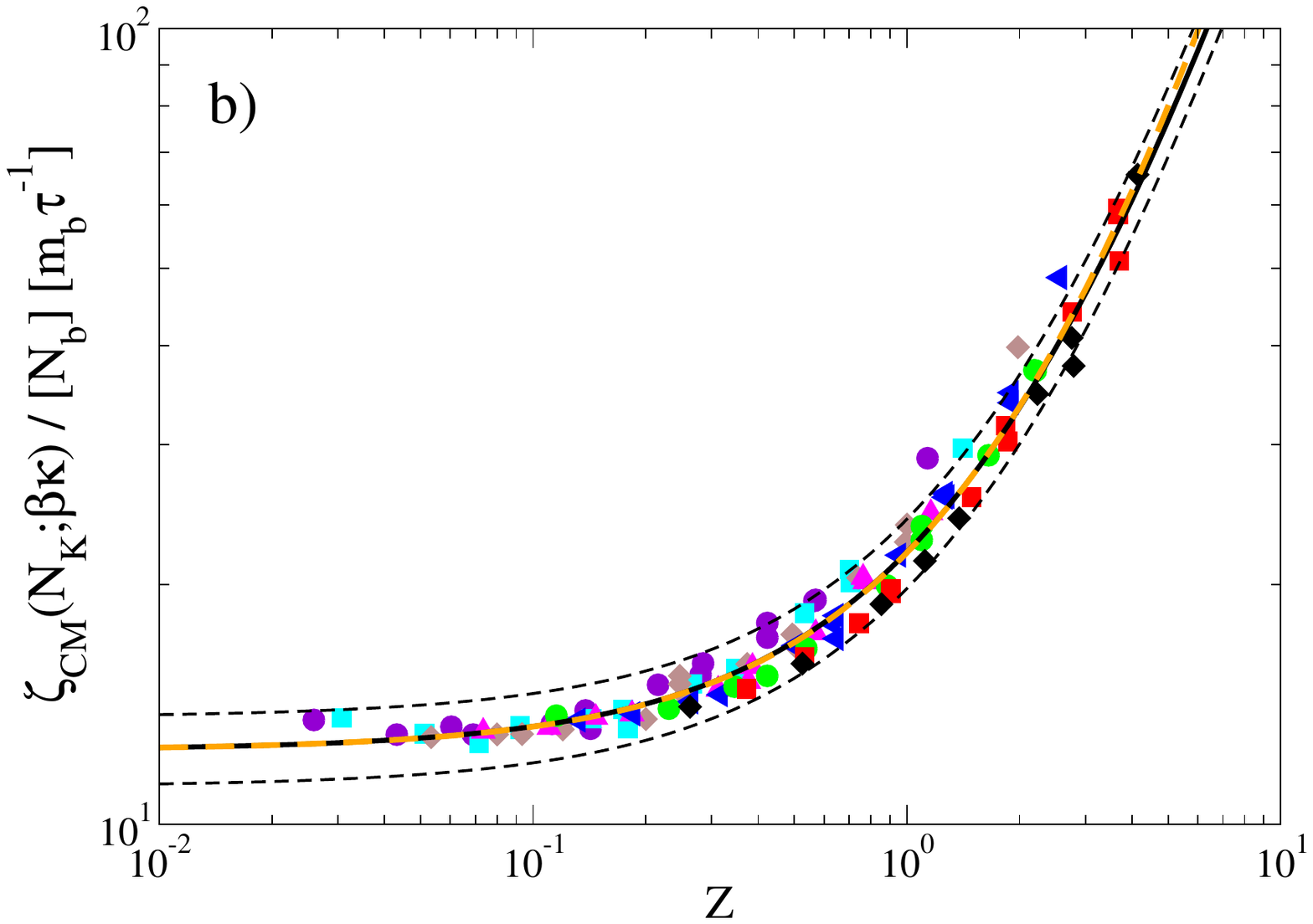}

\caption{\label{fig:Kuhn-friction}
Measured CM friction as function of chain length (a) 
and estimated bead friction as function of number of entanglements (b). Colors indicate chain stiffness.  
Shown is also the Rouse approximation $\zeta_{cm}(N_b)=\zeta_b N_b$ (a: solid black line),
the polynomium  approximant eq. (\ref{eq:f(Z)})  (b: orange dashed line), and the empirical
Pad\'e approximant eq. (\ref{eq:f(Z) Pade})  (b: solid line, dashed line indicate a $10\%$ error of $\zeta_b$). 
}
\end{figure}

Figure~\ref{fig:g3red} shows our results for the CM motion of KG and phantom KG chains.
The different panels display $g_3(t)$ for varying chain lengths for four characteristic values of the reduced bending stiffness, $\kappa$. 
Data for fully interacting KG melts are shown as symbols, while results for corresponding Phantom KG chains are represented as solid lines.
In Figure~\ref{fig:g3red} we plot $g_3(t) N_b/t$, because in this representation data should collapse to a horizontal line, $6 k_BT/\zeta_b$, if the chains CM exhibited simple diffusion with a chain length independent friction per bead, $\zeta_{cm}=N_b \zeta_b$, as assumed by the Rouse model. By construction  this is the behavior shown by the Phantom KG model (Eq.~(\ref{eq:PhantomKGLangevin})) after an initial ballistic regime. The objective of the present and the subsequent section is to justfiy and validate our choice of  the effective bead friction, $\zeta_b$, in single-chain models for KG melts.

In agreement with previous computational\cite{kremer1990dynamics,faller2001chain,bulacu2005effect,hsu2016static} and theoretical results\cite{DoiEdwards86,Wittmer_Meyer_PRL04,Semenov2011PRL}, fully interacting chains in KG melts show a more complex behavior than Phantom KG chains.
At the earliest times, the chains move ballistically in perfect agreement with their phantom counterparts.
At intermediate times, the CM motion is accelerated relative to the phantom chains and relative to their own asymptotic diffusion. In agreement with previous computational and theoretical results\cite{Semenov2012PRE,Semenov2011PRL,Semenov2012PRE2,Semenov2012JPhys}, 
%{\bf results in Semenov papers are not for KG melts, but force capped variations to avoid entanglement effects.}, Well, I think, it's ok not to point this out.
the effect is stronger for longer chains. In the asymptotic regime the order is reversed with a stronger spreading for stiffer, more strongly entangled chains. 

In the following, we focus on the long-time diffusion of unentangled and weakly entangled chains with $N_{K}=1,2,3,$$4,5,8,10,$$15,20,30,40,80$. For completeness, we have also simulated KG monomers and dimers, $N_b=1,2$, whose dynamics is independent of chain stiffness, as well as trimers, $N_b=3$, which are the shortest chains which could exhibit a $\kappa$-dependent dynamics. In all cases, we have obtained the CM diffusion constant, $D_{cm}$, from Eq.~(\ref{eq:cmdiffusion}) by sampling plateau values of $g_3(t)/[6t]$ for log-equidistant times and discarding simulations where the standard deviation of the sampled values exceeded $5\%$ of their average value. The fits are indicated by the dashed lines in Fig.~\ref{fig:g3red}.
Figure \ref{fig:Kuhn-friction}a shows the increase of the CM friction constant, $\zeta_{cm}(N_b)=k_BT/D_{cm}(N_b)$, with chain length. 
For very short oligomers, the apparent bead friction, $\zeta_{cm}(N_b)/N_b$, decreases slightly with increasing chain length:
$\zeta_{cm}(1)/1=17.9 m_b\tau^{-1}$,  $\zeta_{cm}(2)/2=14.3 m_b\tau^{-1}$, while $12.9 m_b\tau^{-1}\le \zeta_{cm}(\kappa)(3)/3\le 13.8m_b\tau^{-1}$.
For longer chains, the apparent bead friction increases with chain length. For our stiffest and most strongly entangled systems the effect sets in already at $N_b=4$. For our most flexible systems, the apparent bead friction plateaus at $\zeta_{cm}(N_b)/N_b = 12.4m_b\tau^{-1}$ around $N_b=10$.

Figure \ref{fig:Kuhn-friction}b shows our results for the apparent bead friction for all chains with $N_b\ge3$ as a function of chain length in entanglement units, $Z=N_b/N_e(\kappa)$. They allow us to draw non-trivial conclusions from three independent obseravations:
\begin{enumerate}
\item Contrary to experiment~\cite{ColbyFettersGraessleyMM1987}, 
% Colby, R. H.; Fetters, L. J.; Graessley, W. W. Macromolecules1987, 20, 2226
$\zeta_{cm}(N_b)/N_b$ essentially reaches a plateau for short chain KG melts. Plausibly, studying the model with purely repulsive interactions at constant density, $\rho_b=0.85\sigma^{-3}$, avoids the need for an iso-free-volume correction~\cite{ColbyFettersGraessleyMM1987} of the data. 
\item The upturn can be safely attributed to entanglement effects, because it occurs for all systems at the same effective chain length, $Z=N_b/N_e(\kappa)$. With $N_e(\kappa)$ being independently derived from PPA, this is additional strong evidence that the results of this  {\em static} analysis of the microscopic topological state is relevant to  the chain {\em dynamics}.
\item The collapse of the data for different values of $\kappa$ suggests a {\em stiffness independent} value of 
\begin{equation}\label{eq:zeta_b}
\zeta_b= 12.4m_b\tau^{-1} 
\end{equation}
for the microscopic bead friction, which we use in theoretical calculations throughout the remainder of the data analysis as well as in the simulations of the Phantom KG chains. The scatter of the data for weakly entangled systems suggests an error bar on the bead friction estimate below $10\%$. 
For comparison, Kremer and Grest~\cite{kremer1990dynamics} obtained $\zeta_b=16\pm2 m_b\tau^{-1}$ for the standard KG model with $\kappa=0$ from the CM motion of chains with $Z\approx1$ and a Rouse mode analysis of their internal dynamics. 
Note, however, that at this point it is difficult to ascertain this value with high precision for stiffer chains, since entanglement effects affect their CM diffusion already for very short chains. 
\end{enumerate}

So far, we have only made use of the fact that the data perfectly collapse, when plotted as a function of $Z$, to conclude, that the effective bead friction is independent of stiffness, Eq.~(\ref{eq:zeta_b}). Also included in Figure ~\ref{fig:Kuhn-friction} is the result of a fit of the function
\begin{equation}\label{eq:f(Z)}
f(Z) \equiv \frac{\zeta_{cm}(Z)}{N_b \zeta_b} = \frac{D_{cm}^{(Rouse)}(Z)}{D_{cm}(Z)} = \frac{\tau_{max}(Z)}{\tau_{max}^{(Rouse)}(Z)} 
\end{equation}
to a second order polynomial, $f(Z) = 1 + a Z + b Z^2$ with $a=0.69$ and $b=0.08$. Asymptotically, we expect $\lim_{Z\rightarrow\infty} f(Z) = 3Z$ (Eq.~(\ref{eq:tau_d})), suggesting a Pad\'e-type interpolation between the two limits,
\begin{equation}\label{eq:f(Z) Pade}
f(Z) = \frac{1+c Z + 3 d Z^2}{1+d Z}\ ,
\end{equation}
where $c=a-b/(a-3)=0.73$ and $d=-b/(a-3)=0.035$. This interpolation is shown as the black solid line in Fig.~\ref{fig:Kuhn-friction} which is both in very good agreement with the simulation data as well as the polynomial within the range of entanglement lengths studied here.

\subsection{Monomer motion}\label{sec:Monomer motion results}

\begin{figure}
\includegraphics[angle=\Angle,width=0.5\columnwidth]{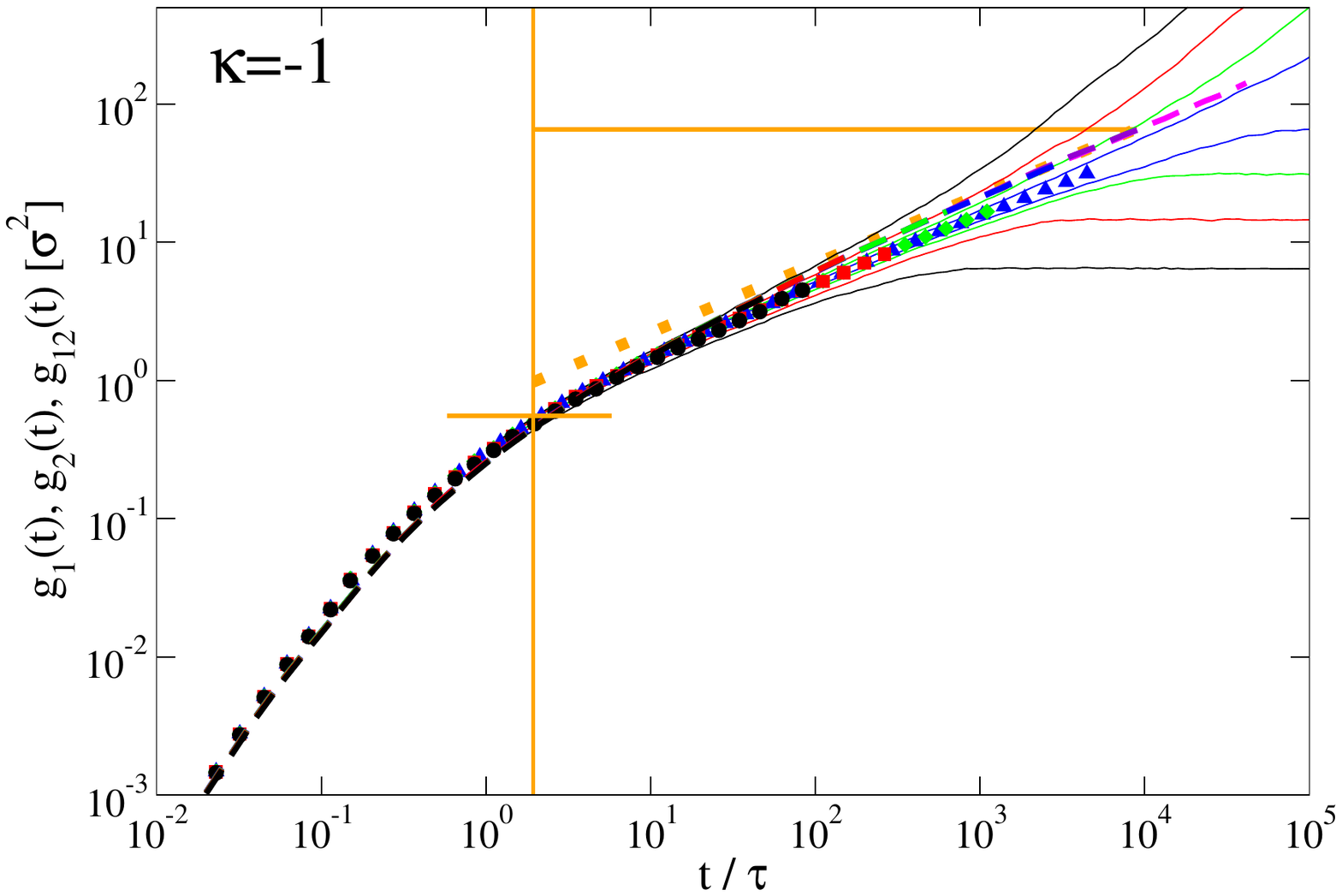}\includegraphics[angle=\Angle,width=0.5\columnwidth]{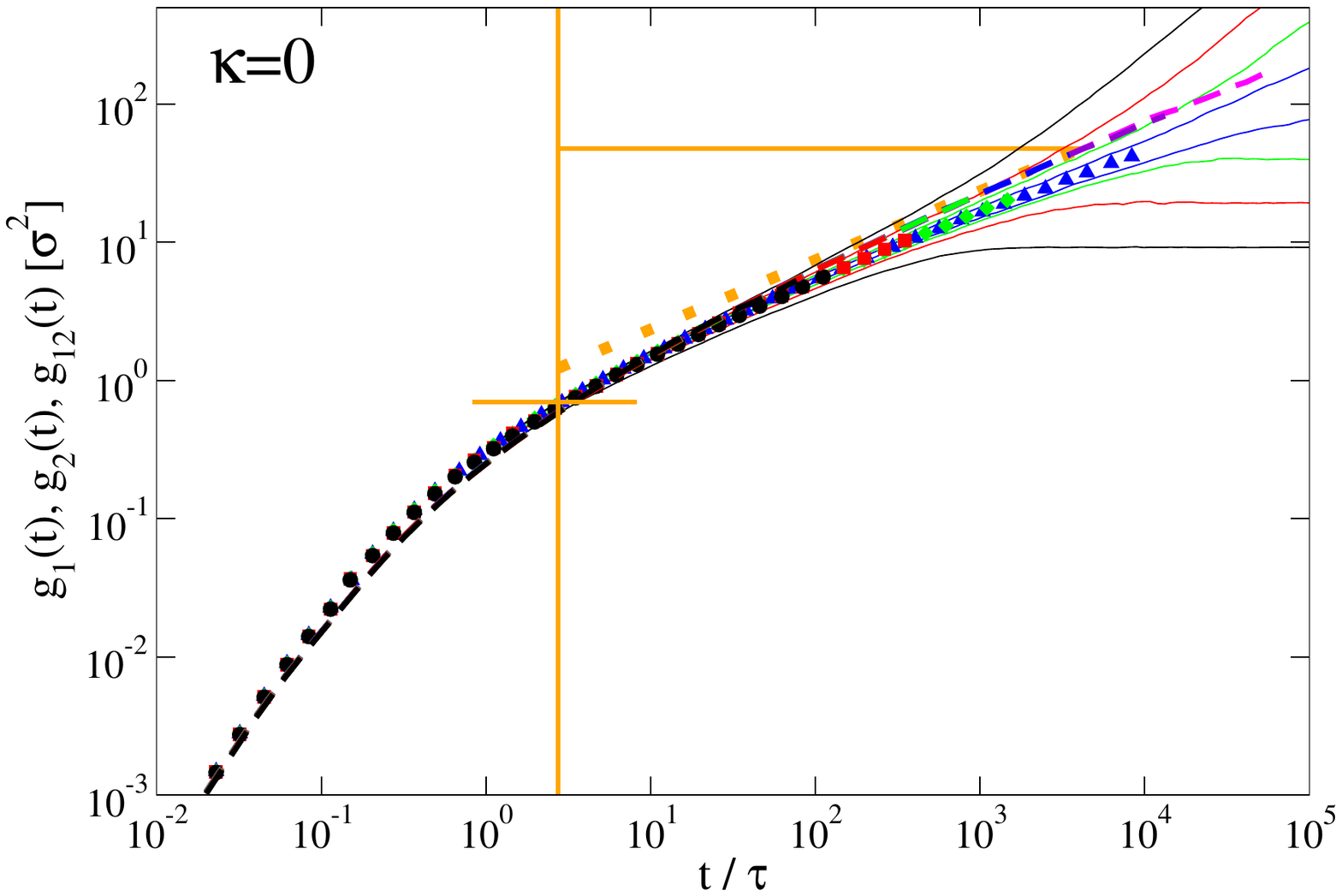}
\includegraphics[angle=\Angle,width=0.5\columnwidth]{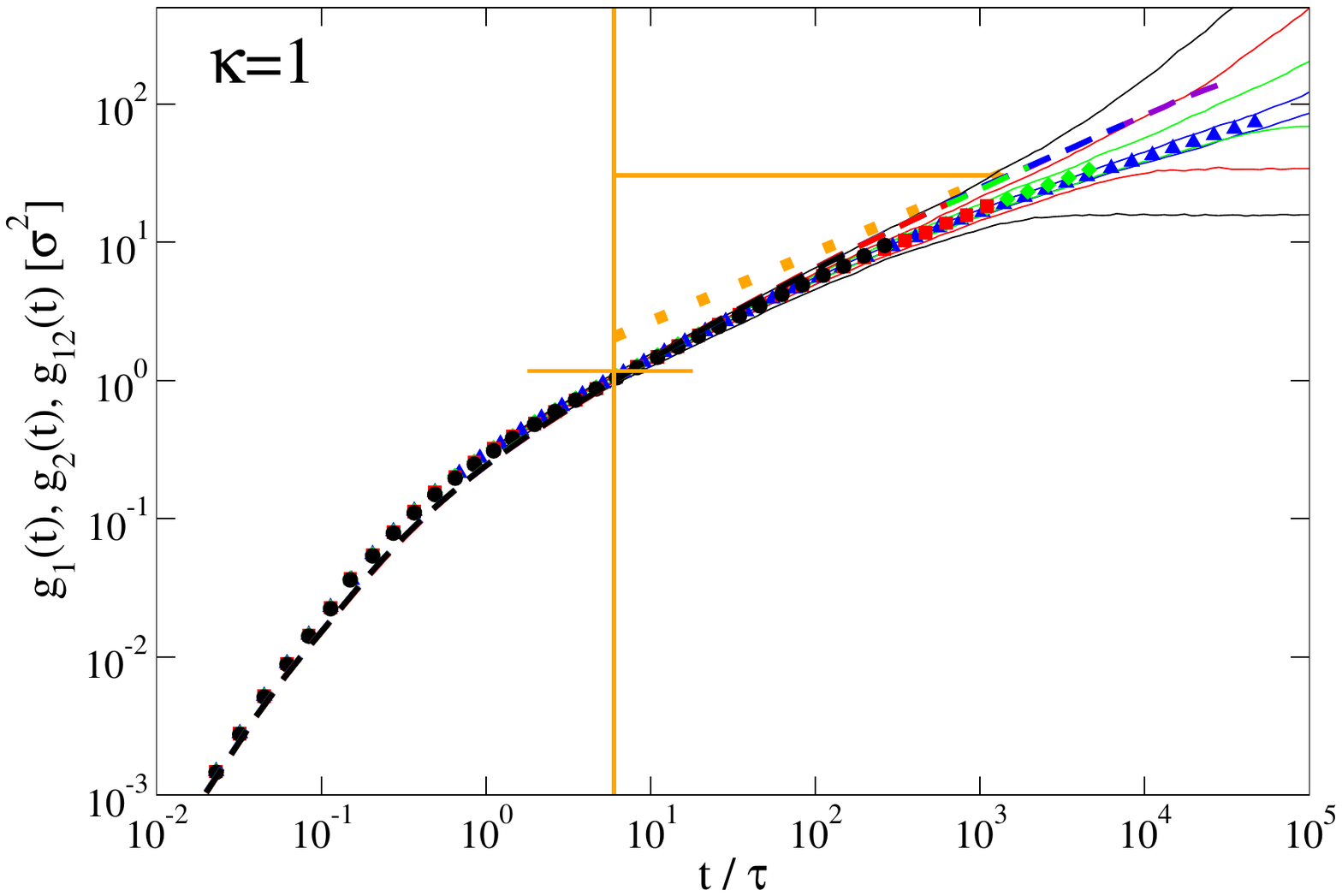}\includegraphics[angle=\Angle,width=0.5\columnwidth]{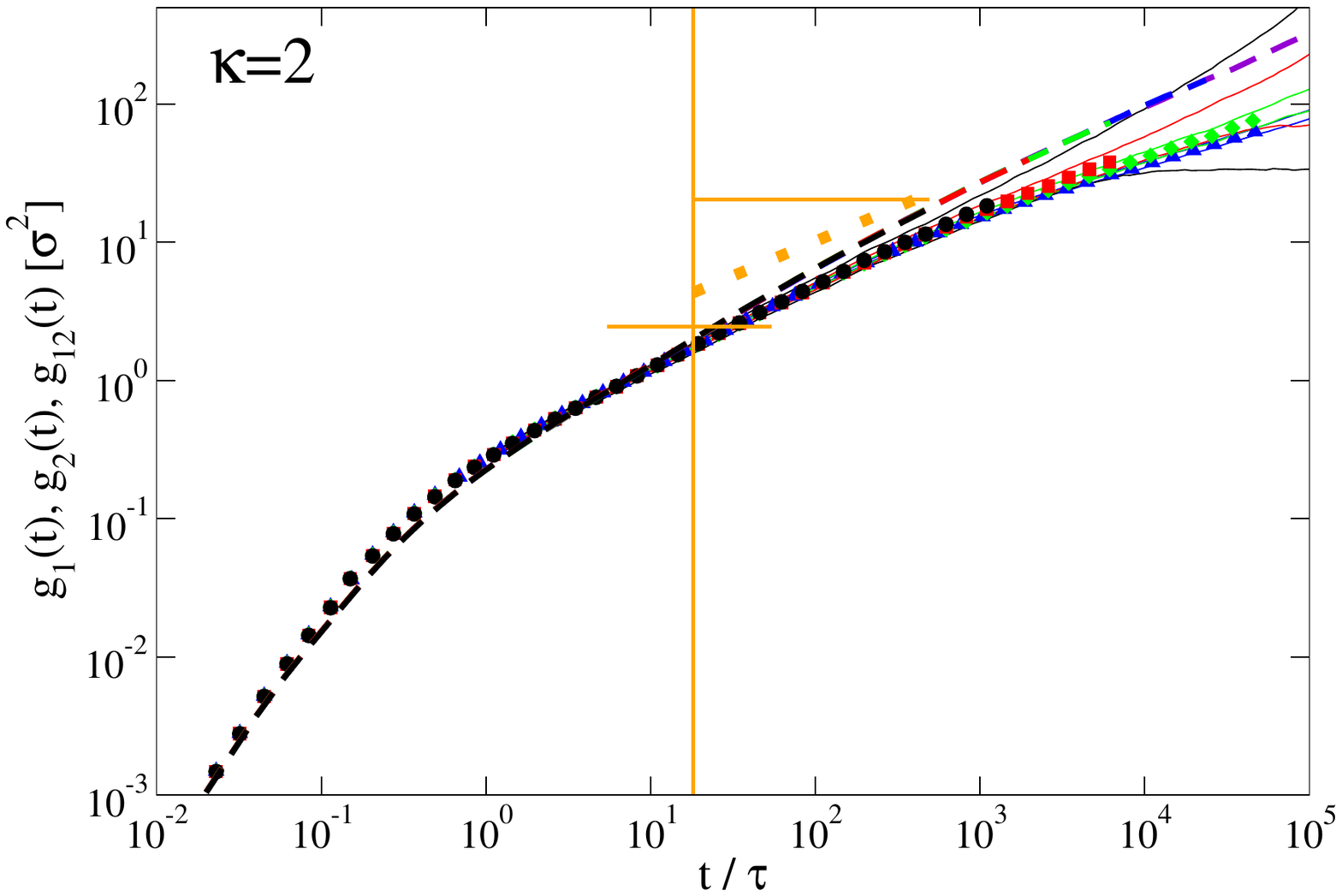}
\caption{\label{fig:g12}
Monomer mean-square displacements $g_1(t)$, $g_2(t)$ (thin lines) and $g_{12}(t)$ eq. (\ref{eq:g12 def}) (symbols) as a function of time for the same systems as in Fig.~\ref{fig:g3red}.
Also shown are $g_{12}(t)$ for Phantom KG chains of varying length (dashed lines) and the theoretical prediction of Rouse theory (orange dotted line).
 Colors indicate chain length, different panels show results for different values of chain stiffness. Also indicated are $\tau_K$ (vertical orange solid line), the displacements $2l_K^2 \langle R^2(N_K)\rangle/\pi^{1.5}$ corresponding to a Kuhn segment $N_K=1$ and also an entanglement segment $N_K=N_{eK}$ (horizontal orange solid lines).
}
\end{figure}

Figure~\ref{fig:g12} shows the monomer motion for the same systems as in Fig.~\ref{fig:g3red}. 
The observation of characteristic features of Rouse dynamics like the sub-diffusive monomer motion, Eq.~(\ref{eq:rouse g2}), 
%or the power-law decay of the shear relaxation modulus, Eq.~(\ref{eq:rouseshearrelaxation}), 
is limited to a time interval $\tau_K \ll t \ll \min(\tau_R,\tau_e)$.
For unentangled chains, $g_2(t)$ levels off on approaching the Rouse time, $\tau_R$, while $g_1(t)=g_2(t)+g_3(t)$ starts to be dominated by the CM diffusion.
Interestingly,  the two deviations almost cancel, if one considers the geometric mean, 
\begin{equation}\label{eq:g12 def}
g_{12}(t)\equiv\sqrt{g_1(t) g_2(t)}\ ,
\end{equation} 
of the two standard measures of monomer diffusion. Below the Rouse time, $g_1(t) \approx g_2(t)$, so that $g_{12}(t)\propto\sqrt{t/\tau_K}$. Beyond the Rouse time, $\sqrt{g_1(t) g_2(t)}\approx \sqrt{R_g^2\, g_3(t)}  \propto\sqrt{t/\tau_K}$ has the same scaling for entirely different reasons. 
Compared to the corresponding results for $g_1(t)$ and $g_2(t)$ (shown as thin lines in corresponding colors),  results for $\sqrt{g_1(t)g_2(t)}$ are significantly easier to interpret. 
Note that we have restricted the latter to times, when $\sqrt{g_1(t)g_2(t)}\le R_g^2$ and where the monomer motion is dominated by the internal dynamics. In this way we can be sure to obtain information, which is independent of the results for $g_3(t)$ for $t>\tau_R$, which we have analysed in the previous section. 

Like the CM motion, the monomer motion is ballistic at the earliest times. In this regime equipartition assures the perfect agreement between KG chains and their Phantom counterparts. The fact that this agreement extends into the diffusive regime is non-trivial.
Firstly, the perfect collapse of melt data for different chain lengths provides further evidence for a chain length independent effective bead friction.
Secondly, the excellent agreement with the Phantom KG simulations for {\em all} values of $\kappa$ validates our choice, Eq.~(\ref{eq:zeta_b}) of a stiffness-independent  effective bead friction in the single-chain Langevin equations, Eqs.~(\ref{eq:overdampedLangevin}) and (\ref{eq:PhantomKGLangevin}). In particular, this validation does not require the chains to reach the asymptotic Rouse regime, before entanglements effects modify their dynamics. We note that the melt $g_{12}(t)$ starts to significantly lag behind the Phantom KG reference curve before the monomers reach the entanglement scale, $g_1(t)\sim g_2(t)\sim \sqrt{g_1(t)g_2(t)}\ll d_T^2$. In particular, this deviation occurs before the Phantom KG chains have reached the asymptotic Rouse regime, Eq.~(\ref{eq:rouse g2 asymptotic}), which we have indicated in the relevant time interval, $\tau_K \le t \le \tau_e \approx N_{eK}^2\tau_K$. 

\section{Discussion}\label{sec:Discussion}

The following discussion of the behavior of KG melts is again organised across increasing length and time scales.
We start at the Kuhn scale with an empirical relation between the Kuhn length and the bending rigidity, $\kappa$, of our chains and an attempt to rationalize the observed dependence (Sec.~\ref{sec:lK discussion}). 
In a second step, we use the information on the bead friction to define the Kuhn units of friction, time and the viscosity at the bead scale (Sec.~\ref{sec:tauK discussion}). 
As a first test of the usefulness of these units, we consider the monomer motion at the Kuhn scale (Sec.~\ref{sec:g2 at Kuhn scale discussion}) and the shear relaxation modulus of unentangled melts (Sec.~\ref{sec:G(t) discussion}).
At the entanglement scale, we start out with an empirical relation between for entanglement length, $N_{eK}(\kappa)$, which we compare to available theoretical predictions (Sec.~ \ref{sec:NeK discussion}).  Next we define the entanglement time, $\tau_e$, as the time, when the monomer mean-square displacements of Phantom KG chains exceed the tube diameter (Sec.~\ref{sec:taue discussion}). 
The we discuss the chain dynamics around the entanglement time (Sec.~\ref{sec:g2 at entanglement scale discussion}). 
Our central result is the demonstration, that the tube model almost quantitatively predicts the magnitude of the monomer displacements in the $t^{1/4}$-regime (Sec.~\ref{sec:g2 at entanglement scale discussion}) given the Kuhn length, PPA input for the entanglement length, and the local friction at the bead scale inferred from {\em oligomer} diffusion data.
%
%In Sec.~\ref{sec:taumax discussion} we briefly discuss an extrapolation of the maximal relaxation time in our systems.

%We close with some general remarks about the Rouse model concerning its usefulness and correctness as well as strategies to determine the appropriate friction coefficients from data (Sec.~\ref{sec:Rouse discussion}).

\subsection{Kuhn length \label{sec:lK discussion}}

\begin{figure}
\includegraphics[angle=\Angle,width=0.5\columnwidth]{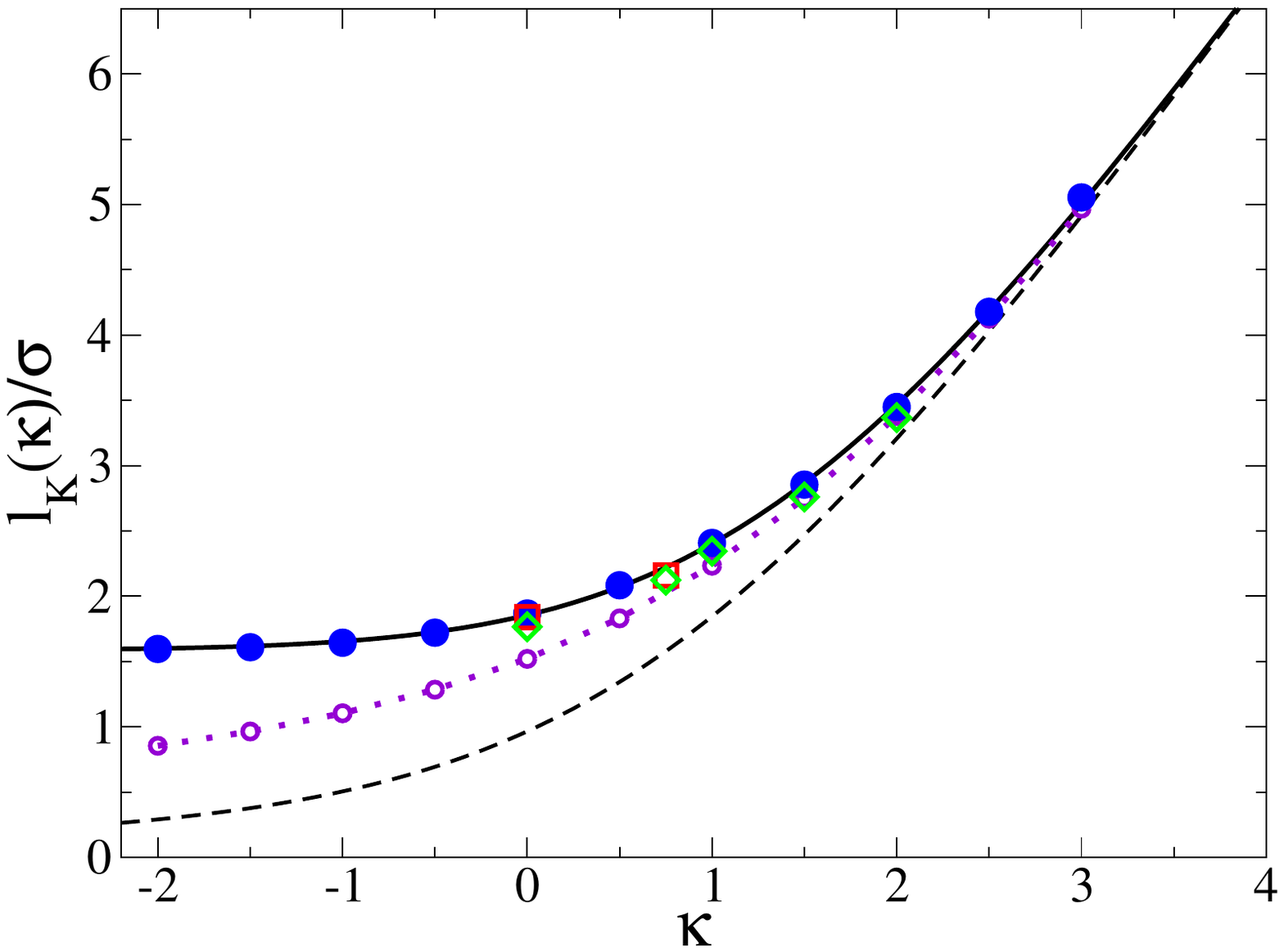}
\caption{\label{fig:Kuhn-length}
Extrapolated Kuhn length $l_{K}$ vs stiffness parameter $\kappa$ (blue filled circles)
and our parameterization eqs. (\ref{eq:lK_bare} and \ref{eq:KuhnParametrization}) (black solid line).
Also shown are the local Kuhn length $l_K^{(1)}$  (open violet circles connected by dotted line),
the bare Kuhn length $l_K^{(0)}(\kappa)$ (black dashed line),
literature data from Hoy et al.\cite{PhysRevE.72.061802}
(red box) and Moreira et al.\cite{moreira2015direct} (green diamond).
}
\end{figure}

Figure \ref{fig:Kuhn-length} summarizes our results for the Kuhn length of KG chains. Our estimates are slightly but systematically higher than those reported previously, because we accounted for long-range bond orientation correlations~\cite{Wittmer_Meyer_PRL04,wittmer2007polymer,wittmer2007intramolecular,beckrich2007intramolecular,semenov2010bond} in our extrapolations to the asymptotic limit. We have performed a block analysis, which shows that the statistical error is smaller than the symbol size.

Overall, the data show the expected monotonous increase of  the Kuhn length with increasing bending stiffness. 
Results for stiff KG chains with $\kappa>2$ are in excellent agreement with the expected bare Kuhn length, Eq.~(\ref{eq:lK_bare}). For these systems the Flory ideality hypothesis\cite{flory49} holds and excluded volume interactions are completely screened.
However, Eq.~(\ref{eq:lK_bare}) significantly underestimates the Kuhn length of the more flexible chains.

As a first step towards taking into account excluded volume effects, we can
devise a local Kuhn length estimate, $l_{K}^{(1)}$, (violet open circles and dashed lines in Fig.~\ref{fig:Kuhn-length} ) by evaluating Eq.~(\ref{eq:lK_definition}) using the actual distribution of bond angles
sampled from our simulation data. This includes the local effects of bead packing and, in particular, effects of the excluded volume between next-nearest neighbor beads along the chain. This approximation breaks down for $\kappa<1$, where pair-interactions and the
correlation hole can no longer be neglected.

The large-scale statistics of the intrinsically most flexible systems are influenced by long-range bond orientation correlations~\cite{Wittmer_Meyer_PRL04,wittmer2007polymer,wittmer2007intramolecular,beckrich2007intramolecular} with
$l_K > l_{K}^{(1)} > l_{K}^{(0)}$. To parameterize the Kuhn length dependence on chain stiffness we fitted the difference between the extrapolated and bare Kuhn lengths, 
\begin{equation}\label{eq:lK(kappa)}
l_K(\kappa) \equiv l_K^{(0)}(\kappa) + \Delta l_K(\kappa)\ ,
\end{equation}
by an empirical formula
\begin{equation}\label{eq:KuhnParametrization}
\frac{\Delta l_K(\kappa)}{\sigma} =   
%0.805994 \left(1 - \tanh\left[-0.0967646 + 0.384872\kappa  + 0.0373285 \kappa^2\right]\right)\ .
0.77 \left(\tanh \left(-0.03 \kappa^2-0.41 \kappa+0.16\right)+1\right)\ .
\end{equation}

We expect this parameterization to be reasonably accurate outside the range of $\kappa$-values for which we have data: systems with $\kappa>2.5$ should be well described by eq.~(\ref{eq:lK_bare}), while the effect of a stronger bias than $\kappa<-2$ towards large bending angles will be limited by the repulsive next-nearest neighbor excluded volume interactions.

The main panel of Fig.~\ref{fig:msid} illustrates the utility (as well as the limits) of the Kuhn length for describing the single-chain statistics. While $l_K$ properly characterizes the scale of the crossover from rigid rod to random walk behavior, the form of the crossover function is influenced by corrections to the Flory theorem, whose importance decreases with chain stiffness. A detailed analysis of the interplay between intrinsic stiffness and the effects of the incomplete screening of excluded volume effects in polymer melts\cite{Wittmer_Meyer_PRL04,wittmer2007polymer,wittmer2007intramolecular,beckrich2007intramolecular,semenov2010bond,Semenov2011PRL,Semenov2012JPhys,Semenov2012PRE,Semenov2012PRE2}  is beyond the scope of the present work. We will return to this point in a separate article.

\subsection{Kuhn time and friction \label{sec:tauK discussion}}

Single-chain models of the polymer {\em dynamics} (Secs.~\ref{sec:single chain Langevin dynamics} to \ref{sec:Rouse Dynamics}) require information on the local viscosity.
In Sections~\ref{sec:CM motion results} and \ref{sec:Monomer motion results} we have inferred a surprisingly simple, stiffness- and essentially chain length independent value for the effective bead friction, $\zeta_b=12.4m_b\tau^{-1}$, in KG melts, Eq.~(\ref{eq:zeta_b}).
Our conclusions are based on the agreement between results for KG melts and for Phantom KG chains, whose effective bending stiffness $\tilde\kappa(\kappa)$ reproduces the Kuhn lengths of the target KG melts.
While the accord is trivial for the initial ballistic regime at times $t\ll m_b/\zeta_b$, this is not at all the case around the Kuhn time, where the monomer dynamics is strongly influenced by intra- and inter-chain interactions. 
Our findings present a remarkable confirmation of 
(i) the intuition of Edwards, Freed and Muthukumar\cite{EdwardsFreed1974,freed1974polymer,freed1978polymer,muthukumar1977huggins}
that the many-chain dynamics in polymer melts may be represented by a single-chain model
and (ii) the quantitative analysis of Semenov {\it et al.}, who found that correlation hole effects~\cite{Semenov2012PRE} and viscoelastic hydrodynamic interactions~\cite{Semenov2011PRL,Semenov2012PRE2,Semenov2012JPhys} cause negligible corrections to the {\em long-term} CM diffusion and the {\em short-term} monomer motion.

In particular, we are now in a position to evaluate the Kuhn friction, time, and viscosity, Eqs.~(\ref{eq:zetaK}) to (\ref{eq:etaK}), as defined in Sec.~\ref{sec: Definition Kuhn time and friction}. Using Eq.~(\ref{eq:zeta_b}) for the bead friction in KG melts, these quantities can be expressed  in simulation units as
%i
\begin{eqnarray}
\zeta_K(\kappa) &\equiv&  \frac{l_K(\kappa)}{l_b} \zeta_b= 12.8 \left(\frac{l_K(\kappa)}{\sigma} \right)m_b \tau^{-1}
\label{eq:zetaK KG}\\
\tau_K(\kappa) &\equiv& \frac{1}{3\pi^2}\frac{\zeta_K(\kappa) l_K^2(\kappa)}{k_B T}= 0.434 \left( \frac{l_K(\kappa)}{\sigma} \right)^3 \tau 
\label{eq:tauK KG}\\
\eta_K(\kappa) &\equiv& \frac{1}{36}\frac{\zeta_K(\kappa)}{l_K(\kappa)} = 0.357\, m_b \sigma^{-1} \tau^{-1} \ .
\label{eq:etaK KG}
\end{eqnarray}
%The rescaling of the friction, Eq.~(\ref{eq:zetaK KG}), from the bead to the Kuhn scale is trivial. 
%The two subsequent relations represent convenient {\em definitions} of the Kuhn time and the viscosity at the Kuhn scale, which are based on the Rouse model (see Sec.~\ref{sec: Definition Kuhn time and friction}). 
The rapid increase of $\tau_K$ with stiffness is illustrated in Fig.~\ref{fig:KuhnEntanglementtimes}, which summarises results for characteristic time scales of KG melts as a function of the ratio, $\left(\frac{l_K(\kappa)}{\sigma} \right)$, of Kuhn length and bead size. As a side remark we note, that with $g_2(t)\propto l_K^2 (t/\tau_K)^{1/2} \ \propto (l_K/\sigma)^{1/2}$ the variation of the monomer diffusion in the Rouse regime is much weaker.

\subsection{Monomer motion around the Kuhn scale  \label{sec:g2 at Kuhn scale discussion}}
%\begin{figure}
%\includegraphics[angle=270,width=0.5\columnwidth]{g12_extrapolation}
%\includegraphics[angle=0,width=0.5\columnwidth]{g12_extrapolation_kuhn}
%\includegraphics[angle=\Angle,width=0.5\columnwidth]{g12_extrapolation_kuhn}

%\caption{\label{fig:g12Kuhn}
%Monomer diffusion $g_{12}(t)$ as a function of time  in Kuhn units for the $N_K=80$ systems
%from Fig.~\ref{fig:g12}. Colors indicate the effective chain
%stiffness: $\kappa=-1$ (black), $\kappa=0$ (red), $\kappa=1$ (green), and $\kappa=2$ (blue). 
%Colored symbols represent the fully interacting KG melts, while colored dashed lines show
%results for the corresponding Phantom KG chains. Thick orange lines indicate the Kuhn time and
%length, while the colored vertical/horizontal lines indicate the entanglement times/lengths of
%the corresponding systems from Fig. \protect\ref{fig:g12KuhnTimeSpace}.
%}
%\end{figure}

\begin{figure}
\includegraphics[angle=\Angle,width=0.3\columnwidth]{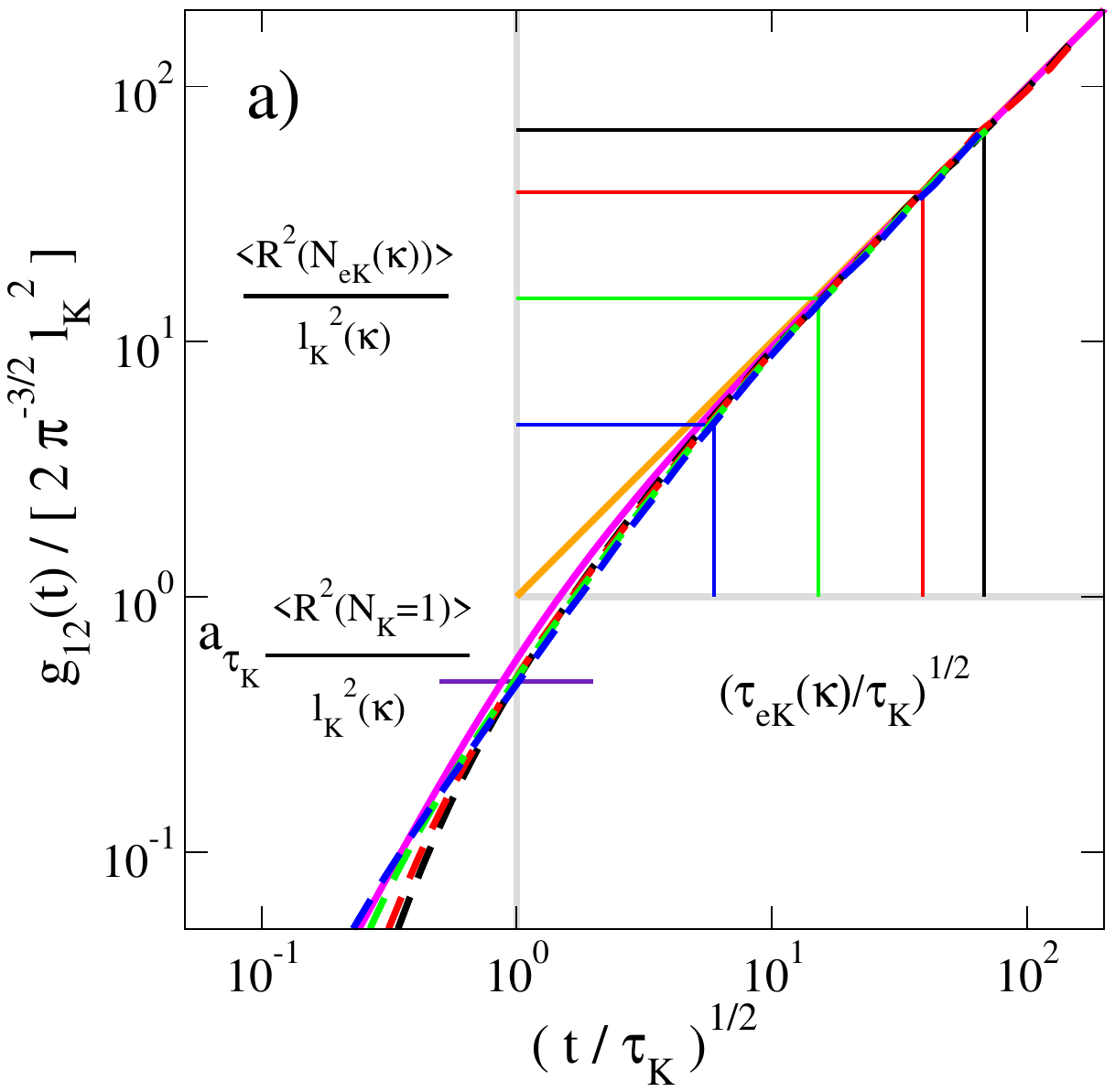}%
\includegraphics[angle=\Angle,width=0.3\columnwidth]{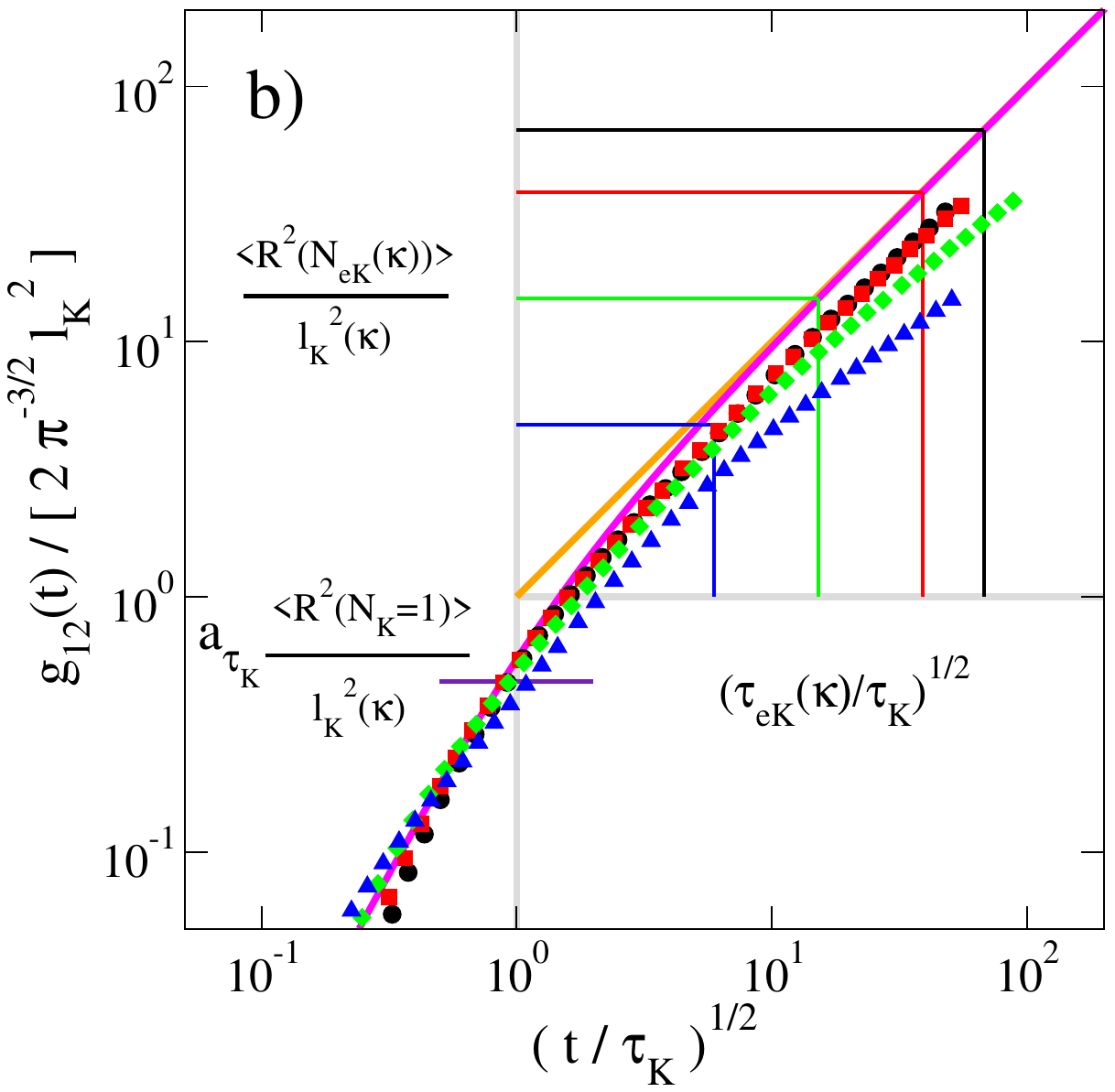}

\includegraphics[angle=\Angle,width=0.3\columnwidth]{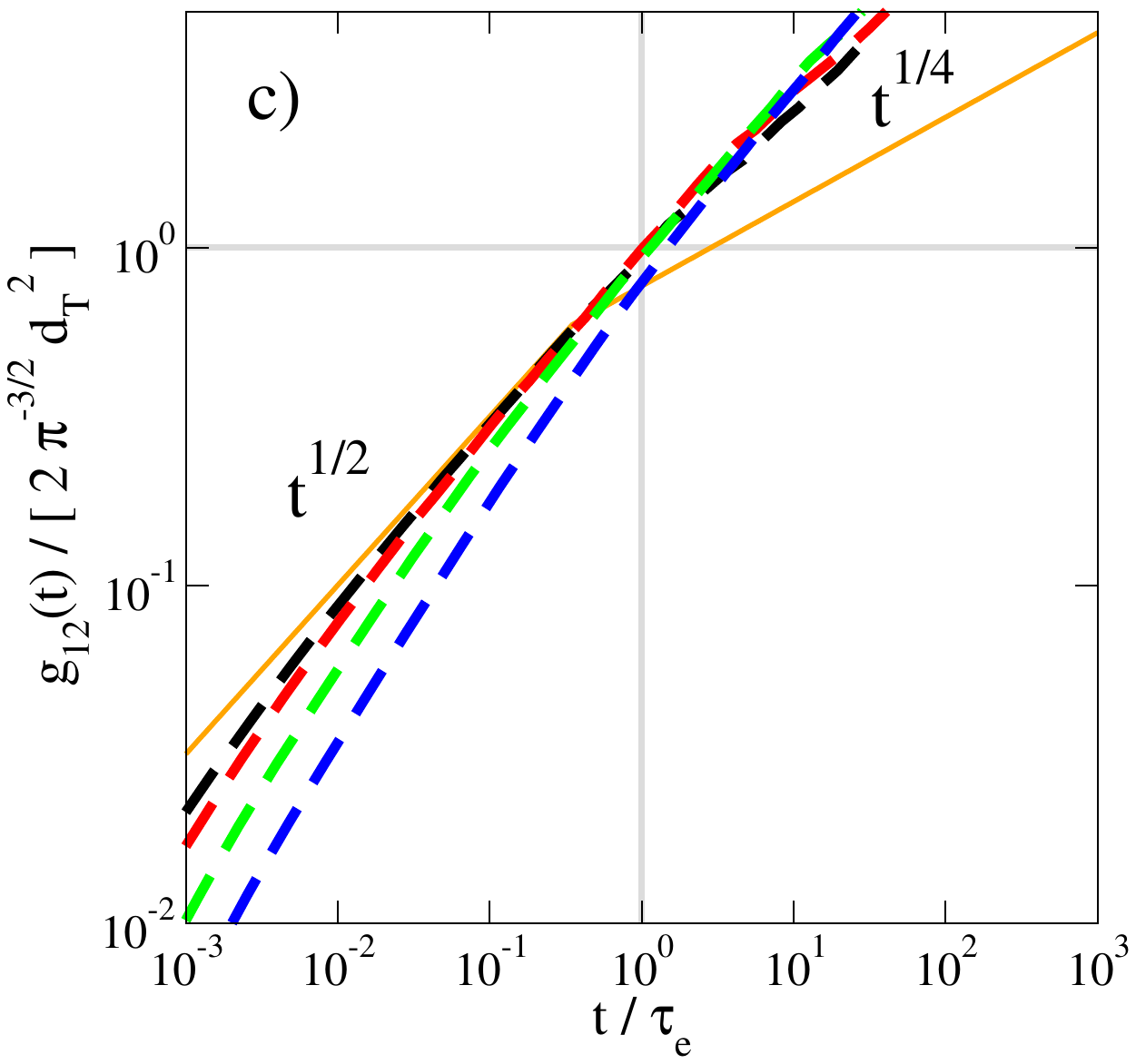}%
\includegraphics[angle=\Angle,width=0.3\columnwidth]{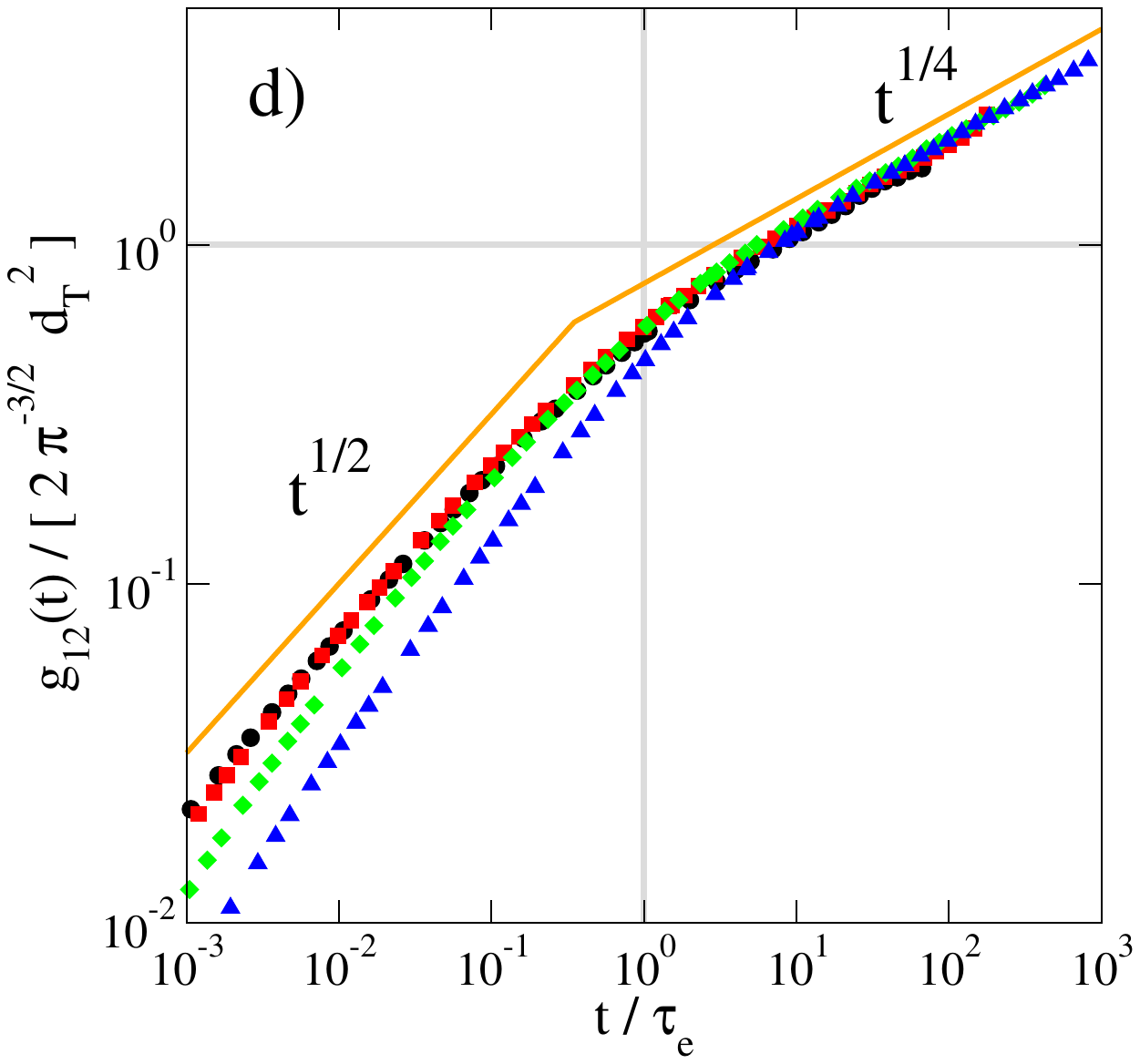}

\caption{\label{fig:g12KuhnTimeSpace}
Monomer motion of Phantom KG chains (l.h.s column, dashed lines) and in fully interacting KG melts (symbols, r.h.s. column) 
plotted in Kuhn units (top row) and entanglement units (bottom row).
Data in (abc) for $N_K=80$ systems as in Fig.~\ref{fig:g12}. Data in (d) for highly entangled chains with $N_b=10000$ beads.
Colors indicate the effective chain stiffness: $\kappa=-1$ (black), $\kappa=0$ (red), $\kappa=1$ (green), and $\kappa=2$ (blue). 
As before, we use dashed lines to indicate results for Phantom KG chains and symbols to show data for fully interacting KG melts.
The thick orange diagonal lines are parameter-free predictions from the Rouse and tube models, Eq.~(\ref{eq:entangledmsd}). 
See the main text for an explanation of the magenta line.
Panel (a): Following the logic of the tube model, we estimate the entanglement time by evaluating, when the monomer mean-square displacements in the topologically {\em unconstrained} Phantom KG chains reach the tube diameter (Eq.~(\ref{eq:taue definition}), $\kappa$-dependent colored horizontal and vertical lines). 
Panel (b): The monomer mean-square displacements in fully interacting KG melts starts to slow down well before reaching the entanglement scale.
Panel (c): In entanglement units, the motion of Phantom KG chains is {\em not} universal on the entanglement scale.
Panel (d): Beyond the entanglement scale, the monomer mean-square displacements in fully interacting KG melts approach the prediction of the tube model.
%
%estimation of entanglement time scale (y axis) based on (i) static PPA input for the tube diameter as the relevant length scale (x axis)
%with (ii) dynamic data for the monomer motion of Phantom KG chains with the $N_K=80$ systems from Fig.~\ref{fig:g12}, which replace the
%Rouse model as the reference system for the polymer dynamics in absence of topological constraints.  {\bf FIX}
}
\end{figure}

Does it really make sense to {\em define} quantities characterising the dynamics at the Kuhn scale by using expressions from the {\em continuum} Rouse model? 
Figure~\ref{fig:g12KuhnTimeSpace}a shows our results for the monomer diffusion of Phantom KG chains with $N_K=80$  from Fig.~\ref{fig:g12}. Instead of individual panels representing results for different system in the ``microscopic'' LJ units, we now use Kuhn units to present all data in a single plot. 
As expected, the behavior is non-universal for times up to the Kuhn time, while the different data sets almost perfectly coincide for $t\ge \tau_K$. 

The characteristic Rouse behavior of flexible chains (the solid orange line in Fig.~\ref{fig:g12KuhnTimeSpace}a) is only fully developed for $t>{\cal O}(100) \tau_K$. This has consequences, when one tries to associate a time scale, $\tau_x$, to a known spatial scale, $R_x$, or chain (contour) length, $N_x$, through the condition, 
\begin{equation}\label{eq:tauX estimator}
g_{12}(\tau_x) \equiv a_{\tau_x} \frac{2}{\pi^{3/2}}  R^2_x  \equiv a_{\tau_x} \frac{2}{\pi^{3/2}}  \left\langle R^2(N_{xK})\right\rangle\ ,
\end{equation}
that the monomer displacements exceed $R^2_x$ at $t=\tau_x$.
Eq.~(\ref{eq:tauX estimator}) and the prefactor $\frac{2}{\pi^{3/2}}$ are directly motivated by the Rouse expression, Eq.~(\ref{eq:rouse g2 asymptotic}), while $a_{\tau_x}$ is a fudge factor, which ideally reduces to $a_{\tau_x}=1$. As an example~\cite{degennes79}, consider the relaxation time of subchains of length $N_K/p$ with $p=2,3, \ldots$ and a mean-square end-to-end distance, $R^2_p=(N_K/p)l_K^2$. 
Evaluating Eq.~(\ref{eq:rouse g2 asymptotic}) for $a_{\tau_x}=1$ assuming Rouse dynamics, Eq.~(\ref{eq:rouse g2 asymptotic}),  yields $\tau_p = \tau_R/p^2$, i.e. the relaxation time of the $p$th Rouse mode. 
Inspection of Fig.~\ref{fig:g12KuhnTimeSpace}a shows, that this procedure can be safely applied for Phantom KG (sub)chains  composed of $N_{xK}>10$ Kuhn segments. 

What happens, if one tries to push this argument to the Kuhn scale as the absolute validity limit of the Rouse and Gaussian chain model?
Insisting on the Gaussian estimate $R^2_K=l_K^2$ entails the need to introduce a fudge factor of $a_{\tau_K}\approx0.5$ to recover an estimate of the Kuhn time from the $g_2(t)$ data, which agrees with our {\em definition}, Eq.~(\ref{eq:Kuhn-time}).
While this is acceptable on a scaling level, we have noted a surprisingly close relation between the statics and dynamics of Phantom KG chains. Considering $N_K$ as a continuous variable, we can invert Eq.~(\ref{eq:taurouse}) to identify the length of a chain, $N_{1K}(t) = \sqrt{t/\tau_K}$, whose relaxation time corresponds to a given time, $t$.  In a second step, we rewrite Eq.~(\ref{eq:g2 Rouse Kuhn units}) in the asymptotically correct form $g_2(t)= \frac{2}{\pi^{3/2}} \langle R^{2}(N_{1K}(t))\rangle$. In the final step, we use the wormlike chain expression, Eq.~(\ref{eq:R2 WLC}), to estimate the consequences of chain stiffness and finite extensibility. The resulting solid magenta line in Fig.~\ref{fig:g12KuhnTimeSpace}a is in much better agreement with the data than the asymptotic Rouse prediction. Without attempting a deeper analysis, we conclude that the Rouse relation for the equilibration length $l_1(t)$  holds in wormlike chains already around the persistence length. The dynamics of semiflexible chains on smaller scales is described by different relations \cite{everaers1999dynamic}, with inertial effects providing additional corrections in the present case.
In practical terms, we can use the intercept
\begin{equation}\label{eq:tauK estimator}
g_{12}(\tau_K^{(est)}) \equiv a_{\tau_K} \frac{2}{\pi^{3/2}}  \left\langle R^2(N_K=1)\right\rangle\ .
\end{equation}
as an estimator for the Kuhn time, where $a_{\tau_K}=0.82$ and $\langle R^2(N_K=1)\rangle / l^2_K = 0.57$ from Eq. (\ref{eq:R2 WLC}). In Figure~\ref{fig:g12KuhnTimeSpace}, the Kuhn time estimator is illustrated by the horizontal violet line.

Panel b of Figure~\ref{fig:g12KuhnTimeSpace} shows corresponding mean-square monomer displacement data for fully interacting chains in KG melts. Up to the Kuhn time, the results are nearly identical to those for the corresponding Phantom KG chains.
As illustrated in Fig.~\ref{fig:KuhnEntanglementtimes}, the estimates of $\tau_K$ based on melt data and Eq.~(\ref{eq:tauK estimator}) are in good agreement with Eq.~(\ref{eq:tauK KG}).
Entanglement effects restrict the time window, $\tau_K\ll t\ll \tau_e$, over which the Rouse model can be expected to apply.  In particular, topological constraints destroy the universality in the monomer motion beyond the Kuhn scales, because the number of Kuhn segments per entanglement length, $N_{eK}$, decreases with stiffness.
In practice, monomer mean-square displacements in the melt never attain the asymptotic Rouse regime, Eq.~(\ref{eq:g2 Rouse Kuhn units}), because entanglements significantly slow down the motion at the scale, where the corresponding Phantom KG chains start to exhibit Gaussian behavior. 
As a consequence, fitting the signature prediction, Eq.~(\ref{eq:rouse g2 asymptotic}), of the Rouse model to data in this range is bound to overestimate the friction. 
For flexible chains ($\kappa\le0$), the effect is weak. They show an extended effective $t^{1/2}$ regime, albeit with an amplitude which is only about $70\%$ of the expected value. 
Since $g_2(t) \propto \sqrt{t/\tau_K}$, corresponding estimates of $\zeta_b$, $\zeta_K$, and $\tau_K$ would exceed our values by a factor of two.
For stiffer chains, the Rouse regime is effectively suppressed by a broad crossover from semi-flexible to entangled dynamics, rendering the extraction of a bead friction delicate.

\subsection{Shear relaxation moduli of unentangled melts \label{sec:G(t) discussion}}

\begin{figure}
\includegraphics[angle=\Angle,width=0.45\columnwidth]{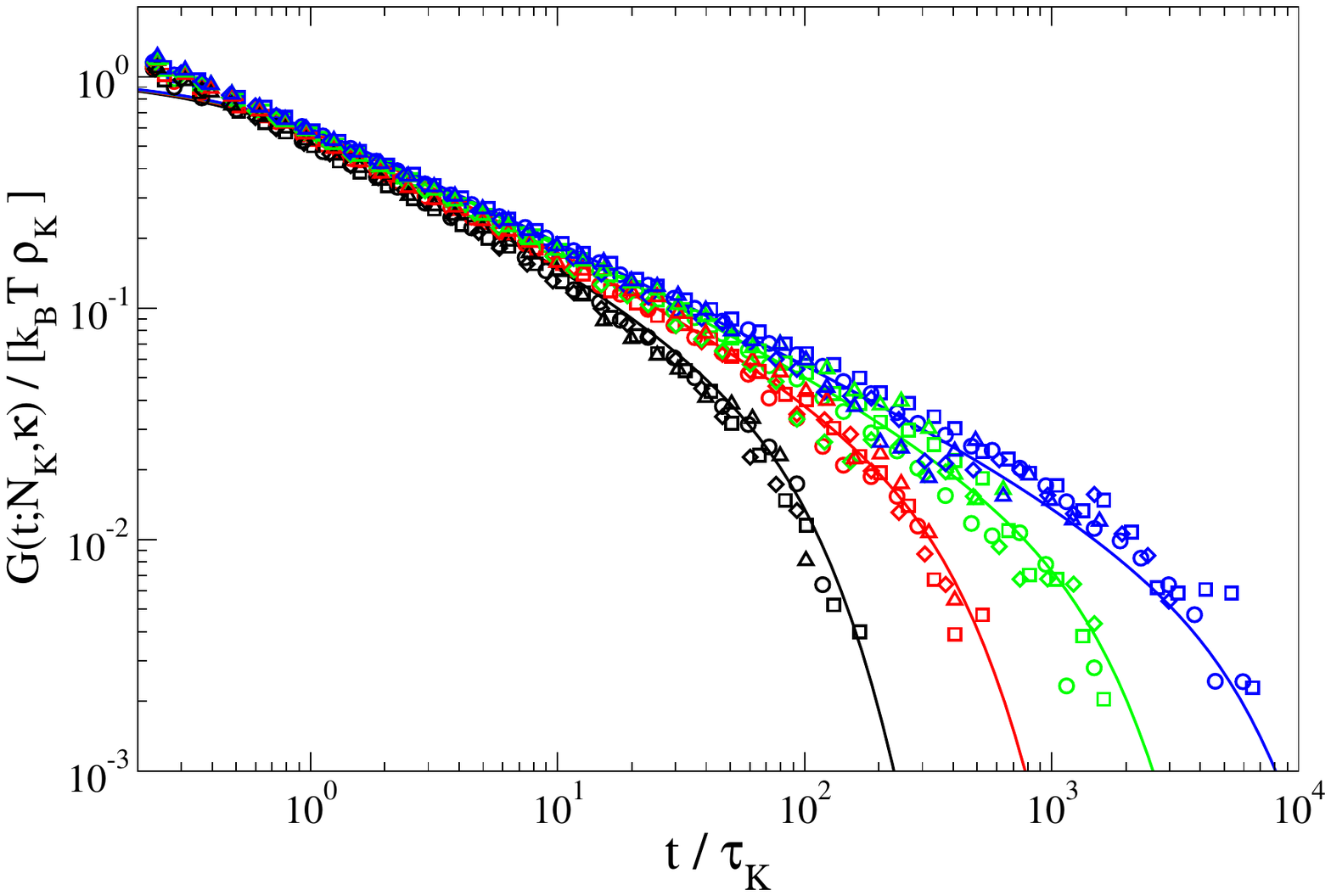}
\caption{\label{fig:gtkgphantom}
Shear-relaxation modulus for phantom KG melts with for stiffness $\kappa=-1, 0, 1,2$ (circle, box, diamond, triangle, respectively) for chain length $N_K=10,20,40,80$ (black, red, green, blue, respectively).  Shown are also the predictions of Rouse theory (lines with colours matching simulation data).
}
\end{figure}

\begin{figure}
\includegraphics[angle=\Angle,width=0.45\columnwidth]{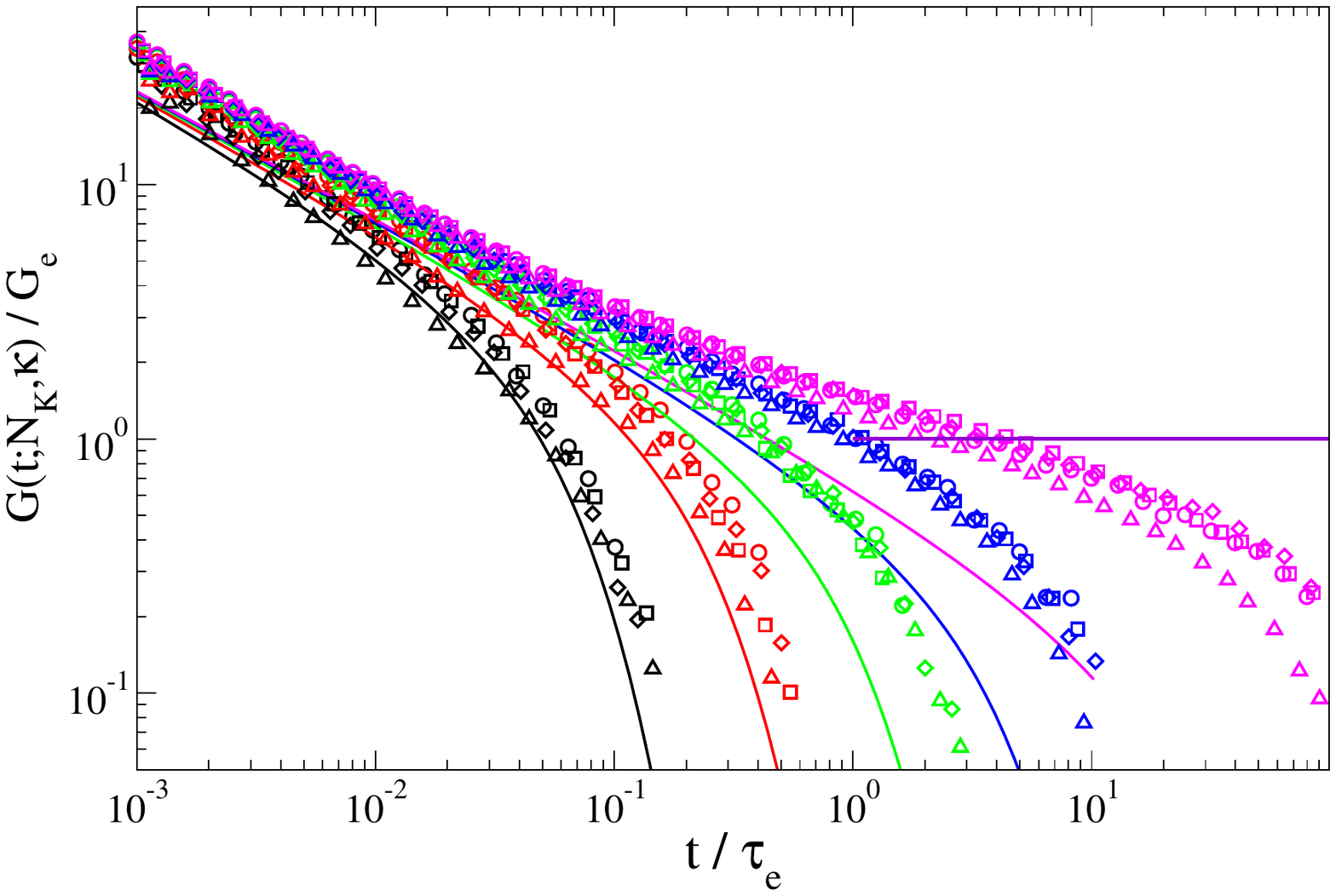}

\caption{\label{fig:gtkg}
Shear-relaxation modulus for KG melts with chain length $Z=0.25,0.5,1,2,5 $ (black, red, green, blue, magenta, respectively) for stiffness $\kappa=-1, 0, 1,2 $ (circle, box, diamond, triangle, respectively) and $N_{eK}(\kappa)=68, 39, 15, 5.2$. Shown are also the predictions of Rouse theory (lines with colours matching simulation data). The horizontal violet line illustrate the entanglement modulus.
}
\end{figure}

The second signature prediction of the Rouse model is the power law decay, Eq.~(\ref{eq:G(t) Rouse Kuhn units}), of the shear relaxation modulus, $G(t)$ \cite{DoiEdwards86}.  In Figure \ref{fig:gtkgphantom}, we show data in Kuhn units for phantom KG chains of different stiffness, $\kappa$, for times $t>\tau$, when high-frequency oscillations due to the bond length relaxation have died off~\cite{likhtman2007linear}.
We observe perfect data collapse for equivalent chain lengths, $N_K$, as well as excellent agreement with the predictions, Eq.~(\ref{eq:G(t) Rouse}), of the Rouse model.

Figure \ref{fig:gtkg} presents a comparison between the Rouse model and simulation data for fully interacting KG melts with $\kappa = -1,0,1,2$. Our results cover the crossover from unentangled to weakly entangled melts. 
In contrast to Fig.~\ref{fig:gtkgphantom} we present the data in entanglement units to show that the terminal relaxation time is governed by the {\em effective} chain length, $Z=0.25,0.5,1,2,5$.
The observed terminal relaxation times for the shortest chains are in good agreement with the Rouse model. The weakly entangled melts show a progressively delayed decay. 
As expected, data for different chain lengths for a given $\kappa$ collapse for short times. Furthermore, they are qualitatively compatible with a power law decay. However, the observed decay is somewhat steeper than predicted by the Rouse model and depends on $\kappa$.  
Likhtman {\em et al.}~\cite{likhtman2007linear} already reported this effect for the standard ($\kappa=0$) KG model and attributed their observation to glassy relaxation modes related to bead-bead interactions, which are not accounted for in the Rouse model. This interpretation seems compatible with our finding, that the absolute stress level is systematically {\em above} the prediction from the Rouse model. 
Contrary to the monomer diffusion, the absolute magnitude of the deviation for $G(t)$ seems to {\em decrease} with stiffness. 
Thus while there is qualitative agreement in the sense that both monomer motion and stress relaxation are slower than predicted by the Rouse model, there is little hope that both quantities could be simultaneously fitted with effective parameters. 

%{\bf Is this something we need to check more? Or plot some magic combination of $g_2(t)$ and $G(t)$, which becomes time-independent for the ideal Rouse model? This sounds a bit crazy, but I would not be surprised, if one could come up with something using inspiration from the analysis of microrheology experirments.}

%The observed terminal relaxation times for the shortest chains are in good agreement with the Rouse model. The weakly entangled melts show a progressively delayed decay. 
%{\bf Reads a bit thin. Replot data in entanglement units?}
%Should we include the LM prediction for infinite chain length into the plot? And did you try to integrate your $G(t)$ to get the viscosity? Can't remember.}

%{\bf In fact, if you look at Fig. 5 or 6 in the Colby, Fetters and Graessley 87 paper, then there are two points: already finding something that resembles the Rouse slope isn't trivial because of the chain-end corrections to density and mobility, which we shouldn't have. And then there is, of course, the crossover to the entangled regime. As in our Fig. 7 we'd have two panels: $\eta(N)$ and then something suitably rescaled in entanglement units. But I think that's for some other paper.}

\subsection{Entanglement length  \label{sec:NeK discussion}}

\begin{figure}
\includegraphics[angle=\Angle,width=0.5\columnwidth]{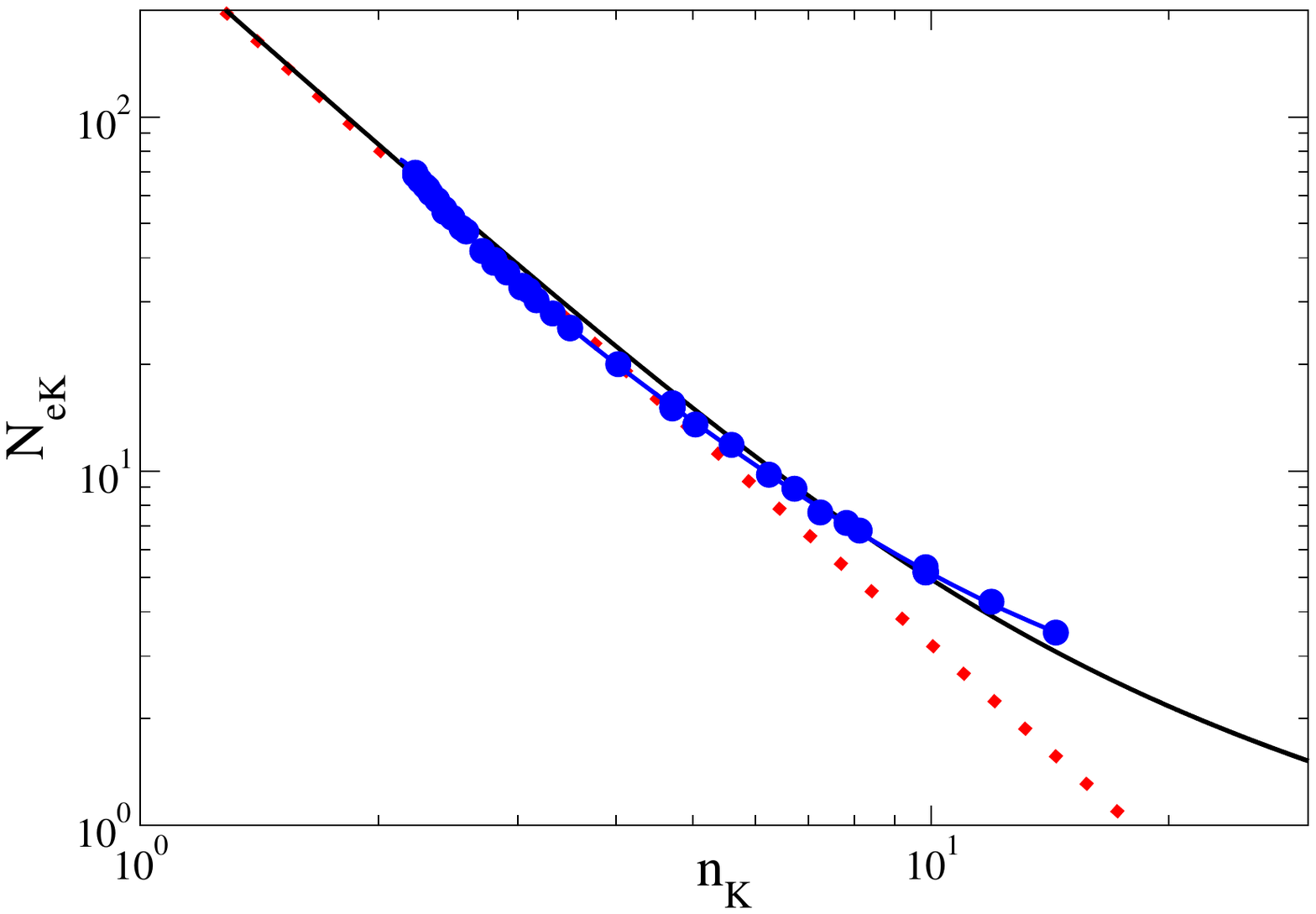}

\caption{\label{fig:nektheory}
Number of Kuhn units between entanglements for KG models with varying stiffness
(blue solid circles), empirical fit eq. (\ref{eq:nekinterpolation}) (thin blue line),
prediction for cross-over eq. (\ref{eq:nekuchida}) (solid black line),
and prediction in the flexible limit eq. (\ref{eq:nekflexible}) (red dashed line)
both with $\alpha=18$.
}
\end{figure}

Our PPA results for the entanglement length in KG melts (Fig.~\ref{fig:Entanglementlength}) can be summarized as

\begin{equation}\label{eq:nekinterpolation}
N_{eK}(\kappa)=
-0.84 \kappa^4+3.14 \kappa^3+3.69 \kappa^2-30.1 \kappa+39.3\ .
%39.130-30.237\kappa+ 4.2834\kappa^2+ 3.2065\kappa^3- 1.2879\kappa^4+ 0.13721\kappa^5\ .
\end{equation}
For the $\kappa$-range we have studied, the number of Kuhn segments per entanglement length decreases by a factor of 20 from $N_{eK}(\kappa=-1)=69$  to $N_{eK}(\kappa=2.5)=3.5$. This underlines the necessity of applying corrections to random walk theory when describing the chain statistics at the entanglement scale of stiffer polymers.

Following the packing argument~\cite{lin87,kavassalis87}, the entanglement length for loosely entangled chain with $N_{eK}\gg 1$ should be given by
\begin{equation}\label{eq:nekflexible}
N_{eK}=\frac{\alpha^2}{n_K^{2}}
\end{equation}
where experiments~\cite{fetters94} and a simple geometrical argument for binary entanglements~\cite{rosa2014ring} suggest $\alpha=19\pm2$. The parameter $\alpha$ can be interpreted as the number of entanglement strands per entanglement volume. 
Uchida {\it et al.}~\cite{uchida2008viscoelasticity} developed a scaling theory to describe the crossover to the tightly entangled regime, suggesting instead
\begin{equation}\label{eq:nekuchida}
N_{eK}=x^\frac{2}{5}\left( 1+ x^\frac{2}{5} +x^2 \right)^{\frac{4}{5}}
\quad\mbox{with}\quad x=\frac{\alpha}{n_K}
\end{equation}

Fig. \ref{fig:nektheory} shows the variation of the number of
Kuhn segments between entanglements as function of the Kuhn number. As
expected from Fig. \ref{fig:Entanglementlength} the number of Kuhn units between
entanglements drops as chains become stiffer, i.e. the Kuhn number grows. 
We observe good agreement between all our simulation results and the prediction
of the Uchida theory eq. (\ref{eq:nekuchida}). For $n_K<4$ the approximation
for flexible chains eq. (\ref{eq:nekflexible}) is also in good agreement
with our simulation data. 
This suggests that we could, in fact, replace our empirical formula eq.
(\ref{eq:nekinterpolation}) with the theoretically motivated extrapolation
eq. (\ref{eq:nekuchida}). However, as can be seen from the figure, this will
be less accurate description of our data for the melts with the most
flexible chains. 

%Perhaps by forcing the KG
%model to be more flexible using a negative angular bending stiffness to
%reduce the effect of the repulsive pair interaction between next-nearest
%beads, locally we are forcing the chains adapt zig-zag conformations which
%could cause deviations from random walk statistics. 

\subsection{Entanglement time \label{sec:taue discussion}}

The primitive path analysis~\cite{PPA} promises to endow the tube model with predictive power by inferring the entanglement {\em length}, $N_{eK}$.
Assuming that the Rouse-like longitudinal dynamics is governed by the same effective bead friction, we can directly read off the entanglement time from the standard relation~\cite{DoiEdwards86}, $\tau_e=N_{eK}^2\tau_K$. 
As a consistency check, we can estimate the entanglement time through the relation\footnote{In Eq.~(\ref{eq:taue definition}) we have neglected the analogue,  $a_{\tau_e}$, of the correction factor in Eq.~(\ref{eq:tauK estimator}) for the Kuhn time, because Phantom KGchains are nearly Gaussian on the entanglement scale (Fig.~\ref{fig:g12KuhnTimeSpace}). Note that when this approximation breaks down, one also needs to reconsider the definition of the tube diameter~\cite{odijk1983statistics}.}
\begin{equation}\label{eq:taue definition}
\lim_{N_K\rightarrow\infty} g^{phantom}_2(\tau_e^{(est)}) \equiv \frac2{\pi^{3/2}}  \langle R^2(N_{eK}) \rangle\ .
\end{equation}
where $d_T^2 = \langle R^2(N_{eK}) \rangle$ ought to be a good approximation for the tube diameter of loosely entangled chains. 
Figure~\ref{fig:g12KuhnTimeSpace}a illustrates the application of Eq.~(\ref{eq:taue definition}) to data for the monomer motion of Phantom KG chains,  which we use as a reference for the dynamics of topologically {\em un}constrained chains.
The entanglement scale is indicated by the combination of 
(i) horizontal lines representing the spatial scale inferred from the primitive-path analysis and the single-chain statistics and
(ii) vertical lines marking the temporal scale inferred from Eq.~(\ref{eq:taue definition}).
Following the discussion in Sec.~\ref{sec:g2 at Kuhn scale discussion}, it is not surprising that the results are in almost perfect agreement with  the standard relation, $\tau_e=N_{eK}^2\tau_K$ (Fig. \ref{fig:KuhnEntanglementtimes}).
Note that with $N_{eK}\sim (\sigma/l_K)^{4}$ predicted by the packing argument, the ratio $\tau_e/\tau_K \sim (\sigma/l_K)^{8}$ decreases extremely rapidly when considered as a function of Kuhn length.
%While our values of $\tau_e$ are {\em predictions} based on PPA results and the dynamics of Phantom KG chains, Fig.~\ref{fig:KuhnEntanglementtimes} also contains literature values inferred from the {\em observed} melt dynamics. 

\subsection{Chain dynamics around the entanglement scale  \label{sec:g2 at entanglement scale discussion}}

%\begin{figure}
%\includegraphics[angle=270,width=0.5\columnwidth]{msd_approx.ps}

%\caption{\label{fig:msdapprox}Comparison between theoretical prediction
%of mean-square displacements with prefactor $\delta_\tau=1$ (solid black lines)
%and simulation results for (from top to bottom) entangled melts with
%$Z=10$ (filled circles), $Z=20$ (filled boxes), and $Z>80$,$N_b=10000$
%(filled diamond) for $\kappa=-1,0,1,2$ (black, red, green, blue,
%respectively). Each system has been shifted a factor of $5$ for increased
%clarity. For the moderately entangled melts, mean-square displacements were
%obtained using only internal beads to avoid effects due to increased
%mobility of the chain ends.  Purple dots denotes the cross-over times
%$\tau_K$,$\tau_e$,$\tau_R$, and $\tau_{max}$ for the different melts.
%}
%\end{figure}
%
\begin{figure}
\includegraphics[angle=\Angle,width=0.95\columnwidth]{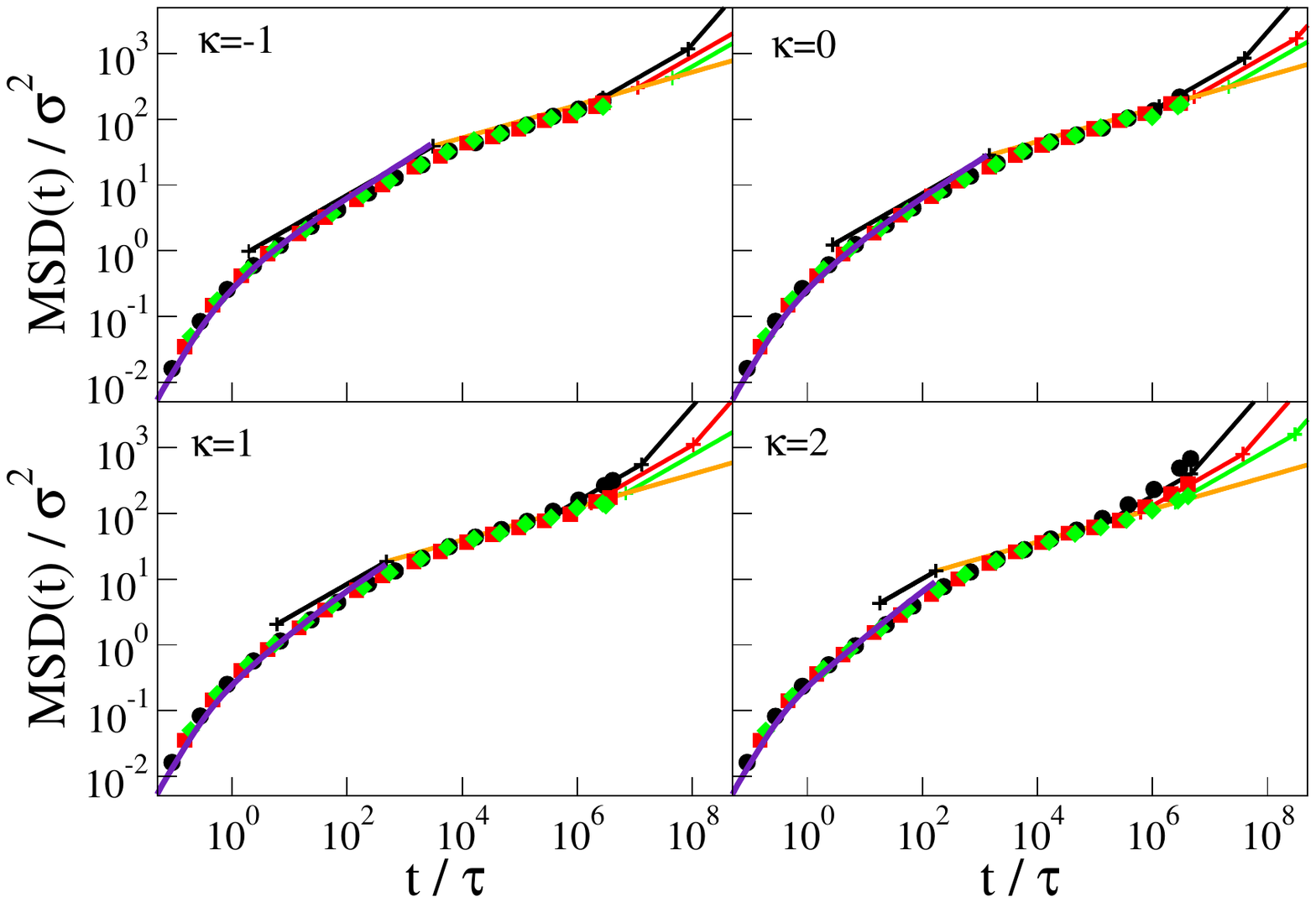}

\caption{\label{fig:msd_g1} 
Monomer mean-square displacement, $g_1(t)$: Comparison between simulation data for KG melts and
parameter-free predictions of the tube model, Eqs.~(\ref{eq:entangledmsd}). MSDs were sampled
melts of $Z=10,20,40$ (black, red, green symbols, respectively) for the middle $25\%$ of the chains
for $\kappa=-1,0,1,2$. $+$ symbols denotes the cross-over times $\tau_K$,$\tau_e^*$,$\tau_R^*$, and
$\tau_{max}$, respectively, for the different melts. $g_1(t)$ sampled for the phantom KG melts is
shown up to $\tau_e^*$ (violet line).
}
\end{figure}

\begin{figure}
\includegraphics[angle=\Angle,width=0.95\columnwidth]{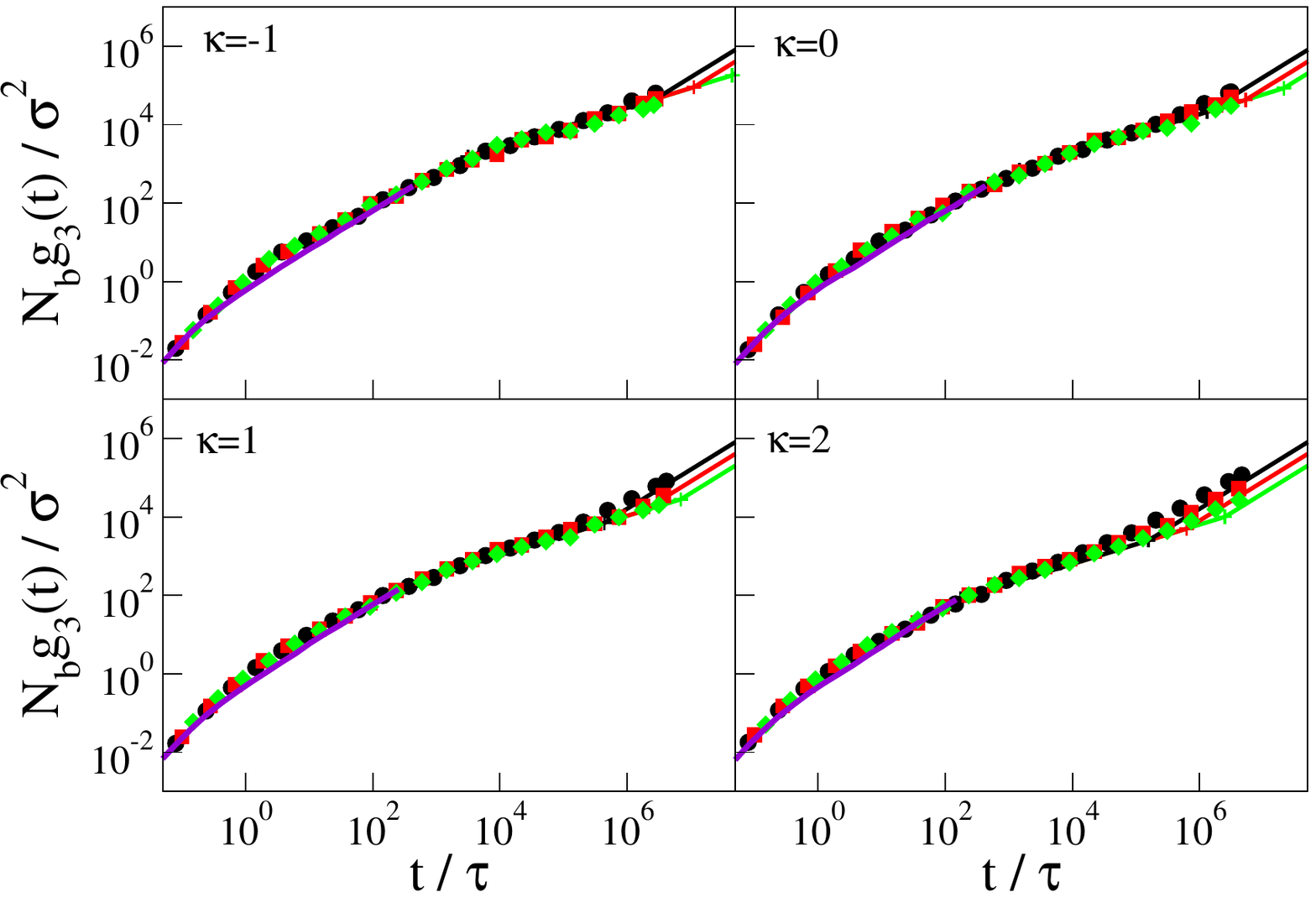}

\caption{\label{fig:msd_g3} 
CM mean-square displacement, $g_3(t)$: Comparison between simulation data for KG melts
and parameter-free predictions of the tube model, Eqs.~(\ref{eq:entangledg3}). See Fig. \protect\ref{fig:msd_g1} for
the meaning of symbols and lines.
}
\end{figure}

\begin{figure}
\includegraphics[angle=\Angle,width=0.5\columnwidth]{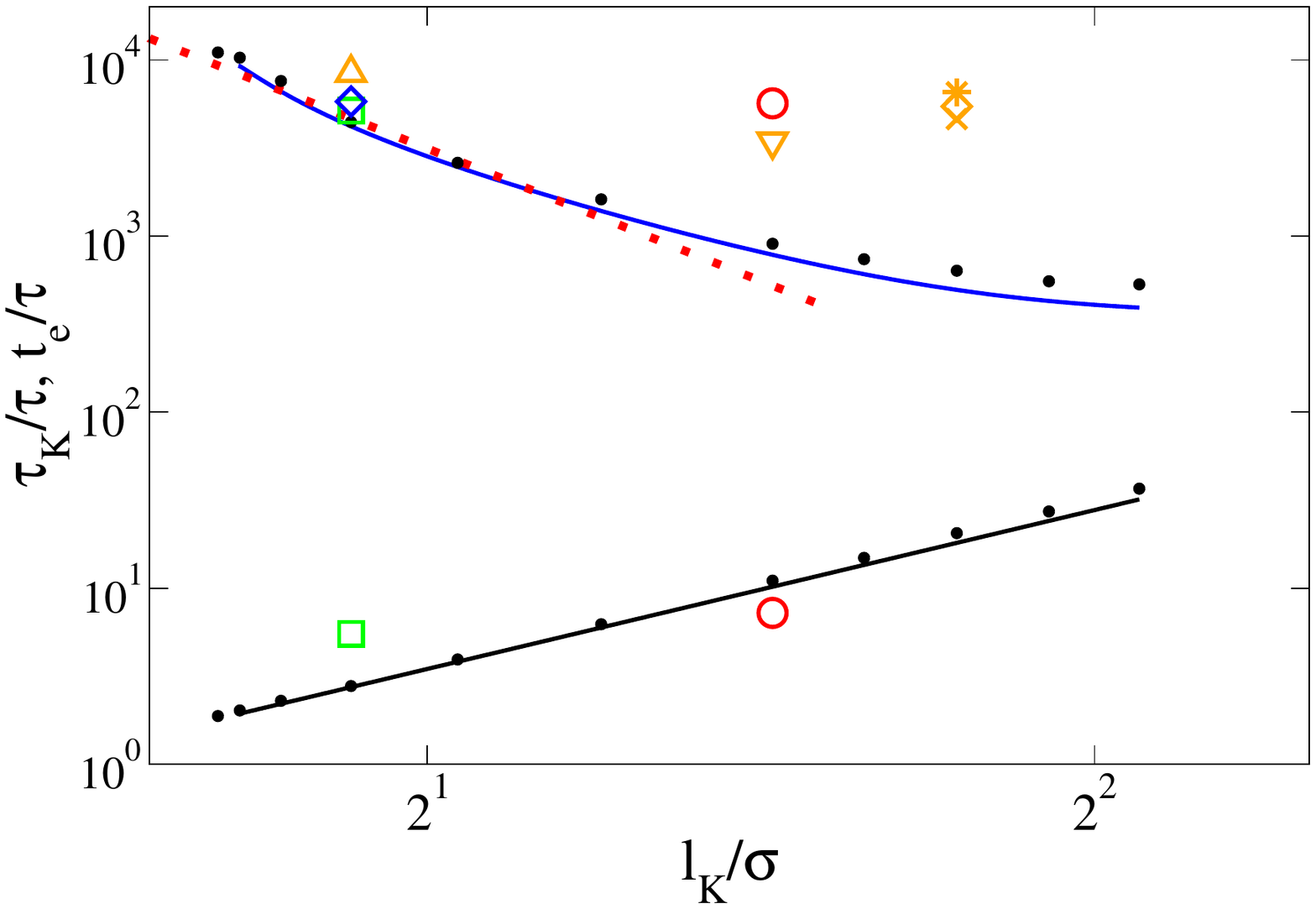}
\caption{\label{fig:KuhnEntanglementtimes}
Characteristic times as functions of the Kuhn length.
Lines denote the estimated Kuhn time using Eq. (\ref{eq:tauK KG}) black line),
estimated entanglement time $\tau_e=\tau_K N_{eK}^2$ using Eqs. (\ref{eq:tauK KG}, \ref{eq:nekinterpolation})
(blue line), and estimated using the packing entanglement length Eqs. (\ref{eq:tauK KG}, \ref{eq:nekflexible}) (red dotted line).
Small filled black circles denotes independent estimations $\tau_K^{(est)}$ and $\tau_e^{(est)}$ using Phantom KG simulations 
Eqs. (\ref{eq:tauK estimator}, \ref{eq:taue definition}).
Big colored symbols denote literature values for the Kuhn and entanglement times from Kremer and Grest\cite{kremer1990dynamics} (green boxes), the entanglement time estimated by Likhtmann et al.\cite{likhtman2007linear}
(blue diamond), the Kuhn and entanglement times estimated Hsu et al.\cite{hsu2016static}
(red circles), and the entanglement times estimated by the reanalysis of cross-over
positions in literature data for $\kappa=0$\cite{wang2012segmental}, $\kappa=1.5$\cite{hsu2016static}, and $\kappa=2.0$
\cite{zhou2006direct,wang2012segmental,wang2008constraint} by Hou\cite{hou2019determine} (orange triangle up, triangle down, 
cross, diamond, star, respectively). (see text for discussion)
}
\end{figure}

Are our entanglement times, which are {\em predictions} based on PPA results and the dynamics of Phantom KG chains, relevant to the actual melt dynamics?
A first piece of evidence is provided by the shear relaxation moduli of weakly entangled melts, Fig.~\ref{fig:gtkg}.
Figure~\ref{fig:msd_g1} shows our results for the monomer diffusion, $g_1(t)$, sampled for beads in the central half of the chain
in entangled KG melts with $Z=10,20,40$.
Corresponding results for the CM diffusion, $g_3(t)$, in the same systems are shown in Fig.~\ref{fig:msd_g3}.
The data in these figures correspond to about $103$ core years of simulation effort.
As previously observed~\cite{kremer1990dynamics,puetz00a,hsu2016static}, 
%{\bf what about FMP? The discussion is on the observation of $t^{1/4}$, where these are the right references.. Yes. But that's the standard model. Maybe FMP showed this behavior for the present {\em family} of wormlike KG models?! We can't be the first to have looked at this.}, 
KG melts show convincing evidence for the initial $t^{1/4}$ and $t^{1/2}$-regimes predicted by the tube model for the monomer and the CM motion. 
The key point to note here is the good quantitative agreement between our data and the predictions, Eqs.~(\ref{eq:entangledmsd}) and Eqs.~(\ref{eq:entangledg3}), of the tube model. 
We emphasize, that this comparison constitutes a parameter-free test of the tube model and provides additional evidence for the power of the primitive path analysis~\cite{PPA} of the microscopic topological state. 
Importantly, the model provides a remarkably consistent description of the chain dynamics using a (PPA) value of $N_{eK}$, which has already been shown to quantitatively predict the plateau modulus of KG melts~\cite{hou2010stress,hsu2016static} and which corresponds to the experimentally defined {\em rheological} entanglement length~\cite{everaers2012topological}.
%{\bf I thought this was worth stressing. Obviously, anybody can throw in a prefactor and define a $\tilde\tau_e = a \tau_e$. Nothing wrong with that, as long as Eq.~(\ref{eq:entangledmsd}) and all other predictions of the tube model are rewritten accordingly. Problems only arise, if one uses a $\tilde\tau_e$ defined in some way in formulas based on a different definition. But then there is THE experimental convention for defining the rheological entanglement length. And that's what we use.}

However, no model is perfect.
In Fig.~\ref{fig:g12KuhnTimeSpace} we (re)plot  $g_1(t)$  in entanglement units to analyse the $t^{1/4}$-regime and the crossover at the entanglement scale in more detail. 
Panel (c) shows data for our Phantom KG reference chains. As expected from Eq.~(\ref{eq:taue definition}),  all $g_1(t)$ cross the tube scale as at $\tau_e$ exhibiting (almost) perfect Rouse behaviour.
%\Comment{In addition, we have evaluated Eq.~(\ref{eq:projection}) for the Phantom KG dynamics to estimate finite stiffness corrections to the  $t^{1/4}$-regime. Interesting but future..}
Panel (d) shows results for highly entangled chains with $N_b=10000$ monomers and $Z>80$, which exhibit a broad and universal $t^{1/4}$-regime between $\tau_e$ and $\tau_R = Z^2 \tau_e$, confirming once more the internal consistency of our choices for the $\kappa$-dependent time- and length scales. At the same time, there are notable deviations from Eq.~(\ref{eq:entangledmsd}). 

Literature values of $\tau_e$ are typically inferred from the intercept~\cite{kremer1990dynamics}  between an (effective) $t^{1/2}$ Rouse regime and the $t^{1/4}$ regime in the monomer diffusion data like those shown in Fig.~\ref{fig:g12KuhnTimeSpace}d. Kremer and co-workers~\cite{kremer1990dynamics,puetz00a,hsu2016static} identified $\tau_e$ on a scaling level with the position of the crossover, $\tau_e=\tau_e^\ast$.  Likhtman and McLeish~\cite{likhtman2002quantitative} started to take into consideration prefactors and suggested $\tau_e= (36/\pi^3) \tau_e^\ast$. We subscribe to Hou's analysis~\cite{hou2017note,hou2019determine}, $\tau_e= (9/\pi) \tau_e^\ast$, which follows from Eq.~(\ref{eq:projection}).
Fig.~\ref{fig:g12KuhnTimeSpace}d points to the additional difficulty, that the crossover is not universal.  Rather we encounter the mirror image of the effect discussed in Sec.~\ref{sec:g2 at Kuhn scale discussion}: The stiffer the chains, the more $g_1(t)$ lags behind the expectation from the continuum Rouse model, and the larger the apparent $\tau_e^\ast$.
%\Comment{ Do you want to put in some colored vertical bars for estimated $\tau_e^\ast$ to mark the point? Not sure that I would want to be very precise. Might as well do this by eye. Please note, that I have ASSEMBLED the following sentences from the text, but if have NOT checked anything.}

For the standard KG model with $\kappa=0$, Kremer and Grest\cite{kremer90} reported a crossover of $\tau_e^*=1800\tau$ in our interpretation, which corresponds to $\tau_e=5160\tau$ when taking the prefactors into account. Similarly, Wang et al.\cite{wang2012segmental} reported $\tau_e^*=2950\tau$, which correspond to $\tau_e=8450\tau$. Finally, Likhtmann\cite{likhtman2007linear} reported $\tau_e(\kappa=0)\approx 5800\tau$ from a direct comparison of KG simulation data to a slip-spring model, where no prefactor corrections should be required since the slip-spring model generates the power-law regimes and cross-overs independent of theory.
After taking prefactors into account, these literature results are roughly consistent, since estimating the precise cross-over position accurately is difficult.
%\Comment{If Alexei compared slip-spring to theory, he should have confirmed the shifts of the cross-overs?? } 
For comparison our entanglement time is estimated by combining primitive path analysis to obtain $N_{eK}$ with our estimate of the bead friction to obtain $\tau_e=4200\tau$ compared to $\tau_e^{(est)}=4440\tau$ which is the time it takes the phantom KG mean-square displacements to reach the tube diameter. 
%\Comment{We can not claim these to be completely independent since the KG phantom simulations relies on the KG bead friction estiamte.}

For KG models of $\kappa=1.5$ and $2.0$, all literature estimates shown in Fig. \ref{fig:KuhnEntanglementtimes} are
based on estimation of cross-over times. 
For $\kappa=1.5$, Hsu et al. reported $\tau_e^*=1980\tau$ while in a reanalysis of their data Hou~\cite{hou2019determine} estimated the cross-over at $\tau_e^*=1200\tau$. This corresponds to $\tau_e=5670\tau$ and $3440\tau$, respectively.
Our analysis would suggest $\tau_e=780\tau$ and $\tau_e^{(est)}=905\tau$.
For $\kappa=2.0$, Hou~\cite{hou2019determine} reanalyzed literature data~\cite{zhou2006direct,wang2012segmental,wang2008constraint} to find
$\tau_e^*=1610, 1900, 2290\tau$, respectively. This corresponds to $\tau_e=4610, 5440, 6560\tau$, when taking into
account the prefactor.
For comparison, our analysis would suggest $\tau_e=490\tau$ and $\tau_e^{(est)}=635\tau$.
The reported values are an order of magnitude larger than our estimates of the entanglement time.

We note that there is an order of magnitude time shift between the dynamics of the flexible and stiff KG melts shown in Fig.~\ref{fig:g12KuhnTimeSpace}d. Hence if we fitted the theoretical prediction our simulation data we would also
over estimate the entanglement times. This suggests that it is in fact impossible to use mean-square internal distances
alone as a basis for estimating the entanglement time of stiff polymer melts.

As a second notable deviation, we find only about 80\% of the expected value for the amplitude of the motion in the $t^{1/4}$-regime, Eq.~(\ref{eq:entangledmsd}), which in turn depends on the absolute values of our target characteristic time and length scales: Kuhn length, $l_K$, the number of Kuhn segments per entanglement length, $N_{eK}$, and the Kuhn and entanglement time, $\tau_K$ and $\tau_e$. 
According to Eq.~(\ref{eq:projection}) this deviation could be explained by
(i) a systematic overestimation of $a_{pp}$ by 20\% and of $N_{eK}$ by a factor of $1.2^2 \approx 1.4$ by the primitive path analysis,
(ii) an effective bead friction for the {\em longitudinal} motion along the tube exceeding our estimate by a factor of $1.2^4\approx2$, or
(iii) a combination of both effects.
Given the excellent agreement with the measured plateau moduli~\cite{hou2010stress,hsu2016static}, (i) seems unlikely,
while a comparable discrepancy between the effective bead frictions for short and long chains was already reported by Kremer and Grest~\cite{kremer1990dynamics,puetz00a} for the standard KG model with $\kappa=0$. 
In our understanding~\cite{ahlrichs2001screening},
a universal effect seems at least qualitatively conceivable, if topological constraints and the screening of hydrodynamic interactions both set in on the same scale. However, we can only speculate, if this idea could be explored in the framework of Refs.~\cite{Semenov2011PRL,Semenov2012JPhys,Semenov2012PRE2,Semenov2012PRE}.

Importantly, there is also evidence to the contrary, because the effective bead friction for the longitudinal motion directly affects the expected crossover time, $\tau_R^\ast = \pi \tau_R$, from the $t^{1/4}$ to the second $t^{1/2}$-regime {\em independently} of the tube geometry.
While our data are insufficient to address this point, we note that Hsu and Kremer\cite{hsu2016static} reported a corresponding crossover time of
$\tau_R(N_b=500,\kappa=1.5)=6.44\times10^5\tau$. Identifying their observation with $\tau_R^\ast$ leads to to an effective bead friction of $\zeta_b=8.7 m_b\tau^{-1}$. Which is also consistent with the reported cross-over of Wang et al.\cite{wang2012segmental} of $\tau_R(n_b=350,\kappa=2) = 3.0\times 10^5\tau$. Identifying their observation with $\tau_R^\ast$ leads to an effective bead friction of $\zeta_b = 6.9 m_b \tau^{-1}$.
A more accurate estimate appears
to be based on directly mapping simulation results to to slip-spring simulations\cite{likhtman2007linear,wang2012segmental}, where the prediction of tube theory Eq. (\ref{eq:entangledmsd}) emerges directly from the simulation results.

\section{Summary and Conclusion\label{sec:Conclusion}}

%systems and overview
In this paper, we have presented a detailed analysis of the behavior of Kremer-Grest~\cite{grest1986molecular,kremer1990dynamics} bead-spring polymer melts as a function of an additional wormlike local chain bending stiffness~\cite{faller1999local,faller2000local,faller2001chain}.
With $-1\le \kappa \le 2.5$ we have varied the stiffness parameter over the relevant range for experimentally available commodity polymer melts~\cite{svaneborgKGmapping}.
The focus of the present work lies on the characteristic time and length scales governing the melt behavior.
In particular, we have (i) obtained reliable interpolations the stiffness dependence of the Kuhn length, $l_K$, the Kuhn friction, $\zeta_K$, the Kuhn time, $\tau_K$, the entanglement length, $N_{eK}$, in units of Kuhn steps, and the entanglement time, $\tau_e$, (ii) shown that these scales are relevant to dynamical observables of KG melts, which are independent from the static and short chain dynamic input data of our determination of the characteristic scales, and (iii) presented parameter-free tests of the Rouse and tube models of polymer dynamics for chains of different stiffness. The total numerical effort for the presented results exceeds $175$ core years.

%explain Characterization

%
To estimate the Kuhn length and the effective friction at the bead or Kuhn scale, we have generated brute force equilibrated ``gold standard'' data.
To obtain the Kuhn length, $l_K$, we analyzed the chain statistics using an estimator, which is sensitive to the slowly converging, large-range bond orientation correlations due to incompressibility effects~\cite{Wittmer_Meyer_PRL04,wittmer2007polymer,wittmer2007intramolecular,beckrich2007intramolecular}.
To avoid finite chain length effects in the estimation of the entanglement length, $N_{eK}$, we have generated very large, highly entangled melts  using a multi-scale equilibration method~\cite{SvaneborgEquilibration2016}. 
In one set of systems composed of $500$ chains of $N_b=10000$ beads the number of entanglements per chains varies between $Z>85$ for the most flexible chains and  $Z=570$ for the stiffest systems.  
As a complement we further generated melts of flexible chains with $1000$ chains with constant number of entanglements $Z=200$ to have highly entangled melts covering the entire range of chain stiffness.
To infer  $N_{eK}$ from the primitive path mesh we used an
% \Comment{I cut out ``new'', because I must have done something like this in the Uchida paper, where Nariya generated data for semi-flexible chains}
 estimator inspired by Ref.~\cite{uchida2008viscoelasticity}, which accounts for the effects of finite chain stiffness. 
For the stiffest investigated systems, we observe small, but notable deviations from the packing prediction, $N_{eK} \sim l_K^{-4}$ and excellent agreement with the predictions from Ref.~\cite{uchida2008viscoelasticity}. 
With respect to dynamics, we started by inferring the effective bead friction, $\zeta_b$,  from our ``gold standard'' data for the long-time CM diffusion in unentangled and weekly entangled melts. The key for validating this estimate was the comparison to the dynamics of Phantom KG chains, which reproduce the stiffness-, inertia- and discretization-induced corrections to the ideal Rouse dynamics. Remarkably, our data are compatible with a chain length and stiffness {\em independent} value of $\zeta_b$. 
Given this value, the Kuhn and entanglement times follow directly from our definition, Eq.~(\ref{eq:Kuhn-time}), and the standard, Rouse relation $\tau_e = \tau_K N_{eK}^2$. 
The two time scales are expected to be well separated in loosely entangled systems with $N_{eK}\gg1$. However, over the experimentally relevant range of effective bending stiffness~\cite{svaneborgKGmapping}, this gap closes rapidly: $\tau_K \sim l_K^3$ grows quickly with the Kuhn length, while $\tau_e = N_{eK}^2 \tau_K \sim l_K^{-5}$ in the initial packing regime. 

To see, if the identified scales are relevant to the dynamics of KG melts, we have studied the shear relaxation moduli of
our weakly entangled systems with $Z=0.25,0.5,1,2,5$ as well the chain dynamics in
moderately entangled melts with $Z=10,20,40$ up to the Rouse time. 
In all cases, we find very good data collapse, when data for KG chains with different stiffness $\kappa$ are represented in Kuhn or entanglement units. 
This validates our results for the characteristic length and time scale in KG melts on a scaling level as being coherent across the studied range of chain stiffness. 

The careful characterization of model polymer systems at the Kuhn scale allows us to perform {\em parameter-free} tests of the predictions of the Rouse and tube models of polymer dynamics.  Given data this is as simple as directly inserting the $\kappa$-dependent length and time scales into the relevant text book expressions.
The Rouse model~\cite{rouse1953theory} for polymer melts~\cite{DoiEdwards86} seems to be a perfect example for G. Box'  aphorism\cite{BoxQuote}, that all models are wrong. 
We see effects of inertia, correlation hole and viscoelastic hydrodynamic effects~\cite{Wittmer_Meyer_PRL04,wittmer2007polymer,wittmer2007intramolecular,beckrich2007intramolecular}. Local liquid-like packing of the beads with remnants of glassy modes~\cite{likhtman2007linear}, chain stiffness and finite extensibility. The crossover to the Gaussian behavior underlying the Rouse model is completed only for times beyond ${\cal O}(100)\tau_K$, corresponding to the Rouse time of a chain of $N_K={\cal O}(10)$ Kuhn segments. This is to be compared to the entanglement lengths, $N_{eK} \in [4,100]$, and the observation that topological constraints affect the chain dynamics well below the tube scale. As a consequence, we have not found a single observable, which {\em quantitatively} follows the predictions of the Rouse model to an extent that would allow us to recover the bead friction from a direct fit of the theory to our data. 
And yet the model is extremely useful\cite{box1979robustness}
due to its simplicity, the insight generated by its qualitatively correct description of the dynamics of unentangled polymers, and for being the starting point for systematic improvements. 
In particular, the model describes the longitudinal dynamics of entangled polymers within the tube, which occurs on scales, where the underlying Gaussian chain model quantitatively applies. 
Remarkably, the tube model almost quantitatively predicts the amplitude of the characteristic $t^{1/4}$-regime in the monomer diffusion given PPA input for the entanglement length and ``bead frictions'' essentially inferred from the center-of-mass diffusion of KG pentamers.
Studies with longer runs for longer chains are required to confirm the small remaining discrepancy in the amplitude of the $t^{1/4}$-regime. In particular, it would be interesting to see, if this amplitude and the crossover to the subsequent reptation regime can be described with a consistent bead friction.

\begin{acknowledgement}
The  simulations  in  this  paper  were  carried  out  using  the
LAMMPS molecular dynamics software.\cite{plimpton1995fast}
Computation/simulation for the work described in this paper was supported
by the DeiC National HPC Center, University of Southern Denmark, Denmark. 

\end{acknowledgement}

\pagebreak

\providecommand{\latin}[1]{#1}
\makeatletter
\providecommand{\doi}
  {\begingroup\let\do\@makeother\dospecials
  \catcode`\{=1 \catcode`\}=2 \doi@aux}
\providecommand{\doi@aux}[1]{\endgroup\texttt{#1}}
\makeatother
\providecommand*\mcitethebibliography{\thebibliography}
\csname @ifundefined\endcsname{endmcitethebibliography}
  {\let\endmcitethebibliography\endthebibliography}{}

\end{document}